\pdfoutput=1
\documentclass[3p]{elsarticle}	
\journal{\ }	

\usepackage{lineno}\modulolinenumbers[5] 
\usepackage[T1]{fontenc} 

\usepackage{algorithm}	
\usepackage[noend]{algorithmic}  
\usepackage{float}	
\newfloat{alg}{tb}{loa}		
\floatname{alg}{Algorithm}		

\usepackage{amsmath} 
\usepackage{amssymb} 
\usepackage{booktabs}
\usepackage{color}
\usepackage{enumitem}	

\usepackage[pdftex,colorlinks,allcolors=blue]{hyperref}	

\newcommand{\bsy}[1]{\boldsymbol{#1}} 
\DeclareMathOperator{\cond}{cond}		
\newcommand{\dbldag}{\raisebox{1pt}{$\sst\dag\!\dag$}} 
\newcommand{\diag}{\mbox{\rm diag}}
\newcommand{\divides}{\,|\,}    
\newcommand{\dividesnot}{\,\notmid\,}    
\newcommand{\El}{\ensuremath{E_{\ell}}}
\newcommand{\eqdef}{\ensuremath{\stackrel{\mbox{\tiny\textsf{def}}}{=}}}  

\newcommand{\Fl}{\ensuremath{F_{\ell}}}
\DeclareMathOperator{\liftsig}{lift\,sig}		
\newcommand{\lowergam}[1]{\ensuremath{\gamma_{\raisebox{-2pt}{$\ssst#1$}}}} 
\newcommand{\notmid}{\mbox{$\hspace{-1.5pt}\not\hspace{2.4pt}\mid\hspace{1.5pt}$}} 

\newcommand{\Psub}[1]{\mbox{$\mathbf{P}_{\! #1}$}}
\newcommand{\Rl}{\ensuremath{R_{\ell}}}
\DeclareMathOperator{\schema}{schema}

\DeclareMathOperator{\sig}{sig}

\newcommand{\ssst}{\scriptscriptstyle}
\newcommand{\sst}{\scriptstyle}

\newcommand{\upchi}{\raisebox{1pt}{$\chi$}} 
\newcommand{\updatechar}[2]{\ensuremath{\upchi_{\raisebox{#1}{#2}}}}	

\newcommand{\wt}[1]{\ensuremath{\widetilde{#1}}}
\newcommand{\zinv}{\mbox{$z^{-1}$}}

\newtheorem{thm}{Theorem}[section]
\newtheorem{cor}[thm]{Corollary}
\newtheorem{lem}[thm]{Lemma}

\newdefinition{defn}{Definition}[section]
\newdefinition{exmp}{Example}[section]
\newcommand{\rem}{\textbf{Remarks.\ }} 

\begin{document}

\begin{frontmatter}
\title{\vspace*{-0.5in} The Causal Complementation Algorithm for  Lifting Factorization of\\Perfect Reconstruction Multirate Filter Banks}

\author[1]{Christopher M.\ Brislawn\corref{cor1}}
\ead{cbrislawn@yahoo.com}
\cortext[cor1]{Corresponding author}
\affiliation[1]{organization={Los Alamos National Laboratory}, 
postcode={Los Alamos}, city={NM 87545--1663}, country={USA}}

\begin{abstract}
An intrinsically causal approach to lifting factorization, called the \emph{Causal Complementation Algorithm}, is developed for arbitrary two-channel perfect reconstruction FIR  filter banks. This addresses an engineering shortcoming of the inherently  noncausal  strategy of Daubechies and Sweldens for factoring discrete wavelet transforms, which was based on the Extended Euclidean Algorithm for  Laurent polynomials. 
The Causal Complementation Algorithm reproduces all lifting factorizations created by the \emph{causal} version of the Euclidean Algorithm approach and generates additional causal factorizations, which are not obtainable via the causal Euclidean Algorithm, possessing degree-reducing properties that generalize those furnished by the Euclidean Algorithm. In lieu of the Euclidean Algorithm, the new approach employs Gaussian elimination in matrix polynomials using  a slight generalization of  polynomial long division.  It is shown that certain polynomial degree-reducing conditions  are both necessary and sufficient for a causal elementary matrix decomposition to be obtainable 
using the Causal Complementation Algorithm,  yielding a formal definition of ``lifting factorization'' that was missing from the work of Daubechies and Sweldens.
\end{abstract}


\end{frontmatter}	

\section{Introduction}\label{sec:Intro}
\begin{figure}[t]
\vspace{-0.15in}
  \begin{center}
    \includegraphics[page=3]{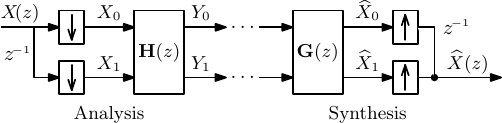}
    \caption{The  polyphase-with-delay filter bank representation.}
    \label{fig:PWD}
  \end{center}
\end{figure}

Multirate digital filter banks and discrete wavelet transforms have proven to be attractive replacements for Fourier transforms in a number of digital communications source coding applications~\cite[\S II]{Bris:13b:TIT}.
The  \emph{Causal Complementation Algorithm} (CCA) constructs  \emph{lifting factorizations}~\cite{Sweldens96,Sweldens:98:SIAM-lifting-scheme}  (a class of elementary matrix decompositions) for causal perfect reconstruction (PR) filter banks by factoring their $2\times 2$ causal polyphase-with-delay transfer matrices~\cite{Daub92,Vaid93,VettKov95,StrNgu96,Mallat99}; see Figure~\ref{fig:PWD}.  
The CCA is an alternative to the approach of Daubechies and Sweldens~\cite{DaubSwel98}, who computed lifting factorizations using the  Extended Euclidean Algorithm (EEA)  over the \emph{Laurent polynomials}  (discrete-time transfer functions $F(z)$ in both $z$ and $\zinv$).
They applied the EEA to polyphase-with-\emph{advance} matrices~\cite{BrisWohl06} with a  \emph{unimodular} normalization,  $|\mathbf{A}(z)|\eqdef\det\mathbf{A}(z)=1$, which leads  to inherently noncausal lifting structures.
The author's previous paper~\cite{Bris:23:Factoring-PRFBs-Causal} introduced the CCA by example and  developed the algebraic tools behind the method.  The CCA factors causal PR matrices into a \emph{cascade} (or product) of causal \emph{lifting steps} (elementary matrices) in a manner that satisfies certain polynomial degree-reducing inequalities.  The present paper uses the  tools from~\cite{Bris:23:Factoring-PRFBs-Causal} to develop the CCA as a formal algorithm and presents results highlighting its advantages.

A finite impulse response (FIR) filter is {causal} if  its impulse response $f(n)$ vanishes for $n<0$ or, equivalently, its Z-transform contains only nonpositive powers of $z$,  $F(z) = \sum_{n=0}^N f(n)z^{-n}$. A linear time-invariant system is traditionally considered  \emph{realizable}  if and only if it is causal and rational.  Lifting decompositions have attracted engineering interest (e.g.,~\cite{ISO_15444_1,ISO_15444_2})  because they significantly reduce a transform's computational complexity and preserve  invertibility  even in the presence of coefficient quantization. For instance,  the (noncausal) unimodular normalization of the 5-tap/3-tap LeGall-Tabatabai piecewise-linear spline wavelet filter bank~\cite{LeGallTabatabai:88:Subband-coding-digital} used in the ISO/IEC JPEG2000 Part~1 image coding standard~\cite{ISO_15444_1} has a (noncausal) linear phase  lifting factorization~\cite[(9)]{Bris:23:Factoring-PRFBs-Causal}, computed using the Laurent EEA,  that reduces the  implementation cost of the filter bank from 2 multiplies per unit input to just 1, a 50\% reduction in multiplicative complexity. 

It is  possible to develop  causal, hence realizable, EEA lifting factorizations for causally normalized polyphase matrices using the polynomial  Euclidean Algorithm, and the causally normalized LeGall-Tabatabai filter bank can be factored using the polynomial EEA into causal linear phase lifting steps~\cite[(23)]{Bris:23:Factoring-PRFBs-Causal} whose  lifting filters  agree modulo delays with the noncausal linear phase lifting filters in the unimodular factorization. Unfortunately, this appears to be an exception attributable to the low order of the LeGall-Tabatabai filter bank.  In~\cite[\S7.8]{DaubSwel98} Daubechies and Sweldens used the Laurent  EEA to factor a noncausal unimodular polyphase-with-advance  matrix for a 5-tap/7-tap linear phase  cubic B-spline wavelet transform; they obtained  lifting  matrices with linear phase lifting filters by exploiting the flexibility afforded by Laurent polynomial division (see~\eqref{CDF75_PWA_anal_lifting} below). 
It is shown in Section~\ref{sec:CubicBSpline:EEA}  that, unlike the LeGall-Tabatabai case, the  polynomial EEA method does \emph{not} produce a causal linear phase analogue in the cubic B-spline case, and one is not obtained merely by  delaying the noncausal lifting filters in the unimodular factorization~\eqref{CDF75_PWA_anal_lifting}.

In contrast to the polynomial EEA method, the CCA  uses  Gaussian elimination to perform row or column reductions on causal polyphase matrices, 
\begin{align}\label{Gaussian_elimination}
\bigl(R_0(z),R_1(z)\bigr) = \bigl(E_0(z),E_1(z)\bigr) - S(z)\,\bigl(F_0(z),F_1(z)\bigr)\,.
\end{align}
The  multipliers $S(z)$ (the lifting filters) are determined using  a slight generalization of the  \emph{causal} polynomial division algorithm that, according to  a  theory of linear Diophantine polynomial equations~\cite{Bris:23:Factoring-PRFBs-Causal}, yields factorizations satisfying degree-reducing properties generalizing those of the EEA. 
In~\cite[\S 6.3]{Bris:23:Factoring-PRFBs-Causal} it is shown that the CCA generates a \emph{causal} linear phase version~\eqref{CDF75_PWD_WSGLS_anal_lifting} of the unimodular factorization~\eqref{CDF75_PWA_anal_lifting} where the corresponding lifting filters differ by simple delays, so the CCA's Gaussian elimination approach is more general than the EEA method and contributes something nontrivial to our understanding of lifting factorzation.

We refer the reader to~\cite{Bris:23:Factoring-PRFBs-Causal} for more background  and  extensive references to related research.

\subsection{Outline of the Paper}\label{sec:Intro:Outline} 
The remainder of the Introduction  defines \hyperref[sec:Intro:StdForm]{standard causal lifting form}.
Section~\ref{sec:PRFB} discusses \hyperref[sec:PRFB:Normalization]{normalization of filter banks} and  \hyperref[sec:PRFB:DegRed]{degree inequalities} that factorizations produced by the CCA  satisfy.  We prove a \hyperref[cor:DRC]{corollary} to a fundamental algebraic result from~\cite{Bris:23:Factoring-PRFBs-Causal} that establishes existence and uniqueness of polynomials satisfying  relevant degree inequalities.  
Section~\ref{sec:PRFB:CCT} 
proves the \hyperref[thm:CCT]{Causal Complementation Theorem}, a generalization of Bezout's Theorem for polynomials using the Slightly Generalized Division Algorithm~\cite[Algorithm~1]{Bris:23:Factoring-PRFBs-Causal} that plays a key role in the CCA.  
Section~\ref{sec:PRFB:Factor} walks  through the general processes involved in the CCA while  Section~\ref{sec:PRFB:CCA} gives a formal presentation of the \hyperref[alg:CCA]{Causal Complementation Algorithm} and defines \hyperref[sec:PRFB:CCA:Schema]{factorization schema}---lists of algorithmic parameters that completely describe  CCA factorization processes.

Section~\ref{sec:CubicBSpline} studies CDF(7,5), the causal 7-tap/5-tap Cohen-Daubechies-Feauveau cubic B-spline filter bank. It is shown that its \hyperref[sec:CubicBSpline:EEA]{causal EEA lifting factorizations} are reproduced by analogous \hyperref[sec:CubicBSpline:CCA]{CCA lifting factorizations}.  Section~\ref{sec:CubicBSpline:Comparison} shows that the computational cost of the CCA is  less than that of the EEA when computing the same lifting factorization.  Section~\ref{sec:CubicBSpline:Other} presents  factorizations of CDF(7,5) obtained using the CCA that cannot be constructed using the causal EEA.  This includes the \hyperref[sec:CubicBSpline:Other:Col1M1]{causal  analogue} of the unimodular linear phase lifting factorization for CDF(7,5) of Daubechies and Sweldens~\cite[\S7.8]{DaubSwel98} mentioned above.  
Expanding upon research  by Zhu and Wickerhauser~\cite{ZhuWicker:12:NearestNeighborLifting} on ill-conditioned lifting factorizations,
an extremely ill-conditioned lifting factorization~\eqref{EEA_col0_cascade_factor} generated by the causal EEA in column~0 of CDF(7,5) is \hyperref[sec:CubicBSpline:Other:Col0M1]{fixed  using the CCA with the Slightly Generalized Division Algorithm} to relocate one of the diagonal delay factors in the  decomposition; this improves the conditioning of the factorization by over eight orders of magnitude.    (\hyperref[app:Condition]{Appendix~A} contains basic facts about condition numbers for PR filter banks.)

Section~\ref{sec:Left} specializes to the case of left lifting  factorizations, i.e., row reductions only.  This simplifies the CCA and ensures that the partial quotients created by left lifting  factorization agree with the \hyperref[lem:partial_prods_quots]{right partial products} of the  decomposition.  A \hyperref[defn:deg_lifting]{left degree-lifting} attribute is defined for right partial products of a lifting cascade and a \hyperref[defn:lifting_sig]{left lifting signature} collects these attributes for all of a cascade's partial products in a list.  The identification of partial products with partial quotients allows us to prove an \hyperref[thm:Equivalence]{Equivalence Theorem} stating that  left lifting factorizations with degree-reducing partial quotients are in one-to-one correspondence with cascades  having  degree-increasing (``degree-lifting'')  right partial products.  
Thus,  having degree-lifting right partial products  can serve as a \emph{definition} of which elementary matrix cascades constitute \emph{(left degree-) lifting} factorizations, independent of the process used to construct the cascade. No such definition distinguishing ``lifting'' factorizations from general elementary matrix decompositions was given in~\cite{DaubSwel98}.
The left degree-lifting property is then used to prove that there are \hyperref[thm:NoIdentityLiftings]{no left degree-lifting factorizations of the identity} and that a left degree-lifting factorization is \hyperref[thm:Conditional]{uniquely determined by its lifting signature}.
The paper concludes with a \hyperref[sec:Left:Daub44]{provably complete tabulation of \emph{all} left degree-lifting factorizations} for the causal  4-tap/4-tap Daubechies paraunitary wavelet filter bank.  

\subsection{Standard Causal Lifting Form}\label{sec:Intro:StdForm}
%
%
\begin{figure}[t]
\vspace{-0.2in}
  \begin{center}
    \includegraphics[page=1]{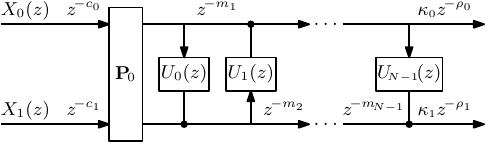}
    \caption{Example of standard causal lifting form for a PR filter bank.  The  initial update characteristic in this example is lower-triangular, $\upchi(\mathbf{U}_0)=1$,  and the number of lifting steps,  $N$, is odd.}
    \label{fig:StdCausalForm}
  \end{center}
\vspace{-0.2in}
\end{figure}
\begin{defn}\label{defn:StdCausalForm}  
If $\mathbf{H}(z)$ is a causal FIR PR matrix then a  cascade decomposition (cf.\ Figure~\ref{fig:StdCausalForm}) of the form
\begin{align}\label{std_causal_form}
\mathbf{H}(z) =
\diag(\kappa_0 z^{-\rho_0},\kappa_1 z^{-\rho_1})\,\mathbf{U}_{\! N-1}(z)\,\bsy{\Lambda}_{N-1}(z)
\cdots\mathbf{U}_1(z)\,\bsy{\Lambda}_{1}(z)\,\mathbf{U}_0(z)\,\Psub{0}\,\diag(z^{-c_0},z^{-c_1})
\end{align}
is in \emph{standard causal  lifting form} (or is a \emph{standard causal lifting cascade}) if it satisfies the following.
\begin{enumerate}
\item\label{scaling_factors}  The  \emph{scaling factors} $\kappa_0,\kappa_1$ are nonzero gain constants, and the delays $z^{-\rho_0},\, z^{-\rho_1}$ (resp., $z^{-c_0},\, z^{-c_1}$) are  common divisors of the rows (resp., columns) of $\mathbf{H}(z)$ for which its corresponding ``coprimification,'' 
\begin{align}\label{def_Q0}
\mathbf{Q}_0(z) \eqdef \diag( z^{\rho_0}, z^{\rho_1})\,\mathbf{H}(z)\,\diag(z^{c_0},z^{c_1}) ,
\end{align}
is causal with coprime rows and columns.  Coprimifications are not necessarily unique; e.g.,
\begin{align*}
\mathbf{H}(z) 
&=
\begin{bmatrix}
z^{-1} & 0\\
0 & 1\vspace{-2pt}
\end{bmatrix}
\hspace{-3pt}
\begin{bmatrix}
1 & 1\\
z^{-2} & \,1+z^{-2}\vspace{-2pt}
\end{bmatrix}
=
\begin{bmatrix}
1 &\, z^{-1}\\
z^{-1} &\, 1+z^{-2}\vspace{-2pt}
\end{bmatrix}
\hspace{-3pt}
\begin{bmatrix}
z^{-1} & 0\\
0 & 1\vspace{-2pt}
\end{bmatrix}.
\end{align*}
Note that~\eqref{std_causal_form} implies a specific choice of coprimification.

\item\label{lifting_matrices}  The $\mathbf{U}_n(z)$ are lower-  or upper-triangular causal unimodular lifting matrices with a causal, nonzero, off-diagonal lifting filter $U_n(z)$.  They are described by the \emph{lifting operators} $\lambda$ and $\upsilon$,
\begin{align}\label{def_lambda_upsilon}
\mathbf{U}(z)
=
\lambda(U(z)) 
\eqdef 
\begin{bmatrix}
1 & 0\\
U(z) & 1\vspace{-2pt}
\end{bmatrix}\text{\ \ or\ \ }
\mathbf{U}(z)
=
\upsilon(U(z))
\eqdef
\begin{bmatrix}
1 & U(z)\\
0 & 1\vspace{-2pt}
\end{bmatrix}.
\end{align}
$\mathbf{U}_n$'s \emph{update characteristic}~\cite[Annex~G]{ISO_15444_2} is the  flag
\begin{equation}\label{lifting_update_char}
\upchi_n\eqdef \upchi(\mathbf{U}_n) \eqdef 
\left\{ \begin{array}{ll}
0 & \mbox{ if $\mathbf{U}_n$ is upper-triangular,}\\
1 & \mbox{ if $\mathbf{U}_n$ is lower-triangular.}
\end{array}\right.
\end{equation}

\item\label{delay_matrices}  The $\bsy{\Lambda}_n(z)$ are diagonal delay matrices with  one  delay, $z^{-m_n}$, $m_n\geq 0$.  
Their update characteristic   is
\begin{equation}\label{delay_update_char}
\upchi(\bsy{\Lambda}_n) \eqdef 
\left\{ \begin{array}{ll}
0 & \mbox{ if $\bsy{\Lambda}_n(z)=\diag(z^{-m_n},1)$,}   \\
1 & \mbox{ if $\bsy{\Lambda}_n(z)=\diag(1,z^{-m_n})$.}   
\end{array}\right.
\end{equation}
%

\item\label{irreducibility}  Cascade~\eqref{std_causal_form} is  \emph{irreducible} (cf.~\cite[Definition~3]{Bris:10:GLS-I}), which means that the update characteristics of $\mathbf{U}_n$ and $\bsy{\Lambda}_n$ agree, $\upchi_n \eqdef \upchi(\mathbf{U}_n) = \upchi(\bsy{\Lambda}_n)$, and strictly alternate, 
$\upchi_n = 1-\upchi_{n\pm 1}$.

\item\label{P_nought}  \Psub{0} is either the identity matrix, $\mathbf I$, or the \emph{swap matrix},
\begin{align}\label{swap_matrix}
\mathbf{J} &\eqdef
\begin{bmatrix}
0 & 1\\
1 & 0\vspace{-1pt}
\end{bmatrix},\quad
\mathbf{J}^{-1} = \mathbf{J}.
\end{align}
\end{enumerate}
\end{defn}
The CCA  provides every causal PR filter bank with multiple decompositions in standard causal  lifting form.

\section{Causal Perfect Reconstruction Filter Banks}\label{sec:PRFB}

\subsection{Normalization of Filters and Filter Banks}\label{sec:PRFB:Normalization}
A common  normalization to impose on a causal FIR filter, $F(z)=\sum f(n)z^{-n}$,  is to  adjust its delay so that the first nonzero impulse response tap is $f(0)$. We say a causal signal or filter $F(z)$  is \emph{left justified} if it is \emph{not} divisible by $z^{-1}$, denoted $z^{-1}\dividesnot F(z)$, which is equivalent to  $f(0)\neq 0$.  A causal filter bank, $\{F_0(z),F_1(z)\}$, and its causal polyphase (analysis or synthesis) matrix,  $\mathbf{F}(z) = \sum \mathbf{f}(n) z^{-n}$, are  \emph{doubly left justified} if both  $F_0$ and $F_1$ are left justified.
%
While double left justification may seem like the most natural way to normalize  a filter bank, not every PR filter bank is equivalent modulo delays to a \emph{doubly} left justified PR filter bank, e.g.,
\begin{equation} 
\begin{array}{ll}
H_0(z) &= 2 + z^{-1} + z^{-2}, \\
H_1(z) &= z^{-1} + z^{-2},
\end{array}
\qquad\qquad
\mathbf{H}(z) =
\begin{bmatrix}
2+z^{-1} & 1\\
z^{-1} & 1 \vspace{-2pt}	
\end{bmatrix}. \label{NonDoubleJust}
\end{equation}
In particular, double left justification is {not} necessarily preserved in  matrix factorizations; i.e., the matrix quotient created by factoring off a lifting step may fail to be doubly justified.  
Instead, coprimality  of polyphase matrix rows and columns is a more useful  normalization for causal lifting.
\begin{lem}[Coprimality]\label{lem:Coprimality}
Let $\mathbf{H}(z)$ be a causal transfer matrix.
\begin{enumerate}[label={\roman*)},ref={\roman*}]		
\item\label{lem:Coprimality:Right}
Suppose $\mathbf{Q}(z)$ is a causal right matrix quotient  of $\mathbf{H}(z)$, $\mathbf{H}(z) = \mathbf{F}(z)\mathbf{Q}(z).$  If the entries in column~$i$ of $\mathbf{H}(z)$, $i\in\{0,1\}$, are coprime then the entries in column~$i$ of $\mathbf{Q}(z)$ are also coprime.

\item\label{lem:Coprimality:Left}
If $\mathbf{Q}(z)$ is a causal left quotient, $\mathbf{H}(z) = \mathbf{Q}(z)\mathbf{F}(z)$,  and row~$i$ is coprime in $\mathbf{H}(z)$ then row~$i$ is  coprime in $\mathbf{Q}(z)$.
\end{enumerate}\end{lem}

\emph{Proof:} 
Suppose that column~0 in $\mathbf{Q}(z)$ is divisible by $G(z)$,
$\mathbf{Q}(z) = \mathbf{Q}'(z)\,\diag(G(z),1).$  This implies that column~0 in $\mathbf{H}(z)$ is also divisible by $G(z)$,
$\mathbf{H}(z) = \mathbf{F}(z)\mathbf{Q}'(z)\,\diag(G(z),1)$.
Therefore, if column~0 in $\mathbf{H}(z)$ is coprime then so is column~0 in $\mathbf{Q}(z)$.
The  other cases are similar.  \qed

\subsection{Degree-Reducing Causal Complementation}\label{sec:PRFB:DegRed}
As emphasized in~\cite{Bris:23:Factoring-PRFBs-Causal} (but not in~\cite{DaubSwel98}),  the degree-reducing aspect of  the polynomial Extended Euclidean Algorithm  differentiates EEA-based lifting factorizations within the much bigger universe of all elementary matrix decompositions.  The starting point for constructing ``degree-reducing'' causal lifting factorizations without using the EEA is the following definition from~\cite{Bris:23:Factoring-PRFBs-Causal}.
%
\begin{defn}\label{defn:CausalComplement}
\cite[Definition~1.1]{Bris:23:Factoring-PRFBs-Causal}
Let  constants $\hat{a}\in\mathbb{C},\;\hat{a}\neq 0,$ and $\hat{d}\in\mathbb{Z},\;\hat{d}\geq 0,$ be given, and let $(F_0,F_1)$ be  causal filters whose greatest common divisor satisfies $\gcd(F_0,F_1) \divides  z^{-\hat{d}}$.
An ordered pair of causal filters $(R_0,R_1)$  is a \emph{causal complement to  $(F_0,F_1)$ for inhomogeneity  $\hat{a}z^{-\hat{d}}$}  if it satisfies the linear Diophantine polynomial equation
\begin{align}\label{complement}
F_0(z)R_1(z) - F_1(z)R_0(z) = \hat{a}z^{-\hat{d}}.
\end{align}
For  $\ell\in\{0,1\}$ a causal complement  $(R_0,R_1)$ to $(F_0,F_1)$ is  \emph{degree-reducing in} \Fl\  if  
\begin{align}\label{deg-reducing}
\deg(\Rl) < \deg(\Fl) - \deg\gcd(F_0,F_1).
\end{align}
\end{defn}

\rem  The hypothesis $\gcd(F_0,F_1)\divides  z^{-\hat{d}}$  is necessary and sufficient for~\eqref{complement} to be solvable~\cite[Theorem~3.2]{Bris:23:Factoring-PRFBs-Causal}.  Another idea  from~\cite{Bris:23:Factoring-PRFBs-Causal} is the concept of \emph{degree-reducing remainders with multiplicities} for polynomial division, which will yield algorithms for factoring off diagonal delay matrices.

\begin{defn}\label{defn:deg_red_mod_M}
(cf.~\cite{Bris:23:Factoring-PRFBs-Causal}, Definition~5.2)
Let $E,F$ be causal filters, $F\neq 0$, and $M\geq 0$. 
For a given quotient $S$, the remainder $R\eqdef E-FS$ has (a pole at 0 of) \emph{multiplicity~$M$} if  $z^{-M} \divides R(z)$.   $R$  is \emph{degree-reducing modulo $M$} if  $R$ has multiplicity~$M$  and  satisfies
\[  \deg(R) < \deg(F) - \deg\gcd(F,z^{-M}) + M.  \]
\end{defn}

Now extend Definition~\ref{defn:CausalComplement}  for  causal complements that are divisible by powers of $z^{-1}$. 
\begin{defn}\label{defn:FB_deg_red_mod_M}
Consider complementary pairs $(F_0,F_1)$ and $(R_0,R_1)$ satisfying~\eqref{complement}.  
Entry $R_j(z)$ has (a pole at 0 of) \emph{multiplicity~$M$}  if $z^{-M}\divides  R_j(z)$; $(R_0,R_1)$ is said to have multiplicity~$M$ if both entries have multiplicity~$M$. For  $\ell\in\{0,1\}$ we say the causal complement  $(R_0,R_1)$ to $(F_0,F_1)$  for inhomogeneity  $\hat{a}z^{-\hat{d}}$ is \emph{degree-reducing modulo~$M$ in \Fl} if  $(R_0,R_1)$ has multiplicity~$M$ and satisfies
\begin{align}\label{deg-reducing-mod-M}
\deg(\Rl) < \deg(\Fl) - \deg\gcd(F_0,F_1) + M.
\end{align}
\end{defn}

We shall generate  incremental reductions in the determinantal degree of a polyphase matrix by deliberately factoring off diagonal delay matrices at chosen points during lifting factorization, rather than waiting  for degree reductions to appear arbitrarily.  Towards this goal   extend~\cite[Theorem~4.5]{Bris:23:Factoring-PRFBs-Causal} to include causal complements that are  degree-reducing modulo~$M$.
\begin{cor}[Linear Diophantine Degree-Reduction Corollary]\label{cor:DRC}
Let  $\hat{a}\in\mathbb{C},\;\hat{a}\neq 0$, and $\hat{d}\in\mathbb{Z}$, $\hat{d}\geq 0$ be given. 
Let  $F_0$ and $F_1$  be causal filters with $\gcd(F_0,F_1)=z^{-d_F}$ where $0\leq d_F\leq\hat{d}$.  
Let $M\in\mathbb{Z}$ satisfy  $0\leq M\leq \hat{d}-d_F$.  
\begin{enumerate}[label={\roman*)},ref={\roman*}]		
\item\label{cor:DRC:1}
If $F_0\neq 0$ then there exists a unique causal complement, $(R_0,R_1)$, to $(F_0,F_1)$ for inhomogeneity $\hat{a}z^{-\hat{d}}$ that is degree-reducing modulo~$M$ in $F_0$,
\begin{align*}
\deg(R_0) < \deg(F_0) - \deg\gcd(F_0,F_1) + M.
\end{align*}

\item\label{cor:DRC:2}
If $F_1\neq 0$ then there  exists a unique causal complement, $(R'_0,R'_1)$, to $(F_0,F_1)$ for  $\hat{a}z^{-\hat{d}}$ that is degree-reducing modulo~$M$ in $F_1$,
\begin{align*}
\deg(R'_1) < \deg(F_1) - \deg\gcd(F_0,F_1) + M.
\end{align*}
\item\label{cor:DRC:3}
Let $F_0,F_1\neq 0$ and let $(R_0,R_1)$ and  $(R'_0,R'_1)$ be the unique causal complements given in~\eqref{cor:DRC:1} and \eqref{cor:DRC:2}, respectively.  These causal complements are the same, $(R_0,R_1)=(R'_0,R'_1)$, if and only if
\begin{equation}\label{cor:DRC:3:dhat_minus_M_small}
\hat{d} < \deg(F_0) + \deg(F_1) - \deg\gcd(F_0,F_1) + M\,.
\end{equation}
%

%
%
\end{enumerate}\end{cor}

\emph{Proof:}  
(\ref{cor:DRC:1}) Since $d_F\leq \hat{d} - M$, by~\cite[Theorem~4.5(i)]{Bris:23:Factoring-PRFBs-Causal} there is a unique causal complement $(G_0,G_1)$  to $(F_0,F_1)$ for  $\hat{a}z^{-(\hat{d}-M)}$ that is degree-reducing in $F_0$, 
\begin{align*}
F_0(z)G_1(z) - F_1(z)G_0(z) =  \hat{a}z^{-(\hat{d}-M)}\;,\quad
\deg(G_0) < \deg(F_0) - \deg\gcd(F_0,F_1)\,.
\end{align*}
Define $R_j(z) \eqdef z^{-M} G_j(z),\;j=0,1$; then 
$F_0(z)R_1(z) - F_1(z)R_0(z) =  \hat{a}z^{-\hat{d}}$
where $(R_0,R_1)$  is degree-reducing modulo~$M$ in $F_0$,
\begin{align*}
\deg(R_0) \;=\; \deg(G_0) + M \;<\; \deg(F_0) - \deg\gcd(F_0,F_1) + M.
\end{align*}

To prove uniqueness, suppose  $(S_0,S_1)$ is another  causal complement to $(F_0,F_1)$ for  $\hat{a}z^{-\hat{d}}$ that is degree-reducing modulo~$M$ in $F_0$. The   complement $(z^MS_0,z^MS_1)$ to $(F_0,F_1)$ for  $\hat{a}z^{-(\hat{d}-M)}$ is causal since $S_0$ and $S_1$ both have multiplicity~$M$.  
Moreover, it is degree-reducing in $F_0$ since 
\begin{align*}
\deg(z^MS_0) \;=\; \deg(S_0) - M \;<\;  \deg(F_0) - \deg\gcd(F_0,F_1) + M - M \;=\; \deg(F_0) - \deg\gcd(F_0,F_1)\,.
\end{align*}
Therefore,  $(z^MS_0,z^MS_1)=(G_0,G_1)$ by  uniqueness  in~\cite[Theorem~4.5(i)]{Bris:23:Factoring-PRFBs-Causal}, so $(S_0,S_1)=(R_0,R_1)$.

(\ref{cor:DRC:2}) Same as the proof of~\eqref{cor:DRC:1}, using~\cite[Theorem~4.5(ii)]{Bris:23:Factoring-PRFBs-Causal}.

(\ref{cor:DRC:3})  Write~\eqref{cor:DRC:3:dhat_minus_M_small} in the form
\begin{align}\label{reduced_ineq}
\hat{d} - M < \deg(F_0) + \deg(F_1) - \deg\gcd(F_0,F_1).
\end{align}
By~\cite[Theorem~4.5(iii)]{Bris:23:Factoring-PRFBs-Causal},
the degree-reducing causal complements $(G_0,G_1)$ and $(G'_0,G'_1)$ from the proofs of clauses~(\ref{cor:DRC:1}) and (\ref{cor:DRC:2}) are equal if and only if~(\ref{reduced_ineq}) is satisfied, and  $(G_0,G_1)=(G'_0,G'_1)$ is equivalent to $(R_0,R_1)=(R'_0,R'_1)$.
\hfill\qed

\subsection{The Causal Complementation Theorem}\label{sec:PRFB:CCT}

As with Corollary~\ref{cor:DRC}, the Causal Complementation Theorem (CCT) can be regarded as a variant of Bezout's Theorem for polynomials~\cite[Theorem~6.1.1]{Daub92}.
The CCT uses a given causal complement $(E_0,E_1)$  to $({F}_0,{F}_1)$ for  $\hat{a}z^{-\hat{d}}$ and the Slightly Generalized Division Theorem (SGDT)~\cite[Theorem~5.3]{Bris:23:Factoring-PRFBs-Causal}  to construct a degree-reducing causal complement  to $({F}_0,{F}_1)$ for  $\hat{a}z^{-\hat{d}}$.  

\begin{thm}[Causal Complementation Theorem]\label{thm:CCT}
Let  $\hat{a}\in\mathbb{C},\;\hat{a}\neq 0$, and $\hat{d}\in\mathbb{Z}$, $\hat{d}\geq 0$ be given. 
Let  $F_0$ and $F_1$  be causal filters with $\gcd(F_0,F_1)=z^{-d_F}$ where $0\leq d_F\leq\hat{d}$.  
Let $M\in\mathbb{Z}$ satisfy  $0\leq M\leq \hat{d}-d_F$. 
Suppose we are given a causal complement $(E_0,E_1)$ to $({F}_0,{F}_1)$ for  $\hat{a}z^{-\hat{d}}$,
\begin{align}\label{causal_compl}
F_0(z) E_1(z) - F_1(z) E_0(z) =  \hat{a}z^{-\hat{d}}.
\end{align}
Define coprime parts $\widetilde{F}_j(z)\eqdef z^{d_F}F_j(z)$, $j=0,1,$
and let  $\ell\in\{0,1\}$ be an index for which $\widetilde{F}_{\ell}(z)$ is left-justified.
It follows that there exists a unique polynomial $S(z)$ whose remainders
\begin{align}\label{gaussian_elimination}
(R_0,R_1)\eqdef (E_0,E_1) - S\bigl(\widetilde{F}_0,\widetilde{F}_1\bigr) 
\end{align}
form the unique causal complement  to $({F}_0,{F}_1)$ for  $\hat{a}z^{-\hat{d}}$ that is degree-reducing modulo~$M$ in \Fl\,.
\end{thm}

\emph{Proof:}  
Dividing~\eqref{causal_compl} by $z^{-d_F}$ we see that $(E_0,E_1)$ is also a causal complement to $\bigl(\widetilde{F}_0,\widetilde{F}_1\bigr)$ for  $\hat{a}z^{-(\hat{d} - d_F)}$,
\begin{align}\label{causal_compl_2}
\widetilde{F}_0(z)E_1(z) -  \widetilde{F}_1(z)E_0(z)   =  \hat{a}z^{-(\hat{d}-d_F)}.
\end{align}
%
If $M=0$ then  $\gcd\bigl(\widetilde{F}_{\ell},z^{-M}\bigr)=1$, and if $M>0$ then any  $\ell\in\{0,1\}$ for which $\widetilde{F}_{\ell}$ is left-justified $\bigl(\text{i.e., for which }z^{-1}\dividesnot\widetilde{F}_{\ell}\bigr)$ satisfies  $\gcd\bigl(\widetilde{F}_{\ell},z^{-M}\bigr)=1$.  
Suppose $M\leq \hat{d}-d_F$ and $\ell\in\{0,1\}$ are such that $\gcd\bigl(\widetilde{F}_{\ell},z^{-M}\bigr)=1$.  
Since $\gcd\bigl(\widetilde{F}_{\ell},z^{-M}\bigr)\divides \El$, the SGDT~\cite[Theorem~5.3]{Bris:23:Factoring-PRFBs-Causal} yields a unique quotient  $S(z)$ whose remainder, $\Rl\eqdef \El - \widetilde{F}_{\ell}S$, is degree-reducing modulo~$M$ (Definition~\ref{defn:deg_red_mod_M}),
\begin{align}\label{deg_red_mod_M}
z^{-M}\divides \Rl(z)\text{\quad and\quad} 
\deg(\Rl) \;<\; \deg\bigl(\widetilde{F}_{\ell}\bigr)+M \;=\;  \deg(\Fl) - d_F + M\,.
\end{align}
Define $R_{\ell'}\eqdef {E}_{\ell'} - \widetilde{F}_{\ell'}S$, where $\ell'\eqdef 1-\ell$, so  $(R_0,R_1)$  is another causal complement to $\bigl(\widetilde{F}_0,\widetilde{F}_1\bigr)$ for  $\hat{a}z^{-(\hat{d}-d_F)}$,
\begin{align}\label{causal_compl_3}
\widetilde{F}_0(z)R_1(z) -  \widetilde{F}_1(z)R_0(z)  =  \hat{a}z^{-(\hat{d}-d_F)}.
\end{align}

We next  show that  $(R_0,R_1)$ has multiplicity~$M$.
$z^{-M}\divides \Rl(z)$ by~\eqref{deg_red_mod_M} and $z^{-M}\divides z^{-(\hat{d}-d_F)}$ by the hypothesis $0\leq M\leq\hat{d} - d_F$ so~\eqref{causal_compl_3} implies  $z^{-M}\divides R_{\ell'}(z)\widetilde{F}_{\ell}(z)$. 
The condition $\gcd\bigl(\widetilde{F}_{\ell},z^{-M}\bigr)=1$ implies   $z^{-M}\divides R_{\ell'}(z)$, giving $(R_0,R_1)$ multiplicity~$M$.  
Multiplying both sides of~\eqref{causal_compl_3} by $z^{-d_F}$ shows that $(R_0,R_1)$ is a causal complement to $({F}_0,{F}_1)$ for  $\hat{a}z^{-\hat{d}}$,
\begin{align}\label{causal_compl_4}
F_0(z) R_1(z) - F_1(z) R_0(z) =  \hat{a}z^{-\hat{d}}.
\end{align}
Inequality~\eqref{deg_red_mod_M} says that the causal complement $(R_0,R_1)$ to $({F}_0,{F}_1)$ for  $\hat{a}z^{-\hat{d}}$ is degree-reducing modulo~$M$ in $\Fl$~\eqref{deg-reducing-mod-M} and is therefore the unique such causal complement by Corollary~\ref{cor:DRC}.  \qed

\rem
If  $(R_0,R_1)$ is degree-reducing modulo~$M$ in \Fl\ but \Rl\  has multiplicity~$M+k$, $k>0$, then  \Rl\ is also degree-reducing modulo~$M+k$ as an SGDT remainder by the remarks following the proof of the SGDT~\cite[Theorem~5.3]{Bris:23:Factoring-PRFBs-Causal}.  
The proof of Theorem~\ref{thm:CCT} shows that $z^{-(M+k)}\divides R_{\ell'}(z)$ if and only if $\hat{d}-d_F\geq M+k$, in which case $(R_0,R_1)$ has multiplicity~$M+k$ and is the unique causal complement that is degree-reducing modulo~$M+k$ in \Fl.

\subsection{Causal Factorizations}\label{sec:PRFB:Factor}
Next, the  theoretical language in which the SGDT and the CCT are presented will  be made computationally effective by employing the Slightly Generalized Division Algorithm (SGDA)~\cite[Algorithm~1]{Bris:23:Factoring-PRFBs-Causal}.
We  walk through the steps  in the Causal Complementation Algorithm prior to a more formal presentation of the CCA in Section~\ref{sec:PRFB:CCA}.
$(F_0,F_1)$  denotes a row or column vector in a  polyphase matrix while $(E_0,E_1)$ or $(R_0,R_1)$ will be a complementary row or column vector.

\subsubsection{Lifting Step Extraction}\label{sec:PRFB:Factor:Lifting}
The effect of a lifting step depends on its  \emph{handedness}---whether it multiplies the matrix being factored on the left or the right. 
A left-handed lifting step is a \emph{row reduction},
$\mathbf{H}(z) = \mathbf{V}(z)\bsy{\Delta}(z)\mathbf{Q}(z)$,
for lifting matrix $\mathbf{V}$, diagonal delay matrix $\bsy{\Delta}$, and coprimified right quotient matrix $\mathbf{Q}$.  For update characteristic $\upchi(\mathbf{V}) = 0$,
\begin{align}\label{upper_triang_left_lift}
\hspace{-6pt}
\begin{bmatrix}
E_0 & E_1\\
F_0 & F_1\vspace{-1pt}
\end{bmatrix}
=
\begin{bmatrix}
1 & S\\
0 & 1\vspace{-2pt}%
\end{bmatrix}
\hspace{-4pt}
\begin{bmatrix}
R_0 & R_1\\
{F}_0 & {F}_1\vspace{-1pt}%
\end{bmatrix}
=
\begin{bmatrix}
1 & S\\
0 & 1\vspace{-2pt}%
\end{bmatrix}
\hspace{-4pt}
\begin{bmatrix}
z^{-m} & 0\\
0 & 1\vspace{-3pt}%
\end{bmatrix}
\hspace{-4pt}
\begin{bmatrix}
\wt{R}_0 & \wt{R}_1\\
{F}_0 & {F}_1\vspace{-1pt}%
\end{bmatrix} \hspace{-2pt}
\end{align}
where $F_0$ and $F_1$ are  coprime, $z^{-m} \eqdef \gcd(R_0,R_1)$,
and $\wt{R}_i(z)\eqdef z^m R_i(z)$, $i\in\{0,1\}$. If $\upchi(\mathbf{V}) = 1$ then
\begin{align}\label{lower_triang_left_lift}
\begin{bmatrix}
F_0 & F_1\\
E_0 & E_1\vspace{-1pt}
\end{bmatrix}
&=
\begin{bmatrix}
1 & 0\\
S & 1\vspace{-2pt}%
\end{bmatrix}
\hspace{-3pt}
\begin{bmatrix}
{F}_0 & {F}_1\\
{R}_0 & {R}_1\vspace{-1pt}%
\end{bmatrix}
=
\begin{bmatrix}
1 & 0\\
S & 1\vspace{-2pt}%
\end{bmatrix}
\hspace{-3pt}
\begin{bmatrix}
1 & 0\\
0 & z^{-m}\vspace{-3pt}%
\end{bmatrix}
\hspace{-3pt}
\begin{bmatrix}
{F}_0 & {F}_1\\
\wt{R}_0 & \wt{R}_1\vspace{-1pt}%
\end{bmatrix}.
\end{align}
A right-handed lifting step is a \emph{column reduction} with a  \emph{left} quotient, $\mathbf{H}(z) = \mathbf{Q}(z)\bsy{\Delta}(z)\mathbf{V}(z)$, e.g.,
\begin{align}
\begin{bmatrix}
F_0 & E_0\\
F_1 & E_1\vspace{-1pt}
\end{bmatrix}
&=
\begin{bmatrix}
F_0 & R_0\\
F_1 & R_1\vspace{-2pt}%
\end{bmatrix}
\hspace{-3pt}
\begin{bmatrix}
1 & S\\
0 & 1\vspace{-2pt}%
\end{bmatrix}
=
\begin{bmatrix}
F_0 & \wt{R}_0\\
F_1 & \wt{R}_1\vspace{-2pt}%
\end{bmatrix}
\hspace{-3pt}
\begin{bmatrix}
1 & 0\\
0 & z^{-m}\vspace{-2pt}%
\end{bmatrix}
\hspace{-3pt}
\begin{bmatrix}
1 & S\\
0 & 1\vspace{-2pt}%
\end{bmatrix}
\text{ for $\upchi(\mathbf{V}) = 0$.}\label{upper_triang_right_lift}
\end{align}
%
We distinguish  between $M$, a minimum multiplicity for $(R_0,R_1)$ specified by the user (illustrated in the examples below), and the actual  multiplicity achieved  by the CCT, $m\eqdef\deg\gcd(R_0,R_1)\geq M$. For division in row or column $\ell$,
\begin{equation}\label{alt_deg-reducing_ineq}
\deg\bigl(\wt{R}_{\ell}\bigr) \;=\; \deg(\Rl) - m 
\;<\; \deg(\Fl) + M - m  \;\leq\; \deg(\Fl)\text{\ \ by~\eqref{deg_red_mod_M}.} 
\end{equation}
%

\subsubsection{Downlifting Factorization}\label{sec:PRFB:Factor:Downlifting}
Begin by coprimifying $\mathbf{H}(z)$, then factoring  $\mathbf{Q}_0(z)$ using  the CCT (Theorem~\ref{thm:CCT}), e.g.,
\begin{align}\label{left-factor_Q0}
\mathbf{Q}_0(z) &=
\begin{bmatrix}
{E}_0 & {E}_1\\
{F}_0 & {F}_1\vspace{-1pt}
\end{bmatrix}
=
\begin{bmatrix}
1 & V_0\\
0 & 1\vspace{-1pt}
\end{bmatrix}
\hspace{-3pt}
\begin{bmatrix}
z^{-m_0} & 0\\
0 & 1\vspace{-1pt}
\end{bmatrix}
\hspace{-3pt}
\begin{bmatrix}
\wt{R}_0 & \wt{R}_1\\
{F}_0 & {F}_1\vspace{-1pt}
\end{bmatrix}
= \mathbf{V}_{\!0}(z)\,\bsy{\Delta}_0(z)\,\mathbf{Q}_1(z)
\end{align}
or
\begin{align}\label{right-factor_Q0}
\mathbf{Q}_0(z) &=
\begin{bmatrix}
E_0 & F_0\\
E_1 & F_1\vspace{-1pt}
\end{bmatrix}
=
\begin{bmatrix}
\wt{R}_0 & F_0\\
\wt{R}_1 & F_1\vspace{-1pt}
\end{bmatrix}
\hspace{-3pt}
\begin{bmatrix}
z^{-m_0} & 0\\
0 & 1\vspace{-2pt}
\end{bmatrix}
\hspace{-3pt}
\begin{bmatrix}
1 & 0\\
V_0 & 1\vspace{-2pt}
\end{bmatrix}
= \mathbf{Q}_1(z)\,\bsy{\Delta}_0(z)\,\mathbf{V}_{\!0}(z),
\end{align}
where $V_0$ is computed using the SGDA~\cite[Algorithm~1]{Bris:23:Factoring-PRFBs-Causal}.
Since $\mathbf{Q}_0(z)$ has coprime rows and columns, if $\mathbf{Q}_1(z)$ is the right-quotient of a row reduction as in~\eqref{left-factor_Q0} then  $\mathbf{Q}_1(z)$ has coprime columns by Lemma~\ref{lem:Coprimality}(\ref{lem:Coprimality:Right}).  Row $(F_0,F_1)$ is  still coprime, and $\bigl(\wt{R}_0,\wt{R}_1\bigr)$ is coprime by construction so $\mathbf{Q}_1(z)$ has coprime rows and columns.  This  also holds  for column reductions~\eqref{right-factor_Q0}.

Continuing  this way, each subsequent quotient matrix is left- or right-factored using the CCT,
\begin{align}
\mathbf{Q}_n &= \mathbf{V}_{\!n}\bsy{\Delta}_n\mathbf{Q}_{n+1}\text{\quad or}\label{nth_left_downlift}\\
\mathbf{Q}_n &= \mathbf{Q}_{n+1}\bsy{\Delta}_n\mathbf{V}_{\!n}, \label{nth_right_downlift}
\end{align}
where each new quotient $\mathbf{Q}_{n+1}$ has coprime rows and columns.  The quotients also satisfy CCT degree-reducing conditions of the form~\eqref{alt_deg-reducing_ineq}, which we now formalize.

\begin{defn}\label{defn:DegRedDownlift}
A left  factorization~\eqref{nth_left_downlift} is called a \emph{degree-reducing left downlift} if there exists a column index $j_n\in\{0,1\}$ such that the element in the downlifted row~$i_n\eqdef \upchi(\mathbf{V}_{\!n})$ and  column~$j_n$ of $\mathbf{Q}_{n+1}(z)$ has strictly lower degree than the element in row~$i'_n\eqdef 1 - i_n$ and  column~$j_n$\,,
\begin{align}\label{defn:DegRedDownlift:left}
\deg\Bigl(Q^{(n+1)}_{i_n,j_n}\Bigr) &< \deg\Bigl(Q^{(n+1)}_{i'_n,j_n}\Bigr).
\end{align}
A right factorization of the form~\eqref{nth_right_downlift} is called a \emph{degree-reducing right downlift} if there exists a row index $i_n\in\{0,1\}$ such that, for $j_n\eqdef 1-\upchi(\mathbf{V}_{\!n})$ and $j'_n\eqdef 1 - j_n$\,,
\begin{align}\label{defn:DegRedDownlift:right}
\deg\Bigl(Q^{(n+1)}_{i_n,j_n}\Bigr) &< \deg\Bigl(Q^{(n+1)}_{i_n,j'_n}\Bigr).
\end{align}
\end{defn}

\rem  For left factorization~\eqref{nth_left_downlift}, the row $\bigl(\wt{R}_0,\wt{R}_1\bigr)$ modified by the lifting step has row index $i_n=\upchi(\mathbf{V}_{\!n})=\upchi(\bsy{\Delta}_{n})$, and~\eqref{defn:DegRedDownlift:left} is the CCT inequality~\eqref{alt_deg-reducing_ineq} with $\ell=j_n$.  For  right factorization~\eqref{nth_right_downlift}, the column $\bigl(\wt{R}_0,\wt{R}_1\bigr)^T$  of remainders has column index $j_n=1-\upchi(\mathbf{V}_{\!n})=\upchi(\bsy{\Delta}_{n})$, and~\eqref{defn:DegRedDownlift:right} is~\eqref{alt_deg-reducing_ineq} with $\ell=i_n$.

\subsubsection{Terminating the Factorization}\label{sec:PRFB:Factor:Terminate}
Factorization terminates with step number $N'$ for which  $\mathbf{Q}_{N'}(z)$  is   diagonal or antidiagonal. By coprimality  the nonzero entries in $\mathbf{Q}_{N'}$ are units, so $\mathbf{Q}_{N'}$ is a constant matrix of the form
\begin{equation}\label{def_QN}
\mathbf{Q}_{N'} = \mathbf{D}_{\kappa_0,\kappa_1}\Psub{0},\mbox{ where } 
\mathbf{D}_{\kappa_0,\kappa_1}\eqdef\diag(\kappa_0,\kappa_1),\;   \kappa_0,\kappa_1\neq 0,\text{ and }
\Psub{0} = \mathbf{I}\mbox{ or }\mathbf{J}.
\end{equation}
Upon termination the downlifting factorization has the  form
\begin{align}\label{factor_to_QN_2}
\mathbf{Q}_0 
\;=\; 
\mathbf{V}_{\!i}\bsy{\Delta}_i\cdots\mathbf{V}_{\!j}\bsy{\Delta}_{j}(\mathbf{D}_{\kappa_0,\kappa_1}\Psub{0})
	\bsy{\Delta}_m\mathbf{V}_{\!m}\cdots\bsy{\Delta}_n\mathbf{V}_{\!n}
\;=\;
\mathbf{D}_{\kappa_0,\kappa_1}\mathbf{V}'_{\!i}\bsy{\Delta}_i\cdots\mathbf{V}'_{\!j}\bsy{\Delta}_{j}
\bsy{\Delta}''_m\mathbf{V}''_{\!m}\cdots\bsy{\Delta}''_n\mathbf{V}''_{\!n}\Psub{0}\,,
\end{align}
where moving $\mathbf{D}_{\kappa_0,\kappa_1}$  to the left end of the cascade creates 
$\mathbf{V}'_k = 
\gamma^{-1}_{\kappa_0,\kappa_1}\!\mathbf{V}_k\eqdef
\mathbf{D}_{\kappa_0,\kappa_1}^{-1}\!\mathbf{V}_k\mathbf{D}_{\kappa_0,\kappa_1}$ 
by~\cite[Formula~(41)]{Bris:23:Factoring-PRFBs-Causal}.
By~\cite[Formula~(43)]{Bris:23:Factoring-PRFBs-Causal}, moving  \Psub{0} to the right end of the cascade leaves 
\begin{equation}\label{dbldag_intertwine}
\bsy{\Delta}''_k\, \mathbf{V}''_k = 
\left\{ \begin{array}{ll}
\bsy{\Delta}_k\mathbf{V}_k & \mbox{ if $\Psub{0}=\mathbf{I}$,}   \\
\bsy{\Delta}_k^{\!\dbldag}\mathbf{V}_k^{\dbldag} & 
	\mbox{ if $\Psub{0}=\mathbf{J}$, where $\mathbf{A}^{\!\dbldag}\eqdef\mathbf{J}\mathbf{A}\mathbf{J}$.} 
\end{array}\right.
\end{equation}

Interestingly, the final  reduction is (nearly) insensitive to user choices.
The CCT on $\mathbf{Q}_{N'-2}$ yields a quotient $\mathbf{Q}_{N'-1}$ with a zero, $\Rl=0$.  Coprimality implies that the entries adjacent to $\Rl$ are \emph{units} (i.e., nonzero constant polynomials), with at most one  non-unit $F_{\ell'}$ diagonally opposite $\Rl$.
$\mathbf{Q}_{N'-1}$ can  be factored by dividing $F_{\ell'}$ by either of the  units.
Depending on which unit is divided into $F_{\ell'}$ the factorization will be  a row or a column reduction,
$\mathbf{Q}_{N'-1} = \mathbf{V}_{N'-1}\mathbf{Q}_{N'}$ or $\mathbf{Q}_{N'-1} =\mathbf{Q}_{N'}\mathbf{V}_{N'-1}$,  
where $\mathbf{Q}_{N'}$ has the form~\eqref{def_QN} and $\bsy{\Delta}_{N'-1}=\mathbf{I}$.  
In fact, these two cases produce identical results.

\begin{lem}\label{lem:termination}
Let $\mathbf{Q}_{N'-1}(z)$ be a PR matrix with coprime rows and columns and a single zero entry diagonally opposite entry $F\neq 0$.  Then dividing $F$ by either of the other two (unit) entries $\kappa_0$ or $\kappa_1$ results in the same factorization of $\mathbf{Q}_{N'-1}(z)$.
\end{lem}

\emph{Proof:}
Suppose  that the  units are on the main diagonal, e.g., $\mathbf{Q}_{N'-1} =\bigl[\begin{smallmatrix} \kappa_0 & F \\
0 & \kappa_1\end{smallmatrix}\bigr]$.
(The lower-triangular case is similar.)
Row reduction (division of $F$ by $\kappa_1$) yields
\begin{align*}
\begin{bmatrix}
\kappa_0 & F \\
0 & \kappa_1 \vspace{-1pt}%
\end{bmatrix}
&=
\begin{bmatrix}
1 & F/\kappa_1 \\
0 & 1 \vspace{-1pt}%
\end{bmatrix}
\hspace{-3pt}
\begin{bmatrix}
\kappa_0 & 0 \\
0 & \kappa_1 \vspace{-1pt}%
\end{bmatrix}
=
\begin{bmatrix}
\kappa_0 & 0 \\
0 & \kappa_1 \vspace{-1pt}%
\end{bmatrix}
\hspace{-3pt}
\begin{bmatrix}
1 & F/\kappa_0 \\
0 & 1 \vspace{-1pt}%
\end{bmatrix}
\end{align*}
using~\cite[Formula~(2.33)]{Bris:23:Factoring-PRFBs-Causal}.  The same  result is obtained by column reduction (division of $F$ by $\kappa_0$).  

Now suppose that  units are on the antidiagonal, e.g., $\mathbf{Q}_{N'-1} =\bigl[\begin{smallmatrix} F & \kappa_0 \\
\kappa_1 & 0 \end{smallmatrix}\bigr]$.
Row reduction gives
\begin{align*}
\begin{bmatrix}
F & \kappa_0 \\
\kappa_1 & 0 \vspace{-1pt}%
\end{bmatrix}
&=
\begin{bmatrix}
1 & F/\kappa_1  \\
0 & 1 \vspace{-1pt}%
\end{bmatrix}
\hspace{-3pt}
\begin{bmatrix}
0 & \kappa_0 \\
\kappa_1 & 0 \vspace{-1pt}%
\end{bmatrix}
=
\begin{bmatrix}
\kappa_0 & 0 \\
0 & \kappa_1 \vspace{-1pt}%
\end{bmatrix}
\hspace{-3pt}
\begin{bmatrix}
1 & F/\kappa_0 \\
0 & 1 \vspace{-1pt}%
\end{bmatrix}
\hspace{-3pt}
\begin{bmatrix}
0 & 1 \\
1 & 0 \vspace{-1pt}%
\end{bmatrix},
\end{align*}
and column reduction produces identical results using~\cite[(43)]{Bris:23:Factoring-PRFBs-Causal},
\begin{align*}
\begin{bmatrix}
F & \kappa_0 \\
\kappa_1 & 0 \vspace{-1pt}%
\end{bmatrix}
&=
\begin{bmatrix}
0 & \kappa_0 \\
\kappa_1 & 0 \vspace{-1pt}%
\end{bmatrix}
\hspace{-3pt}
\begin{bmatrix}
1 & 0  \\
F/\kappa_0 & 1 \vspace{-1pt}%
\end{bmatrix}
=
\begin{bmatrix}
\kappa_0 & 0 \\
0 & \kappa_1 \vspace{-1pt}%
\end{bmatrix}
\hspace{-3pt}
\begin{bmatrix}
1 & 0 \\
F/\kappa_0 & 1 \vspace{-1pt}%
\end{bmatrix}^{\!\dbldag}
\hspace{-2pt}
\begin{bmatrix}
0 & 1 \\
1 & 0 \vspace{-1pt}%
\end{bmatrix}.
\end{align*}
The other antidiagonal case is similar.  \qed

Degree reduction in one row or column  may be offset by degree \emph{growth} in the other row or column.  
For instance, if the filter $F$ in Lemma~\ref{lem:termination} is  a unit (a nonzero constant) then dividing $F$ into  $\kappa_0$ or $\kappa_1$ merely moves the  zero in $\mathbf{Q}$ around  without creating a second zero, and the factorization  fails to terminate. 
\begin{exmp}\label{exmp:reduction_failure}
Assume $F$ is a unit in  the following matrix and divide $\kappa_0$ by $F$ in column~0 to left-factor off an upper triangular lifting step.
\begin{align*}
\begin{bmatrix}
\kappa_0 & 0 \\
F & \kappa_1 \vspace{-1pt}%
\end{bmatrix}
&=
\begin{bmatrix}
1 & \kappa_0/F \\
0 & 1 \vspace{-1pt}%
\end{bmatrix}
\hspace{-3pt}
\begin{bmatrix}
0 & -\kappa_0\kappa_1/F \\
F & \kappa_1 \vspace{-1pt}%
\end{bmatrix}\text{\ \ The zero moves to the upper-left corner of the quotient matrix,}\\
&=
\begin{bmatrix}
1 & \kappa_0/F \\
0 & 1 \vspace{-1pt}%
\end{bmatrix}
\hspace{-3pt}
\begin{bmatrix}
\kappa_0 & -\kappa_0\kappa_1/F \\
0 & \kappa_1 \vspace{-1pt}%
\end{bmatrix}
\hspace{-3pt}
\begin{bmatrix}
1 & 0\\
F/\kappa_1 & 1\vspace{-1pt}
\end{bmatrix}\text{\ \  then to the lower-left after dividing $F$ by $\kappa_1$ in row~1,}\\
%
&=
\begin{bmatrix}
1 & \kappa_0/F \\
0 & 1 \vspace{-1pt}%
\end{bmatrix}
\hspace{-3pt}
\begin{bmatrix}
1 & 0\\
-F/\kappa_0 & 1
\end{bmatrix}
\hspace{-3pt}
\begin{bmatrix}
\kappa_0 & -\kappa_0\kappa_1/F \\
F & 0 \vspace{-1pt}%
\end{bmatrix}
\hspace{-3pt}
\begin{bmatrix}
1 & 0\\
F/\kappa_1 & 1\vspace{-1pt}
\end{bmatrix}\text{\ \ after dividing $\kappa_1$ by $-\kappa_0\kappa_1/F$ in column~1,}\\
&=
\begin{bmatrix}
1 & \kappa_0/F \\
0 & 1 \vspace{-1pt}%
\end{bmatrix}
\hspace{-3pt}
\begin{bmatrix}
1 & 0\\
-F/\kappa_0 & 1
\end{bmatrix}
\hspace{-3pt}
\begin{bmatrix}
\kappa_0 & 0 \\
F & \kappa_1 \vspace{-1pt}%
\end{bmatrix}
\hspace{-3pt}
\begin{bmatrix}
1 & -\kappa_1/F\\
0 & 1
\end{bmatrix}
\hspace{-3pt}
\begin{bmatrix}
1 & 0\\
F/\kappa_1 & 1\vspace{-1pt}
\end{bmatrix}\text{\ \ after dividing $-\kappa_0\kappa_1/F$ by $\kappa_0$  in row~0.}
\end{align*}
\end{exmp}

To avoid such pathologies, once the quotient contains a zero the user should use one of the two equivalent ways in Lemma~\ref{lem:termination}  for reducing the quotient to a unique diagonal or antidiagonal form.

\subsubsection{Simplification to Irreducible Form}\label{sec:PRFB:Factor:Irreducible}
Ordinarily a user will factor off lifting steps so that the  factorization is automatically irreducible (Definition~\ref{defn:StdCausalForm}(\ref{irreducibility})), for instance by performing only row reductions (left factorizations) and alternating the update characteristic $\upchi_n$ of the lifting steps.  If  one employs both row and column reductions then it is possible to generate a decomposition~\eqref{factor_to_QN_2} that is not initially irreducible, in which case one has to multiply adjacent lifting steps with the same update characteristic to reduce the decomposition to irreducible form, e.g.,
%
%
%
\begin{align*}
\begin{bmatrix}
1 & V(z) \\
0 & 1\vspace{-2pt}%
\end{bmatrix}
\hspace{-3pt}
\begin{bmatrix}
z^{-m} & 0\\
0 & 1\vspace{-2pt}%
\end{bmatrix}
\hspace{-3pt}
\begin{bmatrix}
1 & V'(z) \\
0 & 1\vspace{-2pt}%
\end{bmatrix}
\hspace{-3pt}
\begin{bmatrix}
z^{-m'} & 0\\
0 & 1\vspace{-2pt}%
\end{bmatrix}
&=
\begin{bmatrix}
1 & V(z)+z^{-m}V'(z) \\
0 & 1\vspace{-2pt}%
\end{bmatrix}
\hspace{-3pt}
\begin{bmatrix}
z^{-(m+m')} & 0\\
0 & 1\vspace{-2pt}%
\end{bmatrix}\quad\text{or}\\
\begin{bmatrix}
z^{-m} & 0\\
0 & 1\vspace{-2pt}%
\end{bmatrix}
\hspace{-3pt}
\begin{bmatrix}
1 & 0\\
V(z)  & 1\vspace{-2pt}%
\end{bmatrix}
\hspace{-3pt}
\begin{bmatrix}
z^{-m''} & 0\\
0 & 1\vspace{-2pt}%
\end{bmatrix}
\hspace{-3pt}
\begin{bmatrix}
1 & 0\\
V''(z)  & 1\vspace{-2pt}%
\end{bmatrix}
&=
\begin{bmatrix}
z^{-(m+m'')} & 0\\
0 & 1\vspace{-2pt}%
\end{bmatrix}
\hspace{-3pt}
\begin{bmatrix}
1 & 0 \\
z^{-m''}V(z)+V''(z) & 1\vspace{-2pt}%
\end{bmatrix}.
\end{align*}

At the point in~\eqref{factor_to_QN_2} where the left- and right-handed lifting steps meet,
coprimality  implies that either $\bsy{\Delta}_j=\bsy{\Delta}_{N'-1}=\mathbf{I}$ or $\bsy{\Delta}''_m=\bsy{\Delta}_{N'-1}=\mathbf{I}$, as shown in  Section~\ref{sec:PRFB:Factor:Terminate}.
Suppose  $\mathbf{V}'_{\!j}$ and $\mathbf{V}''_{\!m}$ have the same update characteristic; if $\bsy{\Delta}_{j}\bsy{\Delta}''_m$ has the same  characteristic as $\mathbf{V}'_{\!j}$ and $\mathbf{V}''_{\!m}$ then, e.g.,
\begin{align*}
\mathbf{V}'_{\!j}\bsy{\Delta}_{j}\bsy{\Delta}''_m\mathbf{V}''_{\!m} 
&=
\begin{bmatrix}
1 & V(z) \\
0 & 1\vspace{-2pt}%
\end{bmatrix}
\hspace{-3pt}
\begin{bmatrix}
z^{-m} & 0\\
0 & 1\vspace{-2pt}%
\end{bmatrix}
\hspace{-3pt}
\begin{bmatrix}
1 & V''(z) \\
0 & 1\vspace{-2pt}%
\end{bmatrix}
=
\begin{bmatrix}
1 & V(z)+z^{-m}V''(z) \\
0 & 1\vspace{-2pt}%
\end{bmatrix}
\hspace{-3pt}
\begin{bmatrix}
z^{-m} & 0\\
0 & 1\vspace{-2pt}%
\end{bmatrix} . 
\end{align*}
If  $\bsy{\Delta}_{j}\bsy{\Delta}''_m$ has the \emph{opposite}  update characteristic from $\mathbf{V}'_{\!j}$  then we get a  right-handed lifting step,
\begin{align*}
\mathbf{V}'_{\!j}\bsy{\Delta}_{j}\bsy{\Delta}''_m\mathbf{V}''_{\!m} 
&=
\begin{bmatrix}
1 & V(z) \\
0 & 1\vspace{-2pt}%
\end{bmatrix}
\hspace{-3pt}
\begin{bmatrix}
1 & 0\\
0 & z^{-m''}\vspace{-2pt}%
\end{bmatrix}
\hspace{-3pt}
\begin{bmatrix}
1 & V''(z) \\
0 & 1\vspace{-2pt}%
\end{bmatrix}
=
\begin{bmatrix}
1 & 0\\
0 & z^{-m''}\vspace{-2pt}%
\end{bmatrix}
\hspace{-3pt}
\begin{bmatrix}
1 & z^{-m''}V(z)+V''(z) \\
0 & 1\vspace{-2pt}%
\end{bmatrix} . 
\end{align*}
%
%
Such  complications are annoying and can be avoided by partial pivoting, e.g., by factoring off  left lifting steps (row reductions) exclusively; see Section~\ref{sec:Left}.

\subsection{The Causal Complementation Algorithm}\label{sec:PRFB:CCA}
\subsubsection{Factorization Schema}\label{sec:PRFB:CCA:Schema}
We  introduce an accounting device  for tracking user choices  in lifting factorizations.   
Given $\mathbf{Q}_n$, the user  factors off either a left- or a right-handed lifting step  with  multiplicity $M$.  
Handedness is encoded with a letter, $\eta=L$ or $R$.
The \emph{dividend index} $\delta$, the row or column index of the dividends $(E_0,E_1)$,  is encoded with a bit, $\delta=0$ or 1.  
Finally, the user chooses a divisor $\Fl$; the \emph{divisor index} $\ell$ is also encoded with a bit, $\ell=0$ or 1.  
Corollary~\ref{cor:DRC} ensures uniqueness of degree-reducing causal complements so these choices, which are encoded in  a \emph{lifting step schema}, $(\eta_n,M_n,\delta_n,\ell_n)$, determine extraction of a well-defined lifting step. A complete factorization is  specified  by a sequence of such strings, called a \emph{factorization schema}.  By Lemma~\ref{lem:termination}  the final lifting step has no user options, so an  $N$-step lifting factorization is unambiguously specified by a coprimification and an $(N-1)$-step factorization  schema,
%
\begin{align}\label{factorization_schema}
\text{schema} = (\rho_0,\rho_1,c_0,c_1:\,\eta_0,M_0,\delta_0,\ell_0;\,
\ldots;\,\eta_{N-2},M_{N-2},\delta_{N-2},\ell_{N-2}).
\end{align}
%
We only specify a coprimification $(\rho_0,\rho_1,c_0,c_1)$ for filter banks  with \emph{non-}coprime rows or columns.

For instance, the factorization 
of LGT(5,3)  in~\cite[Section~2.2.1]{Bris:23:Factoring-PRFBs-Causal}
reduces row $\delta=0$ by dividing in column $\ell=0$ with $M=0$. The schema for this first  step is  $(\eta_0,M_0,\delta_0,\ell_0)=(L,0,0,0)$.  The next  step reduces row~1 by dividing  in column~0 with $M=0$, so its lifting step schema is $(\eta_1,M_1,\delta_1,\ell_1)=(L,0,1,0)$.   By Lemma~\ref{lem:termination} there are no options for factoring $\mathbf{Q}_2$, which contains a zero, so the factorization schema  is
\begin{align}\label{LGT_col0_basic_schema}
\schema(\text{LGT in col. 0}) = (L,0,0,0;\;L,0,1,0). 
\end{align}

Note that  \emph{schema are not unique}; i.e.,  distinct schema may yield the same factorization. For instance, the remainders $R_0=0$ and $R_1=-z^{-1}$ in the second step of~\cite[Section~2.2.1]{Bris:23:Factoring-PRFBs-Causal}  are not coprime, so  the same lifting step is obtained using the SGDA with $M_1=1$, implying that  $(L,0,1,0)$ and $(L,1,1,0)$ yield the same result. In this case we  abbreviate these equivalent schema as $(L,\{0,1\},1,0)$, placing values that yield the same result in braces,  so~\eqref{LGT_col0_basic_schema} is also specified by
\begin{align}\label{LGT_col0_advanced_schema}
\schema(\text{LGT in col. 0}) = (L,0,0,0;\;L,\{0,1\},1,0).
\end{align}

\subsubsection{Causal Complementation Algorithm}\label{sec:PRFB:CCA:CCA}
The  process  in Section~\ref{sec:PRFB:Factor}  is now formalized as the Causal Complementation Algorithm (CCA). The CCA  reproduces all causal EEA lifting factorizations performed on any row or column of a causal PR transfer matrix and generates other factorizations not computable using the causal EEA, a level of generality that is responsible for some of its complexity.  If one restricts to left lifting factorizations only (Section~\ref{sec:Left}), then line~\ref{alg:CCA:eta} and all control branching based on $\eta_n$ (handedness) disapppears from Algorithm~\ref{alg:CCA}, along with line~\ref{alg:CCA:irreducible}.
\begin{alg}
\caption{\bf(Causal Complementation Algorithm).}\label{alg:CCA}
\textbf{Input:}  Causal FIR PR matrix $\mathbf{H}(z)$. 

\textbf{Output:}  Factorization in standard causal lifting form~\eqref{std_causal_form}.

\algsetup{indent=2em, linenosize=\normalsize, linenodelimiter=.}
\begin{algorithmic}[1]
\STATE \textbf{Choose} an initial coprimification,\label{alg:CCA:coprimify}
\begin{align}\label{alg:CCA:coprimify_H}
\mathbf{H}(z) = \diag(z^{-\rho_0},z^{-\rho_1})\,\mathbf{Q}_0(z)\,\diag(z^{-c_0},z^{-c_1}).
\end{align}

\STATE \textbf{Initialize} $n\leftarrow 0$.

\WHILE{($\mathbf{Q}_n$ is zero-free)}
	
	\STATE  \textbf{Choose} the lifting handedness, $\eta_n \eqdef L$ (or, resp., $\eta_n=R$).\label{alg:CCA:eta}
	
	\STATE  \textbf{Choose} a row (resp., column) of dividends, $(E_0,E_1)$, with dividend index $\delta_n$ and complementary row (resp., column) $(F_0,F_1)$.\label{alg:CCA:delta}

	\STATE  \textbf{Choose} the multiplicity $M_n$ of the causal complement to be computed,  $0\leq M_n\leq \deg|\mathbf{Q}_n|$.
		
	\STATE  \textbf{Choose} a divisor $F_{\ell_n}$ in column (resp., row) $\ell_n\in\{0,1\}$ that satisfies $\gcd\bigl(F_{\ell_n},z^{-M_n}\bigr)=1$. \label{alg:CCA:ell}

	\STATE  Use Theorem~\ref{thm:CCT} (the CCT) with the SGDA to construct the unique lifting filter $S$ and  causal complement $(R_0,R_1)$ to $(F_0,F_1)$ for $|\mathbf{Q}_n|$ that is degree-reducing modulo~$M_n$ in $F_{\ell_n}$.

	\STATE  $R\leftarrow\gcd(R_0,R_1)$ and $\wt{R}_i\leftarrow R_i/R,\; i=0,1$.

	\STATE  Set  update characteristic $\upchi_n\eqdef \delta_n$ (resp., $1-\delta_n$).\label{alg:CCA:chi}

	\STATE  Set $\mathbf{V}_n \eqdef \upsilon(S)$ if $\upchi_n=0$, else  $\mathbf{V}_n \eqdef \lambda(S)$.\label{alg:CCA:V}

	\STATE  Set $\bsy{\Delta}_n \eqdef \diag(R,1)$ if $\delta_n=0$, else  $\bsy{\Delta}_n \eqdef \diag(1,R)$.\label{alg:CCA:Delta}

	\STATE  Define $\mathbf{Q}_{n+1}$: put $\bigl(\wt{R}_0,\wt{R}_1\bigr)$ in row (resp., column) $\delta_n$ and $(F_0,F_1)$ in row (resp., column) $1-\delta_n$.\label{alg:CCA:Qnplus1}
	
	\STATE  Update  factorization~\eqref{alg:CCA:coprimify_H} by replacing $\mathbf{Q}_n$  with 
$\mathbf{V}_{\!n}\bsy{\Delta}_n\mathbf{Q}_{n+1}$ (resp., $\mathbf{Q}_{n+1}\bsy{\Delta}_n\mathbf{V}_{\!n}$).\label{alg:CCA:UpdateQn}

	\STATE  $n\leftarrow n+1$.
	
\ENDWHILE

\STATE Factor  $\mathbf{Q}_n$ per Lemma~\ref{lem:termination} and update factorization~\eqref{alg:CCA:coprimify_H}.  \label{alg:CCA:finalstep}

\STATE  Move $\mathbf{D}_{\kappa_0,\kappa_1}$ to the left end of the factorization using~\cite[Formula~(41)]{Bris:23:Factoring-PRFBs-Causal}.

\STATE  If $\mathbf{P}_0=\mathbf{J}$ then move $\mathbf{P}_0$ to the right end of the factorization using~\cite[Formula~(43)]{Bris:23:Factoring-PRFBs-Causal}.

\STATE  Simplify the factorization to irreducible form per Section~\ref{sec:PRFB:Factor:Irreducible} if necessary.\label{alg:CCA:irreducible}

\STATE  Renumber lifting steps from right to left as in~\eqref{std_causal_form}.

\RETURN  factorization~\eqref{alg:CCA:coprimify_H} in standard causal lifting form.
\end{algorithmic}
\end{alg}

\paragraph*{Notes regarding Algorithm~\ref{alg:CCA}}\mbox{}\newline
\noindent Line~\ref{alg:CCA:coprimify}:  Expression~\eqref{alg:CCA:coprimify_H} will be updated with lifting steps as they are factored off of the quotient matrices $\mathbf{Q}_n$; \eqref{alg:CCA:coprimify_H}  will be in standard causal lifting form upon termination.

\noindent Line~\ref{alg:CCA:eta}:  Several steps in the \textbf{while} loop depend on whether the lifting is left- or, respectively, right-handed.  

\noindent Line~\ref{alg:CCA:delta}:  The dividends $(E_0,E_1)$  come from row $\delta_n$ of $\mathbf{Q}_n$ if $\eta_n = L$ or from  column $\delta_n$ if $\eta_n = R$.  

\noindent Line~\ref{alg:CCA:ell}: Such a choice always exists by Theorem~\ref{thm:CCT}.  If $\eta_n = L$ then  $(F_0,F_1)$ is  a row  in $\mathbf{Q}_n$ so $\ell_n$ is the column index of $F_{\ell_n}$, while $\ell_n$ is the row index of $F_{\ell_n}$ if $\eta_n = R$.

\noindent Line~\ref{alg:CCA:chi}:  As in~\cite{ISO_15444_2}, we  set $\upchi_n\eqdef\upchi(\mathbf{V}_n)=0$ iff $\mathbf{V}_n$ is upper-triangular, regardless of whether $\mathbf{V}_n$ acts on the left or the right, so $\upchi_n= \delta_n$ if $\eta_n = L$ while $\upchi_n= 1-\delta_n$  if $\eta_n = R$.

\noindent Line~\ref{alg:CCA:V}:  $\upsilon$ and $\lambda$ are defined in~\eqref{def_lambda_upsilon}.

\noindent Line~\ref{alg:CCA:Delta}:  Examination of~\eqref{upper_triang_left_lift}--\eqref{upper_triang_right_lift} shows that the update characteristic of $\bsy{\Delta}_n$ is equal to $\delta_n$.

\noindent Line~\ref{alg:CCA:Qnplus1}:  As explained in Section~\ref{sec:PRFB:Factor:Downlifting}, downlifting always leaves $\mathbf{Q}_{n+1}$ with coprime rows and columns.

\noindent Line~\ref{alg:CCA:finalstep}:  $\mathbf{Q}_n=\mathbf{Q}_{N'-1}$ has a zero  so factorization is uniquely determined  (no  options) per Lemma~\ref{lem:termination}.


\section{The  Cubic B-Spline Filter Bank}\label{sec:CubicBSpline} 

We continue the study, begun in~\cite[Section~6]{Bris:23:Factoring-PRFBs-Causal}, of  lifting factorizations for the causal 7-tap/5-tap Cohen-Daubechies-Feauveau linear phase   filter bank, which corresponds to a  cubic B-spline  wavelet transform~\cite[\S 6.A]{CohenDaubFeau92}, \cite[\S 8.3.4]{Daub92}.  The   causal CDF(7,5)  polyphase-with-delay analysis matrix is
\begin{align}\label{CDF75}
\hspace{-1ex}\mathbf{H}(z)
&=
\begin{bmatrix}
\sst  (3 + 5z^{-1} + 5z^{-2} + 3z^{-3})/32 &\,\sst  (-3 + 10z^{-1} - 3z^{-2})/8\\
\sst (1 + 6z^{-1} + z^{-2})/8 &\,\sst  -(1 + z^{-1})/2 
\end{bmatrix},\quad 
|\mathbf{H}(z)| =  -z^{-2}.
\end{align}
We first compute the four lifting factorizations generated by the causal version of the EEA, one of which is extremely ill-conditioned.  This  verifies the claim  in~\cite{Bris:23:Factoring-PRFBs-Causal} that \emph{none} of the causal EEA factorizations of CDF(7,5) have  lifting filters that agree modulo delays with the  linear phase lifting filters obtained using the noncausal  factorization approach of Daubechies and Sweldens~\cite[\S 7.8]{DaubSwel98}. The unimodular polyphase-with-advance representation of the analysis bank for CDF(7,5) and its corresponding linear phase lifting factorization is
\begin{align}
\mathbf{A}(z) 
&=
\begin{bmatrix}
\sst (-3z + 10 - 3z^{-1})/8  &\,\sst  (3z + 5 + 5z^{-1} + 3z^{-2})/32 \\
\sst -(z + 1)/2 &\,\sst (z + 6 + z^{-1})/8
\end{bmatrix}\label{CDF75_PWA_anal_lifting}
=
\begin{bmatrix}
\sst 2 &\sst  0\\
\sst 0 &\sst  1/2
\end{bmatrix}
\hspace{-4pt}
\begin{bmatrix}
\sst 1 &\sst  3(1 + z^{-1})/16\\
\sst 0 &\sst  1
\end{bmatrix}
\hspace{-4pt}
\begin{bmatrix}
\sst 1 &\sst  0\\
\sst -(z + 1) &\sst  1
\end{bmatrix}
\hspace{-4pt}
\begin{bmatrix}
\sst 1 &\sst  -(1 + z^{-1})/4\\
\sst 0 &\sst  1
\end{bmatrix}.
\end{align}
The CCA is used in Section~\ref{sec:CubicBSpline:CCA} to reproduce the factorizations generated by the causal EEA, and these computations are compared in Section~\ref{sec:CubicBSpline:Comparison} to show that the CCA entails a lower factorization cost than the EEA. Section~\ref{sec:CubicBSpline:Other} then generates  CCA factorizations  using the SGDA that are not computable using the causal EEA, including a causal linear phase analogue of~\eqref{CDF75_PWA_anal_lifting}.

\subsection{Factorization via the causal Extended Euclidean Algorithm}\label{sec:CubicBSpline:EEA}
\subsubsection{EEA in Column~0}\label{sec:CubicBSpline:EEA:Col0}  
Initialize  remainders $r_0 \eqdef H_{00}(z) = (3 + 5z^{-1} + 5z^{-2} + 3z^{-3})/32$ and 
$r_1 \eqdef H_{10}(z) = (1 + 6z^{-1} + z^{-2})/8$.
Using the classical  division algorithm, $r_0 = q_0 r_1 + r_2$ where $q_0 = (-13 + 3z^{-1})/4$, $r_2 = (1 + 5z^{-1})/2$, $\deg(r_2)<\deg(r_1)$, and we can write
\begin{align*}
\begin{pmatrix}
r_0\\
r_1
\end{pmatrix}
=
\mathbf{M}_0
\begin{pmatrix}
r_1\\
r_2
\end{pmatrix}
\text{ for\ \ }
\mathbf{M}_0 \eqdef
\begin{bmatrix}
q_0 & 1\\
1 & 0
\end{bmatrix} .
\end{align*}
Next, $r_1 = q_1 r_2 + r_3$ for $q_1 = (29 + 5z^{-1})/100$, $r_3 = -1/50$, and 
\begin{align*}
\begin{pmatrix}
r_1\\
r_2
\end{pmatrix}
=
\mathbf{M}_1
\begin{pmatrix}
r_2\\
r_3
\end{pmatrix}
\text{ where\ \ }
\mathbf{M}_1 \eqdef
\begin{bmatrix}
q_1 & 1\\
1 & 0
\end{bmatrix} .
\end{align*}
Then $r_2 = q_2 r_3 + r_4$ with $q_2 = -25(1 + 5z^{-1})$, $r_4 = 0$, and
\begin{align*}
&\begin{pmatrix}
r_2\\
r_3
\end{pmatrix}
=
\mathbf{M}_2
\begin{pmatrix}
r_3\\
0
\end{pmatrix}
\text{ for\ \ }
\mathbf{M}_2 \eqdef
\begin{bmatrix}
q_2 & 1\\
1 & 0\vspace{-1pt}  
\end{bmatrix} .
\end{align*}
Define $\mathbf{H}'(z) $ by augmenting with unknown filters $a_0$ and $a_1$ as in the EEA examples of~\cite{Bris:23:Factoring-PRFBs-Causal},
\begin{align}
&\mathbf{H}'(z) 
\eqdef
\begin{bmatrix}
r_0 & a_0\\
r_1 & a_1\vspace{1pt}  
\end{bmatrix}
=
\mathbf{M}_0\mathbf{M}_1\mathbf{M}_2
\begin{bmatrix}
r_3 & 0\\
0 & -|\mathbf{H}|/r_3
\end{bmatrix}\label{EEA_col0_def_H'}
=
\begin{bmatrix}
\sst  (3 + 5z^{-1} + 5z^{-2} + 3z^{-3})/32 &\,\sst  -z^{-2}(23 + 22z^{-1} + 15z^{-2})/8\\
\sst (1 + 6z^{-1} + z^{-2})/8 &\,\sst  -z^{-2}(29 + 5z^{-1})/2 
\end{bmatrix} .
\end{align}
$|\mathbf{H}'(z)|=|\mathbf{H}(z)|$  and the matrices agree in column~0, so the  Lifting Theorem~\cite[Corollary~3.6]{Bris:23:Factoring-PRFBs-Causal} says 
$\mathbf{H}(z)=\mathbf{H'}(z)\mathbf{S}(z)$,
\begin{align*}
\begin{bmatrix}
H_{00} & H_{01}\\
H_{10} & H_{11} \vspace{-1pt}
\end{bmatrix}
&=
\begin{bmatrix}
r_0 & a_0\\
r_1 & a_1\vspace{1pt}  
\end{bmatrix}
\hspace{-3pt}
\begin{bmatrix}
1 & S\\
0 & 1 \vspace{-1pt}
\end{bmatrix}
\text{\ iff\ }
\left\{\begin{array}{l}
H_{01} = r_0S + a_0 \\
H_{11} = r_1S + a_1 \vspace{-1pt}
\end{array}\right.
\end{align*}
Solve for $S(z)$ and lift, applying~\cite[Formulas~(41), (43)]{Bris:23:Factoring-PRFBs-Causal} 
to get standard causal lifting form,
\begin{align}
S(z) &= \bigl(H_{01}(z) - a_0(z)\bigr)/r_0(z) = -4 + 20z^{-1},\label{S_CDF75_col0}\\
\mathbf{H}(z) 
&=
\mathbf{M}_0\mathbf{J}^2\mathbf{M}_1\mathbf{M}_2\mathbf{J}^2
\begin{bmatrix}
r_3 & 0\\
0 & -|\mathbf{H}|/r_3
\end{bmatrix}
\mathbf{S}
\;=\;
(\mathbf{M}_0\mathbf{J})(\mathbf{J}\mathbf{M}_1)(\mathbf{M}_2\mathbf{J})
\begin{bmatrix}
-|\mathbf{H}|/r_3 & 0\\
0 & r_3
\end{bmatrix}
\mathbf{S}^{\dbldag}\mathbf{J}\nonumber\\
&=
\begin{bmatrix}
\sst -50 &\sst  0\\
\sst 0 &\sst  -1/50
\end{bmatrix}
\hspace{-4pt}
\begin{bmatrix}
\sst 1 &\sst  (-13 + 3z^{-1})10^{-4}\\
\sst 0 &\sst  1
\end{bmatrix}
\hspace{-4pt}
\begin{bmatrix}
\sst 1 &\sst  0\\
\sst 725 + 125z^{-1} &\sst 1
\end{bmatrix}
\hspace{-4pt}
\begin{bmatrix}
\sst 1 &\sst  -(1 + 5z^{-1})10^{-2}\\
\sst 0 &\sst  1
\end{bmatrix}
\hspace{-4pt}
\begin{bmatrix}
\sst z^{-2} &\sst  0\\
\sst 0 &\sst  1
\end{bmatrix}
\hspace{-4pt}
\begin{bmatrix}
\sst 1 &\sst  0\\
\sst -4 + 20z^{-1} &\sst  1
\end{bmatrix}
\hspace{-4pt}
\begin{bmatrix}
\sst 0 &\sst  1\\
\sst 1 &\sst  0
\end{bmatrix} .\label{CDF75_EEA_col0}
\end{align}

Unfortunately, \eqref{CDF75_EEA_col0}  is  rather ill-conditioned. Use \eqref{lifting_step_cond_nr} and \eqref{S_infty} from Theorem~A.1 in 
\ref{app:Condition} to compute  condition numbers for the individual lifting matrices.
\begin{align*}
\|U_0\|^2_{\infty} &=  24^2 = 576 		&\cond(\mathbf{U}_0)	&= 577.998  \\
\|U_1\|^2_{\infty} &= 0.06^2 = 0.0036 	&\cond(\mathbf{U}_1)	&= 1.06183 \\
\|U_2\|^2_{\infty} &=  850^2 = 722500 	&\cond(\mathbf{U}_2)	&= 722502  \\
\|U_3\|^2_{\infty} &=  2.56\times 10^{-6}	&\cond(\mathbf{U}_3)	&= 1.00160   \\
&&\cond(\mathbf{D}_{\kappa_0,\kappa_1}) 					&= 2500
\end{align*}
The conditioning product~\eqref{condition_number_cascade_bound} for the cascade~\eqref{CDF75_EEA_col0} is
\begin{align}\label{EEA_col0_cascade_factor}
\cond(\mathbf{D}_{\kappa_0,\kappa_1}){\textstyle\prod_i} \cond(\mathbf{U}_i) = 1.1\times 10^{12}.
\end{align}
Thus, one could potentially lose  12 significant figures of precision using this cascade.  
Ill-conditioning of lifting factorizations has been investigated by Zhu and Wickerhauser~\cite{ZhuWicker:12:NearestNeighborLifting}.

\subsubsection{EEA in Column~1}\label{sec:CubicBSpline:EEA:Col1}  
In~\cite[Section~6.1]{Bris:23:Factoring-PRFBs-Causal} it was shown that EEA factorization of~\eqref{CDF75} in column~1 yields
\begin{equation}\begin{split}\label{CDF75_EEA_col1}
\mathbf{H}(z) 
=&
\begin{bmatrix}
\sst -2 &\sst  0\\
\sst 0 &\sst  -1/2
\end{bmatrix}
\hspace{-4pt}
\begin{bmatrix}
\sst 1 &\sst  (-13 + 3z^{-1})/16\\
\sst 0 &\sst  1
\end{bmatrix}
\hspace{-4pt}
\begin{bmatrix}
\sst 1 &\sst  0\\
\sst 1 + z^{-1} &\sst 1
\end{bmatrix}
\hspace{-4pt}
\begin{bmatrix}
\sst 1 & \sst 0\\
\sst 0 &\sst z^{-2}
\end{bmatrix}
\hspace{-4pt}
\begin{bmatrix}
\sst 1 &\sst  -(1 + 5z^{-1})/4\\
\sst 0 &\sst  1
\end{bmatrix}
\hspace{-4pt}
\begin{bmatrix}
\sst 0 &\sst  1\\
\sst 1 &\sst  0
\end{bmatrix} .
\end{split}\end{equation}

\subsubsection{EEA in Row~0}\label{sec:CubicBSpline:EEA:Row0}  
Initialize  remainders
$r_0 \eqdef H_{00}(z) = (3 + 5z^{-1} + 5z^{-2} + 3z^{-3})/32$,
$r_1 \eqdef H_{01}(z) = (-3 + 10z^{-1} - 3z^{-2})/8$.
The classical  division algorithm gives $r_0 = q_0 r_1 + r_2$ for $q_0 = -(5 + z^{-1})/4$,  $r_2 = (-3 + 13z^{-1})/8$,  so
\begin{align*}
(r_0,r_1)
=
(r_1,r_2)\mathbf{M}_0
\text{\, for\ \ }
\mathbf{M}_0 \eqdef
\begin{bmatrix}
q_0 & 1\\
1 & 0
\end{bmatrix} .
\end{align*}
Next, $r_1 = q_1 r_2 + r_3$ where $q_1 = (121 - 39z^{-1})/169$ and  $r_3 = -18/169$, giving
\begin{align*}
(r_1,r_2)
=
(r_2,r_3)\mathbf{M}_1
\text{\, for\ \ }
\mathbf{M}_1 \eqdef
\begin{bmatrix}
q_1 & 1\\
1 & 0\vspace{-1pt}  
\end{bmatrix} .
\end{align*}
Finally, $r_2 = q_2 r_3 + r_4$ where $q_2 = 169(3 - 13z^{-1})/144$,  $r_4 = 0$, and
\begin{align*}
(r_2,r_3)
=
(r_3,0)\mathbf{M}_2
\text{\, for\ \ }
\mathbf{M}_2 \eqdef
\begin{bmatrix}
q_2 & 1\\
1 & 0\vspace{-1pt}  
\end{bmatrix} .
\end{align*}
Augment with causal filters $a_0$ and $a_1$ and define $\mathbf{H}'(z)$ so as to generate $r_0$ and $r_1$,
\begin{align*}
\mathbf{H}'(z) 
&\eqdef
\begin{bmatrix}
r_0 & r_1\\
a_0 & a_1\vspace{1pt}  
\end{bmatrix}
=
\begin{bmatrix}
r_3 & 0 \\
0 & -|\mathbf{H}|/r_3\vspace{-1pt}  
\end{bmatrix}
\mathbf{M}_2\mathbf{M}_1\mathbf{M}_0
=
\begin{bmatrix}
\sst  (3 + 5z^{-1} + 5z^{-2} + 3z^{-3})/32 &\,\sst  (-3 + 10z^{-1} - 3z^{-2})/8\\
\sst -z^{-2}(71 + 74z^{-1} + 39z^{-2})/72 &\,\sst  z^{-2}(-121 + 39z^{-1})/18 
\end{bmatrix} .
\end{align*}
$|\mathbf{H}'|=|\mathbf{H}|$ and the matrices agree in the top row so $\mathbf{H}(z)$ can be left-lifted from $\mathbf{H}'(z)$,  
\begin{align*}
\mathbf{H}(z)
&=
\mathbf{S}(z)\mathbf{H}'(z)
=
\begin{bmatrix}
1 & 0\\
4(3 + 13z^{-1})/9 & 1 \vspace{-1pt}
\end{bmatrix}
\mathbf{H}'(z).
\end{align*}
Use~\cite[(41)]{Bris:23:Factoring-PRFBs-Causal} 
to put the factorization in standard causal lifting form,
\begin{align}   
\mathbf{H}(z) 
&= 
\mathbf{S}(z)
\begin{bmatrix}
r_3 & 0 \\
0 & -|\mathbf{H}|/r_3\vspace{-1pt}  
\end{bmatrix}
(\mathbf{M}_2\mathbf{J})(\mathbf{J}\mathbf{M}_1)(\mathbf{M}_0\mathbf{J})\mathbf{J} \label{CDF75_EEA_row0}\\
&\hspace{-1.5em}=
\begin{bmatrix}
\sst  -18/169 &\,\sst 0 \\
\sst 0 &\,\sst -169/18
\end{bmatrix}
\hspace{-4pt}
\begin{bmatrix}
\sst 1 &\,\sst 0 \\
\sst 144(3 + 13z^{-1})/13^4 &\,\sst 1
\end{bmatrix}
\hspace{-4pt}
\begin{bmatrix}
\sst 1 &\,\sst 0 \\
\sst 0 &\,\sst z^{-2}
\end{bmatrix}
\hspace{-4pt}
\begin{bmatrix}
\sst 1 &\,\sst 169(3 - 13z^{-1})/144 \\
\sst 0 &\,\sst 1
\end{bmatrix}
\hspace{-4pt}
\begin{bmatrix}
\sst 1 &\,\sst 0 \\
\sst (121 - 39z^{-1})/169 &\,\sst 1
\end{bmatrix}
\hspace{-4pt}
\begin{bmatrix}
\sst 1 &\,\sst -(5 + z^{-1})/4 \\
\sst 0 &\,\sst 1
\end{bmatrix}
\hspace{-4pt}
\begin{bmatrix}
\sst 0 &\,\sst 1 \\
\sst 1 &\,\sst 0
\end{bmatrix}.\nonumber
\end{align}
While not as bad as factorization~\eqref{CDF75_EEA_col0},  the cascade~\eqref{CDF75_EEA_row0} is still somewhat ill-conditioned,
\begin{align}
\cond(\mathbf{D}_{\kappa_0,\kappa_1}){\textstyle\prod_i} \cond(\mathbf{U}_i) &= 3.4\times 10^{5}\,.\label{EEA_row0_cascade_factor}
\end{align}
%

\subsubsection{EEA in Row~1}\label{sec:CubicBSpline:EEA:Row1}  
Following  Section~\ref{sec:CubicBSpline:EEA:Row0}, the lifting factorization generated by the  causal EEA in row~1 is
\begin{equation}\begin{split}\label{CDF75_EEA_row1}
\mathbf{H}(z) 
=&
\begin{bmatrix}
\sst -2 &\,\sst 0 \\
\sst 0 &\,\sst -1/2
\end{bmatrix}
\hspace{-4pt}
\begin{bmatrix}
\sst 1  &\,\sst (3 - 13z^{-1})/16 \\
\sst 0 &\,\sst 1
\end{bmatrix}
\hspace{-4pt}
\begin{bmatrix}
\sst z^{-2}  &\,\sst 0 \\
\sst 0 &\,\sst 1
\end{bmatrix}
\hspace{-4pt}
\begin{bmatrix}
\sst 1  &\,\sst 0 \\
\sst 1 + z^{-1} &\,\sst 1
\end{bmatrix}
\hspace{-4pt}
\begin{bmatrix}
\sst 1  &\,\sst -(5 + z^{-1})/4 \\
\sst 0 &\,\sst 1
\end{bmatrix}
\hspace{-4pt}
\begin{bmatrix}
\sst 0  &\,\sst 1 \\
\sst 1 &\,\sst 0
\end{bmatrix} .
\end{split}\end{equation}
We have now shown that 
\emph{none} of the causal EEA factorizations of CDF(7,5) has lifting filters that agree modulo delays with the  linear phase lifting filters in its unimodular  factorization~\eqref{CDF75_PWA_anal_lifting}!

\subsection{Factorization via the Causal Complementation Algorithm}\label{sec:CubicBSpline:CCA}
We now reproduce the EEA factorizations using the CCA (Algorithm~\ref{alg:CCA}).   $\mathbf{H}(z)$ has coprime rows and columns so the delays $\rho_i$ and $c_i$ are  zero and hence are omitted from the factorization schema  for CDF(7,5).

\subsubsection{Division in Column~0}\label{sec:CubicBSpline:CCA:Col0}  
Initialize  $\mathbf{Q}_0(z)\eqdef \mathbf{H}(z)$, set  $E_0\leftarrow H_{00}$  and $F_0\leftarrow H_{10}$,  $\deg(F_0)\leq\deg(E_0)$, and factor
\begin{equation}\begin{split}\label{CDF75_col0_CCA_step0_form}
\hspace{-1ex}\mathbf{Q}_0(z)
&=
\begin{bmatrix}
\sst  (3 + 5z^{-1} + 5z^{-2} + 3z^{-3})/32 &\,\sst  (-3 + 10z^{-1} - 3z^{-2})/8\\
\sst (1 + 6z^{-1} + z^{-2})/8 &\,\sst  -(1 + z^{-1})/2 
\end{bmatrix} 
=
\begin{bmatrix}
E_0 & E_1\\
F_0 & F_1\vspace{-1pt}
\end{bmatrix}
=
\begin{bmatrix}
1 & S\\
0 & 1\vspace{-1pt}%
\end{bmatrix}
\hspace{-3pt}
\begin{bmatrix}
R_0 & R_1\\
F_0 & F_1\vspace{-1pt}%
\end{bmatrix}.
\end{split}\end{equation}
Divide $F_0$ into $E_0$ using  the classical division algorithm, $E_0 = F_0S + R_0$, where
\begin{align}\label{CDF75_col0_CCA_step0}
S(z) = (-13 + 3z^{-1})/4 \text{ and } R_0(z) = (1 + 5z^{-1})/2,\text{ with } \deg(R_0)  < \deg(F_0).
\end{align}
Set $R_1 \leftarrow E_1 - F_1S = -2$.  The first factorization step, with lifting step schema $(L,0,0,0)$, is
\begin{align}
\mathbf{Q}_0(z) 
&=
\begin{bmatrix}
\sst  1 &\sst   (-13 + 3z^{-1})/4\\
\sst  0 &\sst   1%
\end{bmatrix}
\hspace{-4pt}
\begin{bmatrix}
\sst  (1 + 5z^{-1})/2 &\,\sst  -2\\
\sst (1 + 6z^{-1} + z^{-2})/8 &\,\sst  -(1 + z^{-1})/2 
\end{bmatrix}
=
\mathbf{V}_0(z)\mathbf{Q}_1(z). \label{CDF75_col0_CCA_step0_result}
\end{align}
Reset the labels $E_j\leftarrow F_j$ and $F_j\leftarrow R_j$ in $\mathbf{Q}_1(z)$ and factor
\begin{equation}\begin{split}\label{CDF75_col0_CCA_step1_form}
\mathbf{Q}_1(z) 
&=
\begin{bmatrix}
\sst  (1 + 5z^{-1})/2 &\,\sst  -2\\
\sst (1 + 6z^{-1} + z^{-2})/8 &\,\sst  -(1 + z^{-1})/2 
\end{bmatrix}
=
\begin{bmatrix}
F_0 & F_1\\
E_0 & E_1\vspace{-1pt}
\end{bmatrix}
=
\begin{bmatrix}
1 & 0\\
S & 1\vspace{-2pt}%
\end{bmatrix}
\hspace{-3pt}
\begin{bmatrix}
F_0 & F_1\\
R_0 & R_1\vspace{-1pt}%
\end{bmatrix}.
\end{split}\end{equation}
Divide  $F_0$ into  $E_0$ to get $E_0 = F_0S + R_0$, where
\begin{equation}\label{CDF75_col0_CCA_step1}
S(z) = (29 + 5z^{-1})/100\text{ and } R_0 = -1/50.
\end{equation}
Set $R_1 \leftarrow E_1 - F_1S = 2(1 - 5z^{-1})/25$; the second step~\eqref{CDF75_col0_CCA_step1_form}, with schema $(L,0,1,0)$,  is
\begin{align}\label{CDF75_col0_CCA_step1A}
\mathbf{Q}_1(z) 
&=
\begin{bmatrix}
\sst  1 &\sst   0\\
\sst  (29 + 5z^{-1})/100 &\sst   1%
\end{bmatrix}
\hspace{-4pt}
\begin{bmatrix}
\sst  (1 + 5z^{-1})/2 &\,\sst  -2\\
\sst -1/50 &\,\sst 2(1 - 5z^{-1})/25
\end{bmatrix}
=
\mathbf{V}_1(z)\mathbf{Q}_2(z). 
\end{align}
Reset $E_j\leftarrow F_j$, $F_j\leftarrow R_j$  and factor
\begin{align*}
\mathbf{Q}_2(z) 
&=
\begin{bmatrix}
\sst  (1 + 5z^{-1})/2 &\,\sst  -2\\
\sst -1/50 &\,\sst 2(1 - 5z^{-1})/25
\end{bmatrix}
=
\begin{bmatrix}
E_0 & E_1\\
F_0 & F_1\vspace{-1pt}
\end{bmatrix}
=
\begin{bmatrix}
1 & S\\
0 & 1\vspace{-1pt}%
\end{bmatrix}
\hspace{-3pt}
\begin{bmatrix}
R_0 & R_1\\
F_0 & F_1\vspace{-1pt}%
\end{bmatrix}.
\end{align*}
Divide  $F_0$ into  $E_0$ to compute $E_0 = F_0S + R_0$, where
\begin{equation}\label{CDF75_col0_CCA_step2}
S(z) = -25(1 + 5z^{-1})\text{ and }  R_0 = 0.
\end{equation}
Set $R_1 \leftarrow E_1 - F_1S = -50z^{-2}$.  The third step, with  schema $(L,0,0,0)$, is
\begin{align*}
\mathbf{Q}_2(z) 
&=
\begin{bmatrix}
\sst  1 &\sst -25(1 + 5z^{-1})  \\
\sst  0 &\sst   1%
\end{bmatrix}
\hspace{-4pt}
\begin{bmatrix}
\sst z^{-2} &\sst   0\\
\sst  0 &\sst 1%
\end{bmatrix}
\hspace{-4pt}
\begin{bmatrix}
\sst  0 &\,\sst  -50\\
\sst -1/50 &\,\sst 2(1 - 5z^{-1})/25
\end{bmatrix}
=
\mathbf{V}_2(z)\bsy{\Delta}_2(z)\mathbf{Q}_3(z).
\end{align*}
Factor  a diagonal gain matrix and a swap  off of $\mathbf{Q}_3(z)$,
\begin{equation}\begin{split}\label{CDF75_col0_CCA_step3}
\mathbf{Q}_3(z)
&=
\begin{bmatrix}
\sst  -50 &\,\sst  0\\
\sst 0 &\,\sst -1/50
\end{bmatrix}
\hspace{-4pt}
\begin{bmatrix}
\sst  1 &\,\sst  0\\
\sst 4(-1 + 5z^{-1}) &\,\sst 1
\end{bmatrix}
\hspace{-4pt}
\begin{bmatrix}
\sst  0 &\,\sst 1 \\
\sst 1 &\,\sst 0
\end{bmatrix}
= 
\mathbf{D}_{-50,-1/50}\mathbf{V}_3(z)\mathbf{J}.
\end{split}\end{equation}
Putting everything together and moving the diagonal gain matrix using~\cite[(41)]{Bris:23:Factoring-PRFBs-Causal}  gives~\eqref{CDF75_EEA_col0},
\begin{align}\label{CDF75_CCA_col0}
\mathbf{H}(z)
&=
\mathbf{V}_0(z)\mathbf{V}_1(z)\mathbf{V}_2(z)\bsy{\Delta}_2(z)\mathbf{D}_{\ssst -50,-1/50}\mathbf{V}_3(z)\,\mathbf{J}\nonumber\\
&=
-\mathbf{D}_{\ssst 50,1/50}\,\bigl(\lowergam{50,1/50}\!\mathbf{V}_0(z)\bigr)\bigl(\lowergam{50,1/50}\!\mathbf{V}_1(z)\bigr)\bigl(\lowergam{50,1/50}\!\mathbf{V}_2(z)\bigr)\bsy{\Delta}_2(z)\mathbf{V}_3(z)\,\mathbf{J} \nonumber\\
&=
-\mathbf{D}_{\ssst 50,1/50}\mathbf{U}_3(z)\mathbf{U}_2(z)\mathbf{U}_1(z)\bsy{\Lambda}_1(z)\mathbf{U}_0(z)\,\mathbf{J}\text{\ \ after relabeling per~\eqref{std_causal_form}}\nonumber\\
&\hspace{-2em}=
-\begin{bmatrix}
\sst 50 &\sst  0\\
\sst 0 &\sst  1/50
\end{bmatrix}
\hspace{-4pt}
\begin{bmatrix}
\sst 1 &\sst  (-13 + 3z^{-1})10^{-4}\\
\sst 0 &\sst  1
\end{bmatrix}
\hspace{-4pt}
\begin{bmatrix}
\sst 1 &\sst  0\\
\sst 725 + 125z^{-1} &\sst 1
\end{bmatrix}
\hspace{-4pt}
\begin{bmatrix}
\sst 1 &\sst  -(1 + 5z^{-1})10^{-2}\\
\sst 0 &\sst  1
\end{bmatrix}
\hspace{-4pt}
\begin{bmatrix}
\sst z^{-2} &\sst  0\\
\sst 0 &\sst  1
\end{bmatrix}
\hspace{-4pt}
\begin{bmatrix}
\sst 1 &\sst  0\\
\sst -4 + 20z^{-1} &\sst  1
\end{bmatrix}
\hspace{-4pt}
\begin{bmatrix}
\sst 0 &\sst  1\\
\sst 1 &\sst  0
\end{bmatrix} .
\end{align}

The factorization schema for this decomposition is
\begin{align}\label{CDF75_CCA_col0_schema}
\schema(\eqref{CDF75_CCA_col0}) = (L,0,0,0;\;L,0,1,0;\;L,0,0,0).
\end{align}

\subsubsection{Division in Column~1}\label{sec:CubicBSpline:CCA:Col1}  
\cite[Section~6.2]{Bris:23:Factoring-PRFBs-Causal} shows that  CCA factorization of~\eqref{CDF75} in column~1  yields~\eqref{CDF75_EEA_col1}.  In~\cite[Section~6.2.1]{Bris:23:Factoring-PRFBs-Causal} it is shown that division in columns~0 and~1 give the same first lifting step, so the CCA factorization schema is
\begin{align}\label{CDF75_CCA_col1_schema}
\schema(\eqref{CDF75_EEA_col1}) = (L,0,0,\{0,1\};\;L,0,1,1).
\end{align}

\subsubsection{Division in Row~0}\label{sec:CubicBSpline:CCA:Row0}  
Let  $\mathbf{Q}_0(z)\eqdef \mathbf{H}(z)$,  $E_0\leftarrow H_{00}$,  and $F_0\leftarrow H_{01}$.  Find an initial factorization of the form
\begin{equation}\begin{split}\label{CDF75_row0_CCA_step0_form}
\hspace{-1ex}\mathbf{Q}_0(z)
&=
\begin{bmatrix}
\sst  (3 + 5z^{-1} + 5z^{-2} + 3z^{-3})/32 &\,\sst  (-3 + 10z^{-1} - 3z^{-2})/8\\
\sst (1 + 6z^{-1} + z^{-2})/8 &\,\sst  -(1 + z^{-1})/2 
\end{bmatrix} 
=
\begin{bmatrix}
E_0 & F_0\\
E_1 & F_1\vspace{-1pt}
\end{bmatrix}
=
\begin{bmatrix}
R_0 & F_0\\
R_1 & F_1\vspace{-1pt}%
\end{bmatrix}
\hspace{-3pt}
\begin{bmatrix}
1 & 0\\
S & 1\vspace{-1pt}%
\end{bmatrix}
\end{split}\end{equation}
by dividing $F_0$ into $E_0$ using  the classical division algorithm,
\begin{equation}\begin{split}\label{CDF75_row0_CCA_step0}
E_0 = F_0S + R_0,\mbox{ where }S(z) = -(5 + z^{-1})/4,\;
R_0(z) = (-3 + 13z^{-1})/8.
\end{split}\end{equation}
Set $R_1 \leftarrow E_1 - F_1S = -1/2$.  The first factorization step, with schema $(R,0,0,0)$, is
\begin{align}
\mathbf{Q}_0(z) 
&=
\begin{bmatrix}
\sst  (-3 + 13z^{-1})/8 &\,\sst (-3 + 10z^{-1} - 3z^{-2})/8\\
\sst -1/2 &\,\sst  -(1 + z^{-1})/2 
\end{bmatrix}
\hspace{-4pt}
\begin{bmatrix}
\sst  1 &\sst  0\\
\sst  -(5 + z^{-1})/4 &\sst   1%
\end{bmatrix}
=
\mathbf{Q}_1(z)\mathbf{V}_0(z).\label{CDF75_row0_CCA_step0_matrix}
\end{align}
Reset $E_j\leftarrow F_j$, $F_j\leftarrow R_j$  and factor
\begin{equation}\begin{split}\label{CDF75_row0_CCA_step1_form}
\hspace{-0.5ex}\mathbf{Q}_1(z) 
&=
\begin{bmatrix}
\sst  (-3 + 13z^{-1})/8 &\,\sst (-3 + 10z^{-1} - 3z^{-2})/8\\
\sst -1/2 &\,\sst  -(1 + z^{-1})/2 
\end{bmatrix}
=
\begin{bmatrix}
F_0 & E_0\\
F_1 & E_1\vspace{-1pt}
\end{bmatrix}
=
\begin{bmatrix}
F_0 & R_0\\
F_1 & R_1\vspace{-1pt}%
\end{bmatrix}
\hspace{-3pt}
\begin{bmatrix}
1 & S\\
0 & 1\vspace{-2pt}%
\end{bmatrix}.
\end{split}\end{equation}
Divide  $F_0$ into  $E_0$ to get $E_0 = F_0S + R_0$, where
\begin{equation}\label{CDF75_row0_CCA_step1}
S(z) = (121 - 39z^{-1})/169\text{ and }R_0 = -18/169.
\end{equation}
Set $R_1 \leftarrow E_1 - F_1S = -8(3 + 13z^{-1})/169$; then the second lifting step, with schema $(R,0,1,0)$, is
\begin{align*}
\mathbf{Q}_1(z) 
&=
\begin{bmatrix}
\sst  (-3 + 13z^{-1})/8 &\!\sst -18/169\\  
\sst -1/2 &\!\sst  -8(3 + 13z^{-1})/169 
\end{bmatrix}
\hspace{-4pt}
\begin{bmatrix}
\sst  1 &\sst (121 - 39z^{-1})/169\\
\sst 0 &\sst 1
\end{bmatrix}
=
\mathbf{Q}_2(z)\mathbf{V}_1(z).
\end{align*}
Reset $E_j\leftarrow F_j$, $F_j\leftarrow R_j$  and factor
\begin{align*}
\mathbf{Q}_2(z) 
&=
\begin{bmatrix}
\sst  (-3 + 13z^{-1})/8 &\sst -18/169\\  
\sst -1/2 &\sst  -8(3 + 13z^{-1})/169 
\end{bmatrix}
=
\begin{bmatrix}
E_0 & F_0\\
E_1 & F_1\vspace{-1pt}
\end{bmatrix}
=
\begin{bmatrix}
R_0 & F_0\\
R_1 & F_1\vspace{-1pt}%
\end{bmatrix}
\hspace{-3pt}
\begin{bmatrix}
1 & 0\\
S & 1\vspace{-1pt}%
\end{bmatrix}.
\end{align*}
Divide  $F_0$ into  $E_0$ to compute $E_0 = F_0S + R_0$, where
\begin{equation}\label{CDF75_row0_CCA_step3}
S(z) = 169(3 - 13z^{-1})/144\text{ and }  R_0 = 0.
\end{equation}
Set $R_1 \leftarrow E_1 - F_1S = -169z^{-2}/18$.  The third step, with schema $(R,0,0,0)$,  is
\begin{align*}
\mathbf{Q}_2(z) 
&=
\mathbf{Q}_3(z)\bsy{\Delta}_2(z)\mathbf{V}_2(z)
=
\begin{bmatrix}
\sst  0 &\sst -18/169\\  
\sst -169/18 &\sst  -8(3 + 13z^{-1})/169 
\end{bmatrix}
\hspace{-4pt}
\begin{bmatrix}
\sst  z^{-2} &\sst   0\\
\sst  0 &\sst 1%
\end{bmatrix}
\hspace{-4pt}
\begin{bmatrix}
\sst  1 &\sst 0\\
\sst 169(3 - 13z^{-1})/144 &\sst 1
\end{bmatrix}.
\end{align*}

Factor
\begin{align*}
\mathbf{Q}_3(z) 
&=
\begin{bmatrix}
\sst  -18/169 &\,\sst  0\\
\sst 0 &\,\sst -169/18
\end{bmatrix}
\hspace{-4pt}
\begin{bmatrix}
\sst  1 &\sst  0\\
\sst 144(3 + 13z^{-1})/169^2 &\sst 1
\end{bmatrix}
\hspace{-4pt}
\begin{bmatrix}
\sst  0 &\,\sst 1 \\
\sst 1 &\,\sst 0
\end{bmatrix}
= 
\mathbf{D}_{1/\kappa,\,\kappa}\mathbf{V}_3(z)\mathbf{J},\text{ where }\kappa \eqdef -169/18.
\end{align*}
Putting everything together,
\begin{align*}
\mathbf{H}(z) = \mathbf{D}_{1/\kappa,\, \kappa}\mathbf{V}_3(z)\mathbf{J}\bsy{\Delta}_2(z)\mathbf{V}_2(z)\mathbf{V}_1(z)\mathbf{V}_0(z),
\end{align*}
and using~\cite[(43)]{Bris:23:Factoring-PRFBs-Causal} to move $\mathbf{J}$ to the right end of the cascade yields~\eqref{CDF75_EEA_row0}, with
\begin{align}\label{CDF75_CCA_row0_schema}
\schema(\eqref{CDF75_EEA_row0}) = (R,0,0,0;\;R,0,1,0;\;R,0,0,0).
\end{align}
%
%
%
%

\subsubsection{Division in Row~1}\label{sec:CubicBSpline:CCA:Row1}  
Let us use this case to highlight the non-uniqueness of schema. Referring to~\eqref{CDF75_row0_CCA_step0_form}, we see that
\begin{align*}
\deg\begin{vmatrix}
R_0 & F_0\\
R_1 & F_1\vspace{-1pt}%
\end{vmatrix}
&=
\deg|\mathbf{Q}_0| = 2  < \deg(F_0) + \deg(F_1) = 3.
\end{align*}
By Corollary~\ref{cor:DRC}(\ref{cor:DRC:3})  the  complement $(R'_0,R'_1)$ to $(F_0,F_1)$ for  $-z^{-2}$ given by division in row~1 of $\mathbf{Q}_0$ is the same as  $(R_0,R_1)$  computed  in~\eqref{CDF75_row0_CCA_step0},  yielding the same  lifting step~\eqref{CDF75_row0_CCA_step0_matrix}, with schema $(R,0,0,\{0,1\})$.  Per~\eqref{CDF75_row0_CCA_step1_form},
\begin{align}\label{CDF75_CCA_row1_step1_LDDRC_test}
\deg\begin{vmatrix}
F_0 & R_0\\
F_1 & R_1\vspace{-1pt}%
\end{vmatrix}
=
\deg|\mathbf{Q}_1| = 2 > \deg(F_0) + \deg(F_1)  = 1.
\end{align}
By Corollary~\ref{cor:DRC}(\ref{cor:DRC:3})  division in row~1 yields a \emph{different} causal complement to $(F_0,F_1)$ for  $-z^{-2}$ than~\eqref{CDF75_row0_CCA_step1}. Dividing $F_1$ into $E_1$ gives
$E_1 = F_1S + R_1$ for $S(z) = 1 + z^{-1},\; R_1 = 0$.
Set $R_0\leftarrow E_0 - F_0S = -2z^{-2}$ so that
\begin{align}\label{CDF75_CCA_row1_step1}
\mathbf{Q}_1(z) 
&=
\begin{bmatrix}
\sst  (-3 + 13z^{-1})/8 &\sst -2\\  
\sst -1/2 &\sst  0 
\end{bmatrix}
\hspace{-4pt}
\begin{bmatrix}
\sst  1 &\sst 0\\
\sst 0 &\sst z^{-2}
\end{bmatrix}
\hspace{-4pt}
\begin{bmatrix}
\sst  1 &\sst 1 + z^{-1}\\
\sst 0 &\sst 1
\end{bmatrix}
=
\mathbf{Q}_2(z)\bsy{\Delta}_1(z)\mathbf{V}_1(z).
\end{align}
$R_0$ and $R_1$ are both divisible by $z^{-2}$ so $(R_0,R_1)$ is also the unique causal complement to $(F_0,F_1)$ for $-z^{-2}$ that is degree-reducing modulo $M_1=2$ in $F_1$, i.e., $\deg(R_0) < \deg(F_1) + M_1$.
This means  the lifting step schema for the second step can be given as $(R,\{0,1,2\},1,1)$.
Now factor
\begin{align*}
\mathbf{Q}_2(z)
&=
\begin{bmatrix}
\sst  -2 &\sst 0\\
\sst 0 &\sst -1/2
\end{bmatrix}
\hspace{-4pt}
\begin{bmatrix}
\sst  1 &\sst (3 - 13z^{-1})/16\\
\sst 0 &\sst 1
\end{bmatrix}
\hspace{-4pt}
\begin{bmatrix}
\sst  0 &\sst 1\\
\sst 1 &\sst 0
\end{bmatrix}
=
\mathbf{D}_{\kappa^{-1}, \kappa}\mathbf{V}_2(z)\mathbf{J},\; \;\kappa \eqdef -1/2.
\end{align*}
Putting it all together,
$\mathbf{H}(z) = \mathbf{D}_{\kappa^{-1}, \kappa}\mathbf{V}_2(z)\mathbf{J}\bsy{\Delta}_1(z)\mathbf{V}_1(z)\mathbf{V}_0(z)$
and~\cite[(43)]{Bris:23:Factoring-PRFBs-Causal} yields~\eqref{CDF75_EEA_row1} with 
\begin{align}\label{CDF75_CCA_row1_schema}
\schema(\eqref{CDF75_EEA_row1}) = (R,0,0,\{0,1\};\;R,\{0,1,2\},1,1).
\end{align}

The  reader may have noticed that when $M_1=2$ the inequality~\eqref{CDF75_CCA_row1_step1_LDDRC_test} is reversed,
\[  \deg|\mathbf{Q}_1| = 2 < \deg(F_0) + \deg(F_1) + M_1  = 3.  \]
In this case Corollary~\ref{cor:DRC}(\ref{cor:DRC:3}) implies that we obtain the same causal complement, $(R_0,R_1)=(-2z^{-2},0)$, if we divide in row~0 of $\mathbf{Q}_1$, which the reader can verify by dividing $F_0$ into $E_0$ using the SGDA with $M_1=2$. This furnishes yet another valid schema for computing the second lifting step~\eqref{CDF75_CCA_row1_step1}, namely, $(R,2,1,0)$.

\subsection{Computational Complexity Comparison of CCA vs.\ EEA}\label{sec:CubicBSpline:Comparison}
As with LGT(5,3)~\cite[\S 2]{Bris:23:Factoring-PRFBs-Causal}, all four  lifting factorizations of CDF(7,5) generated by the EEA are reproduced by the CCA.  We now compare the computational cost of  the CCA and  EEA factorization methods.    Rather than counting multiplies or adds we  quantify  complexity in terms of  \emph{polynomial} operations, ignoring  efficiencies due to filter symmetries, the better to compare and contrast the  differences between the two approaches.

\subsubsection{Division in Column~0}\label{sec:CubicBSpline:Comparison:Col0}
The  operations involved in running the EEA in column~0 (Section~\ref{sec:CubicBSpline:EEA:Col0}) are as follows.
%
\begin{itemize}
\item 
Two polynomial divisions (P~divs), $r_i = q_i r_{i+1} + r_{i+2}$, to compute $q_0$ and $q_1$.

\item  One scalar-polynomial multiply (SP~mult) to compute  $q_2=r_2/r_3$ since $r_3$ is a unit.

\item
Multiply out $\mathbf{H}'(z)$ as in~\eqref{EEA_col0_def_H'},
\begin{align*}
\mathbf{H}'(z) 
&=
\begin{bmatrix}
q_0q_1q_2 + q_2 + q_0 & \;q_0q_1+1\\
q_1q_2+1 & q_1\vspace{1pt}  
\end{bmatrix}
\hspace{-3pt}
\begin{bmatrix}
r_3 & 0\\
0 & z^{-2}/r_3
\end{bmatrix}.
\end{align*}
This  costs three polynomial-polynomial multiplies (PP~mults), two polynomial-polynomial adds (PP~adds), and four scalar-polynomial multiplies (SP~mults). We ignore polynomial orders unless  they are monomials (e.g., $z^{-2}/r_3$), which are treated as scalars, and we ignore the cost of scalar-polynomial adds.

\item
Compute the final lifting step  $\mathbf{H}(z)=\mathbf{H'}(z)\mathbf{S}(z)$ as in~\eqref{S_CDF75_col0}: one PP~add and one P~div.

\item
The intertwining operations using~\cite[(41)]{Bris:23:Factoring-PRFBs-Causal}  in~\eqref{CDF75_EEA_col0} to put the factorization in standard form: three SP~mults. We ignore the cost of the single scalar-scalar divide, $\kappa_0/\kappa_1$.
\end{itemize}
%
Similarly, the operation count for the column~0 CCA factorization  in Section~\ref{sec:CubicBSpline:CCA:Col0} is as follows.
%
\begin{itemize}
\item 
Two P~divs, \eqref{CDF75_col0_CCA_step0} and \eqref{CDF75_col0_CCA_step1}, and one SP~mult~\eqref{CDF75_col0_CCA_step2} that are identical to  corresponding  EEA operations.

\item 
Two PP~mults, one SP~mult, and two PP~adds (since $E_1=-2$ in $\mathbf{Q}_2$) to complete calculation of causal complements, $R_1 \leftarrow E_1 - F_1S$.

\item
One SP~mult to factor  the gain scaling matrix and the last lifting step as in~\eqref{CDF75_col0_CCA_step3}.

\item
Three intertwining operations~\cite[(41)]{Bris:23:Factoring-PRFBs-Causal} (SP~mults)   to put the factorization in standard form. 
\end{itemize}
%
\begin{table}[tb]		
\vspace{-0.15in}
\caption{Complexity of EEA and CCA Factorizations for CDF(7,5).}\label{tab:CDF75_complexity}
\begin{center}
\begin{tabular}{lcccc}
Factorization	& PP~adds	& SP~mults	& PP~mults	& P~divs \\
\midrule		
col.~0 EEA 	& 3			& 8			& 3			& 3\\
col.~0 CCA	& 2			& 6			& 2			& 2\\
\midrule		
col.~1 EEA 	& 1			& 6			& 1			& 2\\
col.~1 CCA	& 2			& 4			& 2			& 1\\
\midrule		
row~0 EEA 	& 2			& 6			& 3			& 3\\
row~0 CCA	& 2			& 3			& 2			& 2\\
\midrule		
row~1 EEA 	& 1			& 5			& 1			& 2\\
row~1 CCA	& 2			& 2			& 2			& 1\\
\end{tabular}
\end{center}
\vspace{-2em}
\end{table}
%
See Table~\ref{tab:CDF75_complexity} for complexity results for EEA and CCA calculations in both rows and columns.

\subsubsection{Analysis of Complexity Comparisons}\label{sec:CubicBSpline:Comparison:Analysis}
These comparisons reveal the price the EEA  pays for  dividing in  one row or column  while ignoring the other. Once the EEA reaches a gcd, $r_n$, we form the matrix $\mathbf{H}'(z)$  defined  by the quotients $q_i$ and the augmentation factor $|\mathbf{H}|/r_n$; $\mathbf{H}'(z)$ agrees with $\mathbf{H}(z)$ in one row or column and has the same determinant.  This forces us to  evaluate the matrix product for $\mathbf{H}'(z)$ and use the Lifting Theorem, which involves another P~div, to obtain the final lifting step.
In contrast, by extracting a reduced-degree causal complement  at each step, the CCA spreads its computational cost  more evenly  across the factorization process.  By Lemma~\ref{lem:termination}, the final lifting step is obtained by  an SP~mult instead of a more costly P~div, which explains why the CCA consistently costs one P~div less than the EEA.   The asymptotic  cost for  an $N$-step lifting factorization is as follows; intertwining costs to put the factorizations into standard causal lifting form are similar.
\begin{itemize}
\item Extended Euclidean Algorithm
\begin{itemize}
\item  Compute  $N-1$ EEA quotients $q_i$: $N-2$ P~divs, 1 SP~mult.
\item  Evaluate $\mathbf{H}'(z)$:  multiply $N-1$ of the $\mathbf{M}_i$ matrices ($N-2$ times, roughly, two PP~mults plus two PP~adds) and multiply by a diagonal augmentation matrix (four SP~mults).  
\item  Compute the final lifting matrix: one PP~add and one P~div.
\item  Polynomial operation cost:  5 SP~mults, $2N-3$ PP~adds, $2N-4$ PP~mults, $N-1$ P~divs.
\end{itemize}
\item
Causal Complementation Algorithm
\begin{itemize}
\item  Compute  $N-1$ lifting filters: $N-2$ P~divs, 1 SP~mult.
\item  Compute $N-1$ causal complements: $N-2$ PP~mults, 1 SP~mult, and $N-2$ PP~adds.
\item  Compute the final lifting matrix: one SP~mult.
\item  Polynomial operation cost:  3 SP~mults, $N-2$ PP~adds, $N-2$ PP~mults, $N-2$ P~divs.
\end{itemize}
\end{itemize}
Evaluating $\mathbf{H}'(z)$ makes the marginal  cost of factoring off an additional lifting step using the EEA  one P~div, two PP~mults, and two PP~adds, which grows significantly faster than the marginal cost using the CCA (one P~div, one PP~mult, and one PP~add) as the number of  lifting steps increases.

\subsection{Other CCA Factorizations of  CDF(7,5)}\label{sec:CubicBSpline:Other}
We now use the CCA to derive lifting factorizations that are not  obtainable using the causal EEA.  The first,  a causal version  of the unimodular WS group lifting factorization~\eqref{CDF75_PWA_anal_lifting}, was presented in~\cite[Section~6.3]{Bris:23:Factoring-PRFBs-Causal} and uses the SGDA  with $M_0=1$.

\subsubsection{SGDA in Column~1 with $M_0=1,\,M_1=0$}\label{sec:CubicBSpline:Other:Col1M1}  
As shown in~\cite[Section~6.3]{Bris:23:Factoring-PRFBs-Causal}, factoring~\eqref{CDF75} using the CCA and performing generalized division with $M_0=1,\; M_1=0$ in the SGDA  produces the lifting factorization
\begin{align}
\mathbf{H}(z) 
&=
\begin{bmatrix}
\sst 2 &\sst  0\\
\sst 0 &\sst  1/2
\end{bmatrix}
\hspace{-4pt}
\begin{bmatrix}
\sst 1 &\sst  3(1 + z^{-1})/16\\
\sst 0 &\sst  1
\end{bmatrix}
\hspace{-4pt}
\begin{bmatrix}
\sst  z^{-1} &\sst   0\\
\sst  0 &\sst   1
\end{bmatrix}
\hspace{-4pt}
\begin{bmatrix}
\sst 1 &\sst  0\\
\sst -(1 + z^{-1}) &\sst  1
\end{bmatrix}
\hspace{-4pt}
\begin{bmatrix}
\sst  1 &\sst   0\\
\sst  0 &\sst   z^{-1}
\end{bmatrix}
\hspace{-4pt}
\begin{bmatrix}
\sst 1 &\sst  -(1 + z^{-1})/4\\
\sst 0 &\sst  1
\end{bmatrix}
\hspace{-4pt}
\begin{bmatrix}
\sst  0 &\sst   1\\
\sst  1 &\sst   0
\end{bmatrix}.\label{CDF75_PWD_WSGLS_anal_lifting}
\end{align}
The lifting filters in~\eqref{CDF75_PWD_WSGLS_anal_lifting} are 
causally delayed versions of the noncausal linear phase lifting filters in the unimodular factorization~\eqref{CDF75_PWA_anal_lifting}.  This factorization was \emph{not} produced by the causal EEA  in Section~\ref{sec:CubicBSpline:EEA}, which can be predicted by observing that~\eqref{CDF75_PWD_WSGLS_anal_lifting} contains two distinct, first-order diagonal delay matrices, a result that cannot occur using the causal EEA.  
It is  well-conditioned: the   conditioning product~\eqref{condition_number_cascade_bound} for~\eqref{CDF75_PWD_WSGLS_anal_lifting} is 
$\cond(\mathbf{D}_{\kappa_0,\kappa_1})\textstyle{\prod}_i  \cond(\mathbf{U}_i) = 28$. 
Some left-handed (row reduction) factorization schema for~\eqref{CDF75_PWD_WSGLS_anal_lifting} are 
\[  \schema = (L,1,0,\{0,1\};\;L,\{0,1\},1,1)\text{\ \ and\ \ }(L,1,0,\{0,1\};\;L,1,1,0).  \]
The  reader can confirm that~\eqref{CDF75_PWD_WSGLS_anal_lifting} is also obtained using the  SGDA  with $M_0=1$ in row~1 of~\eqref{CDF75}.

\subsubsection{SGDA in Column~0 with $M_0=1,\,M_1=0$}\label{sec:CubicBSpline:Other:Col0M1}  
For further evidence of the power of the SGDA,  we now  use  the SGDA to  address the ill-conditioning  in factorization~\eqref{CDF75_EEA_col0},~\eqref{CDF75_CCA_col0}.  As in~\eqref{CDF75_col0_CCA_step0_form}  divide in column~0 of~\eqref{CDF75} to get an initial factorization of the form
%
\begin{align}
\hspace{-1ex}\mathbf{Q}_0(z)
&=
\begin{bmatrix}
\sst  (3 + 5z^{-1} + 5z^{-2} + 3z^{-3})/32 &\,\sst  (-3 + 10z^{-1} - 3z^{-2})/8\\
\sst (1 + 6z^{-1} + z^{-2})/8 &\,\sst  -(1 + z^{-1})/2 
\end{bmatrix} \label{CDF75_again}
=
\begin{bmatrix}
E_0 & E_1\\
F_0 & F_1\vspace{-1pt}
\end{bmatrix}
=
\begin{bmatrix}
1 & S\\
0 & 1\vspace{-1pt}%
\end{bmatrix}
\hspace{-3pt}
\begin{bmatrix}
R_0 & R_1\\
F_0 & F_1\vspace{-1pt}%
\end{bmatrix}.
\end{align}
This time,  use the SGDA with $M_0=1$ to divide $E_0 = F_0S + R_0$ 
with  $R_0$ divisible by $z^{-1}$,
\begin{align*}
S(z) = 3(1 + z^{-1})/4,\text{\ \ where\ \ }
R_0 &= -z^{-1}(1+z^{-1})/2 \text{\ \ with\ \ } \deg(R_0) < \deg(F_0) + M_0 = 3.
\end{align*}

Define $R_1\leftarrow E_1 - F_1S = 2z^{-1}$. Note that  $z^{-1}$ divides $R_1(z)$ as well as $R_0(z)$, as guaranteed by the  Causal Complementation Theorem (Theorem~\ref{thm:CCT}), so the first lifting step can  be written
\begin{align}
\mathbf{Q}_0(z)
&=
\begin{bmatrix}
\sst  1 &\sst   3(1 + z^{-1})/4\\
\sst  0 &\sst   1\vspace{-1pt}%
\end{bmatrix}
\hspace{-4pt}
\begin{bmatrix}
\sst  z^{-1} &\sst   0\\
\sst  0 &\sst   1\vspace{-1pt}%
\end{bmatrix}
\hspace{-4pt}
\begin{bmatrix}
\sst  -(1 + z^{-1})/2 &\sst  2\\
\sst (1 + 6z^{-1} + z^{-2})/8 &\,\sst  -(1 + z^{-1})/2 
\end{bmatrix}
=
\mathbf{V}_0(z)\bsy{\Delta}_0(z)\mathbf{Q}_1(z)\label{CDF75_col0_SGDA_step0}.
\end{align}
Based on the bound $\deg |\mathbf{Q}_0| = 2 < \deg(F_0) + \deg(F_1) + M_0 = 4$, Corollary~\ref{cor:DRC}(\ref{cor:DRC:3}) implies that we would obtain the same lifting step~\eqref{CDF75_col0_SGDA_step0} by dividing  in column~1 using the SGDA with $M=1$.

For the next step, the classical division algorithm in column~0 of  $\mathbf{Q}_1$ gives
\begin{align*}
\mathbf{Q}_1(z)
&=
\begin{bmatrix}
\sst  -(1 + z^{-1})/2 &\,\sst  2\\
\sst (1 + 6z^{-1} + z^{-2})/8 &\,\sst  -(1 + z^{-1})/2 
\end{bmatrix}
=
\begin{bmatrix}
F_0 & F_1\\
E_0 & E_1\vspace{-1pt}
\end{bmatrix}
=
\begin{bmatrix}
1 & 0\\
S & 1\vspace{-1pt}%
\end{bmatrix}
\hspace{-3pt}
\begin{bmatrix}
F_0 & F_1\\
R_0 & R_1\vspace{-1pt}%
\end{bmatrix},
\end{align*}
where $S(z) = -(5 + z^{-1})/4$ and $R_0 = -1/2$.
Set $R_1\leftarrow E_1 - F_1S = 2$; the second lifting step is 
\begin{align}\label{CDF75_col0_SGDA_step1}
\mathbf{Q}_1(z)
=
\begin{bmatrix}
\sst  1 &\,\sst  0\\
\sst -(5 + z^{-1})/4 &\,\sst 1 
\end{bmatrix}
\hspace{-4pt}
\begin{bmatrix}
\sst  -(1 + z^{-1})/2 &\,\sst  2\\
\sst -1/2 &\,\sst  2 
\end{bmatrix}
=
\mathbf{V}_1(z)\mathbf{Q}_2(z).
\end{align}
%
Finally, factor $\mathbf{Q}_2 = \bigl[\begin{smallmatrix} E_0 & E_1\\ F_0 & F_1\end{smallmatrix}\bigr]$ by dividing $F_0$ into $E_0$ using the classical division algorithm,
\begin{align}
\mathbf{Q}_2(z)\label{CDF75_col0_SGDA_step2}
&=
\begin{bmatrix}
\sst  1 &\,\sst  1 + z^{-1} \\
\sst 0 &\,\sst 1 
\end{bmatrix}
\hspace{-4pt}
\begin{bmatrix}
\sst  z^{-1} &\,\sst  0\\
\sst 0 &\,\sst  1 
\end{bmatrix}
\hspace{-4pt}
\begin{bmatrix}
\sst  0 &\,\sst  -2\\
\sst -1/2 &\,\sst  2 
\end{bmatrix}
=
\mathbf{V}_2(z)\bsy{\Delta}_2(z)\mathbf{Q}_3,\text{ where} \\
\mathbf{Q}_3\label{CDF75_col0_SGDA_step3}
&=
-\begin{bmatrix}
\sst  2 &\,\sst  0\\
\sst 0 &\,\sst  1/2 
\end{bmatrix}
\hspace{-4pt}
\begin{bmatrix}
\sst  1 &\, \sst 0 \\
\sst -4 &\,\sst 1 
\end{bmatrix}
\hspace{-4pt}
\begin{bmatrix}
\sst  0 &\,\sst  1\\
\sst 1 &\,\sst  0 
\end{bmatrix}
=
-\mathbf{D}_{\ssst 2,1/2}\mathbf{V}_3\,\mathbf{J}.
\end{align}
%
Combine~\eqref{CDF75_col0_SGDA_step0}--\eqref{CDF75_col0_SGDA_step3} and put in standard causal lifting form,
\begin{align}
\mathbf{H}(z) \label{CDF75_col0M1_SGDA}
&=
-\mathbf{V}_0(z)\bsy{\Delta}_0(z)\mathbf{V}_1(z)\mathbf{V}_2(z)\bsy{\Delta}_2(z)\mathbf{D}_{\ssst 2,1/2}\mathbf{V}_3(z)\,\mathbf{J}\nonumber\\
&=
-\mathbf{D}_{\ssst 2,1/2}\,\bigl(\lowergam{2,1/2}\!\mathbf{V}_0(z)\bigr)\bsy{\Delta}_0(z)\bigl(\lowergam{2,1/2}\!\mathbf{V}_1(z)\bigr)\bigl(\lowergam{2,1/2}\!\mathbf{V}_2(z)\bigr)\bsy{\Delta}_2(z)\mathbf{V}_3(z)\,\mathbf{J}\text{\ \ by~\cite[(41)]{Bris:23:Factoring-PRFBs-Causal}}\nonumber\\
&=
-\begin{bmatrix}
\sst 2 &\sst  0\\
\sst 0 &\sst 1/2
\end{bmatrix}
\hspace{-4pt}
\begin{bmatrix}
\sst 1 &\sst  3(1 + z^{-1})/16\\
\sst 0 &\sst  1
\end{bmatrix}
\hspace{-4pt}
\begin{bmatrix}
\sst  z^{-1} &\sst   0\\
\sst  0 &\sst   1
\end{bmatrix}
\hspace{-4pt}
\begin{bmatrix}
\sst 1 &\sst  0\\
\sst -(5 + z^{-1}) &\sst  1
\end{bmatrix}
\hspace{-4pt}
\begin{bmatrix}
\sst 1 &\sst  (1 + z^{-1})/4\\
\sst 0 &\sst  1
\end{bmatrix}
\hspace{-4pt}
\begin{bmatrix}
\sst  z^{-1} &\sst   0\\
\sst  0 &\sst   1
\end{bmatrix}
\hspace{-4pt}
\begin{bmatrix}
\sst 1 &\sst  0\\
\sst -4 &\sst  1
\end{bmatrix}
\hspace{-4pt}
\begin{bmatrix}
\sst  0 &\sst   1\\
\sst  1 &\sst   0
\end{bmatrix}.
\end{align}
As with~\eqref{CDF75_PWD_WSGLS_anal_lifting},  factorization~\eqref{CDF75_col0M1_SGDA} is \emph{not} generated by  the causal EEA in \emph{any} row or column of~\eqref{CDF75}.  

Using  \eqref{lifting_step_cond_nr}, \eqref{S_infty}  as before, the condition numbers and conditioning product for~\eqref{CDF75_col0M1_SGDA} are
\begin{align}
\|U_0\|^2_{\infty}	&= 16	&\cond(\mathbf{U}_0)	&= 17.9443\nonumber \\
\|U_1\|^2_{\infty}	&= 1/4	&\cond(\mathbf{U}_1)	&= 1.64039\nonumber \\
\|U_2\|^2_{\infty}	&= 36	&\cond(\mathbf{U}_2)	&= 37.9737\nonumber \\
\|U_3\|^2_{\infty}	&= 9/64	&\cond(\mathbf{U}_3)	&= 1.45185\nonumber \\
&&\cond(\mathbf{D}_{\kappa_0,\kappa_1}) 			&= 4\nonumber \\
&&\cond(\mathbf{D}_{\kappa_0,\kappa_1}){\textstyle\prod_i} \cond(\mathbf{U}_i) &= 6.5\times 10^{3}.\label{CCA_col0_cascade_factor}
\end{align}
Compared to  factorization~\eqref{CDF75_EEA_col0}, \eqref{CDF75_CCA_col0}, factoring off a single diagonal delay factor of $z^{-1}$  with the first lifting step has  reduced the ill-conditioning~\eqref{EEA_col0_cascade_factor} by over eight orders of magnitude. This demonstrates a form of ``algebraic rigidity'' on the part of the matrix $\mathbf{H}(z)$ against having an initial left lifting step factored off with no concomitant reduction in the determinantal degree.
The factorization schema for~\eqref{CDF75_col0M1_SGDA} is
\begin{align}\label{CDF75_CCA_col0M1_schema}
\schema = (L,1,0,\{0,1\};\;L,0,1,0;\;L,\{0,1\},0,0)).
\end{align}
Notation for the first step, $(L,1,0,\{0,1\})$, indicates that we get the same initial  step whether we divide in column $\ell_0=0$ or  $1$, per Corollary~\ref{cor:DRC}(\ref{cor:DRC:3}).  This fails for the next step. The last step yields the same result whether $M_2=0$ or 1 by uniqueness of degree-reducing causal complements, but the  complement depends on whether we divide in column~0 or~1.

\subsubsection{SGDA in Row~0 with $M_0=1$}\label{sec:CubicBSpline:Other:Row0M1}  
Now apply the CCA to the poorly conditioned factorization~\eqref{CDF75_EEA_row0} using  the SGDA with $M_0=1$ in factorization schema  $(R,1,0,\{0,1\};\;R,0,1,0;\;R,\{0,1\},0,0).$
The resulting factorization is
\begin{equation}\begin{split}\label{CDF75_row0_M1_lifting}
\mathbf{H}(z) 
&=
\begin{bmatrix}
\sst  -2 &\sst  0\\
\sst  0 &\sst  -1/2%
\end{bmatrix}
\hspace{-4pt}
\begin{bmatrix}
\sst  1 &\sst  0\\
\sst  16/3 &\sst   1 %
\end{bmatrix}
\hspace{-4pt}
\begin{bmatrix}
\sst  1 &\sst  0\\
\sst  0 &\sst   z^{-1}%
\end{bmatrix}
\hspace{-4pt}
\begin{bmatrix}
\sst  1 &\,\sst -3(1 + z^{-1})/16\\
\sst 0 &\,\sst  1
\end{bmatrix}
\hspace{-4pt}
\begin{bmatrix}
\sst  1 &\,\sst 0\\
\sst (13 - 3z^{-1})/3 &\,\sst  1 
\end{bmatrix}
\hspace{-4pt}
\begin{bmatrix}
\sst  1 &\sst  0\\
\sst  0 &\sst   z^{-1}%
\end{bmatrix}
\hspace{-4pt}
\begin{bmatrix}
\sst  1 &\sst  -(1 + z^{-1})/4 \\
\sst 0 &\sst   1%
\end{bmatrix}
\hspace{-4pt}
\begin{bmatrix}
\sst  0 &\sst 1 \\
\sst  1 &\sst  0 %
\end{bmatrix}.
\end{split}\end{equation}
It is over an order of magnitude better conditioned than~\eqref{CDF75_EEA_row0},
\begin{align}
\cond(\mathbf{D}_{\kappa_0,\kappa_1}){\textstyle\prod_i} \cond(\mathbf{U}_i) &= 8.8\times 10^{3}\,.\label{CCA_row0_cascade_factor}
\end{align}

\section{Left Degree-Lifting CCA Factorizations}\label{sec:Left}
Now impose partial pivoting by restricting  to left lifting (or ``left CCA'') factorizations in which $\eta_n=L$ for all $n$ (i.e., row reductions only).  A left CCA factorization always \emph{left}-factors  the previous quotient, alternating the update characteristic  so the factorization is irreducible.  The partial quotients  in a left CCA factorization can  therefore be reconstructed  from the right partial products of the factored cascade.   Algorithm~\ref{alg:CCA}  undergoes several simplifications for left CCA factorizations: steps~\ref{alg:CCA:eta} and~\ref{alg:CCA:irreducible} disappear, as does the branching on $\eta_n$ in  Steps~\ref{alg:CCA:delta}, \ref{alg:CCA:ell}, \ref{alg:CCA:chi}, \ref{alg:CCA:Qnplus1}, \ref{alg:CCA:UpdateQn}, and \ref{alg:CCA:finalstep}. 

One goal of this paper is to  \emph{define}  ``lifting decompositions,'' a feature missing from~\cite{DaubSwel98}. While~\cite{DaubSwel98}  factored example  matrices, no general concept of  ``lifting decompositions'' was  formally defined. Needless to say, it is  difficult to prove theorems  about  undefined concepts. We now show that a standard causal lifting cascade can be obtained as a left CCA factorization if and only its right partial products satisfy a degree-increasing property that mirrors the left degree-reducing property~\eqref{defn:DegRedDownlift:left} in Definition~\ref{defn:DegRedDownlift}.  
For  cascades that satisfy this criterion  we  define a \emph{left lifting signature} that will be shown to determine the  cascade uniquely.
One can then determine whether a standard causal lifting cascade is a left CCA factorization  using  properties intrinsic to the cascade, which justifies using this characterization to \emph{define} (left degree-) lifting cascades.  Similar results, which we do not pursue here,  hold for \emph{right} lifting factorizations.
Analogous characterizations would be very awkward to state for CCA decompositions containing \emph{both} row and column reductions.

\subsection{Left Degree-Lifting Matrix Cascades}\label{sec:Left:Cascades}

\subsubsection{The Left Degree-Lifting Property and Left Lifting Signatures}\label{sec:Left:Cascades:Sigs}
Assume we are given a causal  PR matrix with a factorization in standard causal lifting form~\eqref{std_causal_form}.
As with factorization schema (Section~\ref{sec:PRFB:CCA:Schema}),  we focus on the coprimified matrix~\eqref{def_Q0},
\begin{align}\label{Q0_factorization}
\mathbf{Q}_0 = 
\mathbf{D}_{\kappa_0,\kappa_1}\mathbf{U}_{\! N-1}\,\bsy{\Lambda}_{N-1}\,\mathbf{U}_{\! N-2}\cdots\mathbf{U}_{1}\,\bsy{\Lambda}_1\,\mathbf{U}_0\,\Psub{0} .
\end{align}
Denote the $n^{\rm th}$ right partial product, $1\leq n\leq N$, of~\eqref{Q0_factorization} as
\begin{equation}\label{partial_products} 
\Psub{n}(z) \eqdef \mathbf{U}_{n-1}(z)\bsy{\Lambda}_{n-1}(z)\Psub{n-1}(z)
= \left[P^{(n)}_{ij}(z)\right]_{i,j}.
\end{equation}
The row vector in $\Psub{n}(z)$ that has  been updated by the $(n-1)^{\rm th}$ lifting step is 
$\Bigl(P^{(n)}_{\raisebox{1pt}{$\sst \chi$}_{\raisebox{-1pt}{$\ssst n-1$}}, 0} \,,\,
P^{(n)}_{\raisebox{1pt}{$\sst \chi$}_{\raisebox{-1pt}{$\ssst n-1$}}, 1}\Bigr)$,
whose row index  is  the update characteristic  $\updatechar{-1pt}{$\sst n-1$}$ of $\mathbf{U}_{n-1}$ and $\bsy{\Lambda}_{n-1}$.

\begin{defn}[Left Degree-Lifting Cascades]\label{defn:deg_lifting} 
$\Psub{n}(z)$ is \emph{left degree-lifting in column $j$} for $j\in\{0,1\}$ if it has coprime rows and the degree
of its entry in row~$\updatechar{-1pt}{$\sst n-1$}$, column~$j$, is  greater than that of the other entry in 
column~$j$:
\begin{equation}\label{deg_lifting}
\deg \Bigl(P^{(n)}_{\raisebox{1pt}{$\sst \chi$}_{\raisebox{-1pt}{$\ssst n-1$}},\, j}\Bigr)    
>
\deg\Bigl(P^{(n)}_{\raisebox{1pt}{$\sst \chi$}'_{\ssst n-1},\, j}\Bigr),\quad\upchi'_{n-1}\eqdef 1 - \updatechar{-1pt}{$\sst n-1$}.  
\end{equation}
The \emph{left degree signature} of $\Psub{n}(z)$ is the set of indices~$j$ for the columns in which it is left degree-lifting,
$\sig(\Psub{n})\subset\{0,1\}$.
$\Psub{n}$ is \emph{left degree-lifting} if $\sig(\Psub{n})\neq\emptyset$.
A causal lifting cascade~\eqref{Q0_factorization}  is called \emph{left degree-lifting}  if  $\Psub{n}$ is  degree-lifting ``through $\Psub{N-1}$,'' i.e., for $0\leq n\leq N-1$.  
Note that $\sig(\Psub{N})$ may be empty even if the cascade is left degree-lifting through $\Psub{N-1}$, e.g., for equal-length filter banks like Daub(4,4) (see \S \ref{sec:Left:Daub44}). 
More generally,~\eqref{std_causal_form} is  \emph{left degree-lifting} if the factorization of its coprimification, $\mathbf{Q}_0$, is left degree-lifting.
\end{defn}

\rem
Henceforth, ``degree-lifting'' (``degree signature'') will  mean \emph{left} degree-lifting (\emph{left} degree signature).
$\sig(\Psub{0})$ is defined via~\eqref{deg_lifting} with $\updatechar{-1pt}{$\sst -1$} \eqdef 1 - \updatechar{-1pt}{$\sst 0$}$ and is always nonempty.  If $\Psub{0}=\mathbf{I}$ and $\updatechar{-1pt}{$\sst -1$}=0$ 
then $P^{(0)}_{00}=1>P^{(0)}_{10}=0$ so  $\Psub{0}$  is degree-lifting in column~0.  In general,
\begin{equation}\label{sig_P0}
\sig(\Psub{0}=\mathbf{I}) = \{\updatechar{-1pt}{$\sst -1$}\},\quad \sig(\Psub{0}=\mathbf{J}) = \{\updatechar{-1pt}{$\sst 0$}\}.
\end{equation}
$\Psub{1}(z)$ involves just one lifting matrix and is always  degree-lifting in the column with the zero,
\begin{align}\label{P1}
\Psub{1}(z) &\eqdef   \mathbf{U}_0(z)\,\Psub{0}
=
\begin{bmatrix}
1 & U_0(z) \\
0 & 1
\end{bmatrix}
\text{ or }
\begin{bmatrix}
U_0(z) & 1 \\
1 & 0
\end{bmatrix}.
\end{align}
It is also  degree-lifting in the column with the lifting filter if and only if the  filter's degree is  positive.

Directly comparing the polyphase order-increasing property \cite[Definition~10]{Bris:10:GLS-I} for unimodular polyphase-with-advance liftings to~\eqref{deg_lifting} for causal PWD liftings is tricky, but~\eqref{deg_lifting}  is generally weaker than \cite[Definition~10]{Bris:10:GLS-I} since a causal lifting step can be degree-lifting without being polyphase order-increasing, as  in~\eqref{P1}  when $\deg \bigl(U_0\bigr)=0$.  Moreover, Definition~\ref{defn:deg_lifting} only requires that a cascade be degree-lifting through $\Psub{N-1}$ whereas  \cite[Theorem~1]{Bris:10:GLS-I}  requires that the \emph{final} lifting step  also be  polyphase order-increasing. This limits \cite[Theorem~1]{Bris:10:GLS-I} to \emph{unequal}-length filter banks, a limitation that does not apply to  the present paper.

\begin{defn}[Left Lifting Signatures]\label{defn:lifting_sig} 
The \emph{left lifting signature} of a left degree-lifting factorization
in standard causal lifting form~\eqref{std_causal_form}
is the following list of degree signatures  (Definition~\ref{defn:deg_lifting}),  determinantal degrees, and initial update characteristic:
\begin{align}\label{lifting_sig} 
\hspace{-0.5em}\liftsig \eqdef \bigl[\rho_0,\rho_1,c_0,c_1:\sig(\Psub{N-1});\deg|\bsy{\Lambda}_{N-2}|,
\sig(\Psub{N-2});\ldots\deg|\bsy{\Lambda}_{1}|,\sig(\Psub{1});\,\sig(\Psub{0}) : \updatechar{-1pt}{$\sst 0$}\bigr].\hspace{-0.5em}
\end{align}
\end{defn}

\rem  
Square brackets  distinguish signatures from schema. If $N=0$ then $\liftsig = [\rho_0,\rho_1,c_0,c_1]$, and  we omit $\rho_0,\,\rho_1,\,c_0,\,c_1$  if  they are all zero. $\deg|\bsy{\Lambda}_{N-1}|$ is redundant and is thus omitted from the signature,
\begin{equation}\label{deg_det_Lambda_Nminus1}
\deg|\bsy{\Lambda}_{N-1}|  = \deg|\mathbf{Q}_0| -  {\textstyle \sum_{i=1}^{N-2}}\deg|\bsy{\Lambda}_i|.
\end{equation}
The final product, $\Psub{N}(z) = \mathbf{D}^{-1}_{\kappa_0,\kappa_1}\mathbf{Q}_0(z)$, agrees with $\mathbf{Q}_0(z)$ up to gain normalizations,  and the proof of Theorem~\ref{thm:Conditional} (below) will show that the factors in a left degree-lifting cascade \eqref{Q0_factorization} are completely determined by its lifting signature~\eqref{lifting_sig}; i.e., that different cascades always have different signatures. For example, the four-step cascade~\eqref{CDF75_col0M1_SGDA}  is degree-lifting through $\Psub{3}$, and Theorem~\ref{thm:Conditional}  implies that  \eqref{CDF75_col0M1_SGDA} is the only left degree-lifting factorization of CDF(7,5) with  lifting signature
%
\begin{align*}
\liftsig(\eqref{CDF75_col0M1_SGDA}) = \bigl[\{0,1\};\,0,0;\,1,0;\,1 : 1\bigr] .
\end{align*}

\subsubsection{Equivalent Polyphase Matrix Decompositions}\label{sec:Left:Cascades:Equivalent}
Let  $\mathbf{Q}_0(z)$ have an irreducible left downlifting factorization
\begin{align}
\mathbf{Q}_0
&=
\mathbf{V}_{\!0}\,\bsy{\Delta}_0\cdots\mathbf{V}_{\!N-2}\,\bsy{\Delta}_{N-2}\,\mathbf{V}_{\!N-1}\,\mathbf{D}_{\kappa_0,\kappa_1}\,\Psub{0}\label{left_downlifting}\\
&=
\mathbf{D}_{\kappa_0,\kappa_1}\,\gamma^{-1}_{\kappa_0,\kappa_1}\!\mathbf{V}_{\!0}\,\bsy{\Delta}_0\,
\gamma^{-1}_{\kappa_0,\kappa_1}\!\mathbf{V}_{\!1}\,\bsy{\Delta}_1
\cdots\gamma^{-1}_{\kappa_0,\kappa_1}\!\mathbf{V}_{\!N-2}\,\bsy{\Delta}_{N-2}\,\gamma^{-1}_{\kappa_0,\kappa_1}\!\mathbf{V}_{\!N-1}\,\Psub{0}.\label{left_factored_form}
\end{align}
Equate~\eqref{left_factored_form} with~\eqref{Q0_factorization}: let $n'\eqdef N-1-n$ and identify
\begin{align}
\mathbf{U}_{n}(z)&= \gamma^{-1}_{\kappa_0,\kappa_1}\!\mathbf{V}_{\!n'}(z)\mbox{ for }0\leq n'\leq N-1,\label{def_Ui}\\
\bsy{\Lambda}_{n}(z)&=\bsy{\Delta}_{n'}(z)\mbox{ for }0\leq n'\leq N-2.\label{def_Li}
\end{align}
Substitution transforms~\eqref{left_factored_form} into a standard causal lifting cascade,
\begin{align}\label{lifting_form}
\mathbf{Q}_0 =
\mathbf{D}_{\kappa_0,\kappa_1}\,\mathbf{U}_{N-1}\,\bsy{\Lambda}_{N-1}\cdots\mathbf{U}_{1}\,\bsy{\Lambda}_{1}\,\mathbf{U}_0\,\Psub{0}.
\end{align}
Reversing the steps  transforms~\eqref{lifting_form} into  downlifting form~\eqref{left_downlifting}, creating a one-to-one correspondence between irreducible left downlifting factorizations~\eqref{left_downlifting} and irreducible left lifting cascades~\eqref{lifting_form}.

\begin{defn}\label{defn:Equivalence}
A left downlifting factorization~\eqref{left_downlifting} and a  standard causal  lifting cascade~\eqref{lifting_form} for $\mathbf{Q}_0$ are \emph{equivalent} if they have the same number of factors and satisfy~\eqref{def_Ui}--\eqref{def_Li}.
\end{defn}
\begin{lem}\label{lem:partial_prods_quots}
Let $\mathbf{Q}_0(z)$ be coprimified and let $\mathbf{Q}_n(z)$ be the right partial quotients~\eqref{nth_left_downlift} for a left downlifting factorization~\eqref{left_downlifting} of $\mathbf{Q}_0$.
If $\Psub{n}(z)$ are the right partial products~\eqref{partial_products} of the equivalent standard causal lifting cascade~\eqref{lifting_form} then
\begin{align}\label{partial_prods_quots}
\mathbf{Q}_{N-n}(z) = \mathbf{D}_{\kappa_0,\kappa_1}\mathbf{P}_{n}(z)\mbox{ for } 0\leq n\leq N.
\end{align}
\end{lem}

\emph{Proof:}  Induction on $n$. The case $n=0$ is~\eqref{def_QN}.

\emph{Case: $0<n\leq N$.} Assume $\mathbf{Q}_{N-(n-1)}=\mathbf{D}_{\kappa_0,\kappa_1}\mathbf{P}_{n-1}$:
\begin{align*}
\hspace{1.5in}\mathbf{Q}_{N-n}(z) 
&= 
\mathbf{V}_{\!N-n}(z)\bsy{\Delta}_{N-n}(z)\mathbf{Q}_{N-(n-1)}(z)\mbox{ by  \eqref{nth_left_downlift}}\\
&=
\mathbf{V}_{\!N-n}(z)\bsy{\Delta}_{N-n}(z)\mathbf{D}_{\kappa_0,\kappa_1}\mathbf{P}_{n-1}(z)\text{ by the induction hypothesis}\\
&=
\mathbf{D}_{\kappa_0,\kappa_1}\gamma^{-1}_{\kappa_0,\kappa_1}\!\mathbf{V}_{\!N-n}(z)\bsy{\Delta}_{N-n}(z)\Psub{n-1}(z)
	\mbox{ by~\cite[(41)]{Bris:23:Factoring-PRFBs-Causal}}\\
&=
\mathbf{D}_{\kappa_0,\kappa_1}\mathbf{U}_{n-1}(z)\bsy{\Lambda}_{n-1}(z)\Psub{n-1}(z)\mbox{ by~\eqref{def_Ui}--\eqref{def_Li}} \\
&= \mathbf{D}_{\kappa_0,\kappa_1}\Psub{n}(z)\text{ by~\eqref{partial_products}. \hspace{2.7in}\qed}
\end{align*}

We now show that  left CCA \emph{degree-reducing} downlifting factorizations, whose right partial quotients $\mathbf{Q}_{n+1}(z)$ all satisfy~\eqref{defn:DegRedDownlift:left}, are equivalent to left \emph{degree-lifting} cascades (Definition~\ref{defn:deg_lifting}).  

%
\begin{thm}[Equivalence Theorem]\label{thm:Equivalence}
Let $\mathbf{Q}_0(z)$ be coprimified.  Each left  downlifting factorization~\eqref{left_downlifting} of $\mathbf{Q}_0(z)$ with  degree-reducing right partial quotients $\mathbf{Q}_{n+1}(z)$ satisfying~\eqref{defn:DegRedDownlift:left} for $0\leq n\leq N-1$ is equivalent to a left degree-lifting cascade.  Conversely, every left degree-lifting  cascade is equivalent to a left  degree-reducing CCA factorization.
\end{thm}

\emph{Proof:}  Let~\eqref{left_downlifting} be degree-reducing with equivalent causal lifting cascade~\eqref{lifting_form}.  To prove that~\eqref{lifting_form} is degree-lifting we need to show that there exist indices $j=j(n)$ such that~\eqref{deg_lifting} is satisfied for $0\leq n\leq N-1$.

Set $\upchi_n\eqdef\upchi(\mathbf{U}_{n})$ and $i_n\eqdef\upchi(\mathbf{V}_{\!n})$, $0\leq n\leq N-1$.  By~\eqref{def_Ui}, $\mathbf{U}_{n-1}(z)=\gamma^{-1}_{\kappa_0,\kappa_1}\!\mathbf{V}_{\!N-n}(z)$ so $\upchi_{n-1}=i_{N-n}$.  Let $n'\eqdef N-1-n$ for $0\leq n,n'\leq N-1$; then Lemma~\ref{lem:partial_prods_quots} says  
\begin{align}\label{partial_product_identity}
\mathbf{P}_{n}(z) = \mathbf{D}_{\kappa_0,\kappa_1}^{-1}\mathbf{Q}_{N-n}(z) = \mathbf{D}_{\kappa_0,\kappa_1}^{-1}\mathbf{Q}_{n'+1}(z).
\end{align}
Since the cascades are irreducible,
$\upchi_{n-1} = i_{N-n} = i_{n'+1}  = i'_{n'} \eqdef 1-i_{n'}$ and $i_{n'} = \upchi'_{n-1}$.
For $0\leq n,n'\leq N-1$ set $j(n)\eqdef j_{n'}$, which is the column index in the degree-reducing property~\eqref{defn:DegRedDownlift:left} satisfied by $\mathbf{Q}_{n'+1}(z)$.  By~\eqref{partial_product_identity} and~\eqref{defn:DegRedDownlift:left}, for $0\leq n,n'\leq N-1$
%
\begin{align}
\label{deg-red_implies_deg-lift}
\deg \bigl(P^{(n)}_{\raisebox{1pt}{$\sst \chi$}_{\raisebox{-1pt}{$\ssst n-1$}} ,\, j(n)}\bigr) 
=
\deg\bigl(Q^{(n'+1)}_{i'_{n'} ,\, j_{n'}}\bigr) 
> \deg\bigl(Q^{(n'+1)}_{i_{n'} ,\, j_{n'}}\bigr)
=
\deg\bigl(P^{(n)}_{\raisebox{1pt}{$\sst \chi$}'_{\raisebox{-1pt}{$\ssst n-1$}} ,\, j(n)}\bigr).  
\end{align}
%
This establishes~\eqref{deg_lifting}, proving that~\eqref{lifting_form} is degree-lifting.

Conversely,  suppose  we are given a left degree-lifting cascade of the form~\eqref{lifting_form} and its equivalent  downlifting form~\eqref{left_downlifting}, with  $\mathbf{V}_{\!n}(z)$ and $\bsy{\Delta}_n(z)$ defined by~\eqref{def_Ui}--\eqref{def_Li}. Set $\mathbf{Q}_N\eqdef \mathbf{D}_{\kappa_0,\kappa_1}\Psub{0}$ and  use~\eqref{nth_left_downlift} to define $\mathbf{Q}_{N-1}(z),\ldots,\mathbf{Q}_1(z)$ recursively.  This ensures that the partial products $\Psub{n}(z)$ of~\eqref{lifting_form} and the ``quotient matrices'' $\mathbf{Q}_{n}(z)$ satisfy the hypotheses of Lemma~\ref{lem:partial_prods_quots}.

To prove that $\mathbf{Q}_{n+1}$ is degree-reducing we need  indices $j_n$ such that~\eqref{defn:DegRedDownlift:left} is satisfied for $0\leq n\leq N-1$.  The argument mirrors that used to establish~\eqref{deg-red_implies_deg-lift}.  Let $j(n)$ be the column indices for which the partial products $\Psub{n}$ satisfy~\eqref{deg_lifting},  set $j_n\eqdef j(n')$,  and $i_n\eqdef\upchi(\mathbf{V}_{\!n})$, $0\leq n,n'\leq N-1$.  By Lemma~\ref{lem:partial_prods_quots}, 
$\mathbf{Q}_{n+1} = \mathbf{D}_{\kappa_0,\kappa_1}\mathbf{P}_{n'}$.
Systematically interchange $n$ and $n'$  in~\eqref{deg-red_implies_deg-lift} and apply the degree-lifting hypothesis~\eqref{deg_lifting} to get
%
\begin{align}
\label{deg-lift_implies_deg-red}
\deg\bigl(Q^{(n+1)}_{i'_{n} ,\, j_{n}}\bigr) 
=
\deg \bigl(P^{(n')}_{\raisebox{1pt}{$\sst \chi$}_{\raisebox{-1pt}{$\ssst n'-1$}} ,\, j(n')}\bigr) 
>
\deg\bigl(P^{(n')}_{\raisebox{1pt}{$\sst \chi$}'_{\raisebox{-1pt}{$\ssst n'-1$}} ,\, j(n')}\bigr)  
=
\deg\bigl(Q^{(n+1)}_{i_{n} ,\, j_{n}}\bigr).
\end{align}
%
This proves~\eqref{defn:DegRedDownlift:left}. To prove that this equivalent decomposition~\eqref{left_downlifting} is identical to a left downlifting factorization generated by the CCA, consider the leftmost lifting step in~\eqref{left_downlifting}, $\mathbf{Q}_0 = \mathbf{V}_{\!0}\,\bsy{\Delta}_0\,\mathbf{Q}_1$, e.g.,
\begin{align*}
\mathbf{Q}_0 
&= 
\begin{bmatrix}
{E}_0 & {E}_1\\
{F}_0 & {F}_1\vspace{-1pt}
\end{bmatrix}
=
\begin{bmatrix}
1 & V_0\\
0 & 1\vspace{-1pt}
\end{bmatrix}
\hspace{-2pt}
\begin{bmatrix}
z^{-m_0} & 0\\
0 & 1\vspace{-1pt}
\end{bmatrix}
\hspace{-2pt}
\begin{bmatrix}
\wt{R}_0 & \wt{R}_1\\
{F}_0 & {F}_1\vspace{-1pt}
\end{bmatrix}
\end{align*}
for the case where $i_0\eqdef\upchi(\mathbf{V}_{\!0})=0$.  We have
$\deg({F}_{j_0}) = \deg(Q^{(1)}_{i'_0,\,j_0})  > \deg(Q^{(1)}_{i_0,\,j_0}) = \deg(\wt{R}_{j_0})$ by~\eqref{deg-lift_implies_deg-red} since $\mathbf{Q}_1$ is degree-reducing in column~$j_0$.  Thus,
\[  \deg(R_{j_0}) < \deg({F}_{j_0}) - \deg\gcd({F}_0,{F}_1) + m_0,  \]
where $\gcd({F}_0,{F}_1)=1$, so $(R_0,R_1)$ is a causal complement to $({F}_0,{F}_1)$ for $\hat{a}z^{-\hat{d}_0}$ that is degree-reducing modulo~$m_0$ in ${F}_{j_0}$.  By Corollary~\ref{cor:DRC} such a complement is unique.  Applying this argument to each lifting step in turn, we conclude that~\eqref{left_downlifting}  is identical to the degree-reducing downlifting factorization provided by Theorem~\ref{thm:CCT} in which the $n^{th}$  partial quotient is degree-reducing modulo $m_n$ in column~$j_n$.\hfill\qed

\rem  Since left degree-lifting cascades (Definition~\ref{defn:deg_lifting}) are in one-to-one correspondence with the left degree-reducing factorizations generated by the CCA, we regard Definition~\ref{defn:deg_lifting} as an intrinsic definition of \emph{causal  (left degree-) lifting factorizations},  which includes all column-wise causal EEA factorizations.

\subsection{Uniqueness Properties of Degree-Lifting Factorizations}\label{sec:Left:Unique}
Using Definition~\ref{defn:deg_lifting}, we can now prove  basic  results about  left degree-lifting cascades.  Previous work produced infinitely many unimodular lifting factorizations of the identity, e.g.,~\cite[Proposition~1 and Example~1]{Bris:10:GLS-I}, \cite[Example~1]{Bris:13b:TIT}, but as discussed in~\cite[\S V-A]{Bris:10:GLS-I}  the identity cannot be factored as a unimodular \emph{polyphase order-increasing}  cascade.  We now prove that the identity cannot be factored into a causal \emph{left degree-lifting}  cascade. Then we  prove that a left degree-lifting cascade is completely determined by its lifting signature, although the same filter bank may have other left liftings with different  signatures.

\begin{thm}\label{thm:NoIdentityLiftings}
There are no nontrivial left degree-lifting factorizations of the identity.
\end{thm}

\emph{Proof:}   Suppose we are given a left degree-lifting factorization of $\mathbf{I}$ in standard causal lifting form~\eqref{lifting_form}.   $\deg|\mathbf{I}|=0$  so the cascade  has no  diagonal delay matrices, implying it can be written
\begin{align}\label{factor_Id}
\mathbf{I} &= \mathbf{D}_{\kappa_0,\kappa_1}\mathbf{U}_{N-1}(z)\,\Psub{N-1}(z).
\end{align}
If $N=1$ then, since $\Psub{N-1}(z) = \Psub{0} = \mathbf{I}$ or $\mathbf{J}$, the product of the matrices in~\eqref{factor_Id} will contain \emph{three} nonzero  entries unless the lifting filter is $U_0=0$, implying 
$\mathbf{U}_0 = \Psub{0} = \mathbf{D}_{\kappa_0,\kappa_1} = \mathbf{I}.$

Now let $N\geq 2$.  Suppose that $\mathbf{U}_{N-1}(z)$ is upper-triangular, i.e., $\updatechar{-2pt}{$\sst N-1$}=0$. By~\eqref{factor_Id},
\begin{align*}
\begin{bmatrix}
1 & 0\\
0 & 1\vspace{-2pt}
\end{bmatrix} 
&=
\begin{bmatrix}
\kappa_0 & 0\\
0 & \kappa_1\vspace{-1pt}%
\end{bmatrix}
\hspace{-2pt}
\begin{bmatrix}
1 & U_{N-1} \\
0 & 1\vspace{-2pt}%
\end{bmatrix} 
\hspace{-2pt}
\begin{bmatrix}
{R}_0 & {R}_1\\
{F}_0 & {F}_1\vspace{-1pt}%
\end{bmatrix}
=
\begin{bmatrix}
1 & \kappa_0\kappa_1^{-1}U_{N-1} \\
0 & 1\vspace{-2pt}%
\end{bmatrix} 
\hspace{-2pt}
\begin{bmatrix}
\kappa_0 & 0\\
0 & \kappa_1\vspace{-1pt}%
\end{bmatrix}
\hspace{-2pt}
\begin{bmatrix}
{R}_0 & {R}_1\\
{F}_0 & {F}_1\vspace{-1pt}%
\end{bmatrix}\mbox{ by \cite[(41)]{Bris:23:Factoring-PRFBs-Causal}}\\
&=
\begin{bmatrix}
1 & \kappa_0\kappa_1^{-1}U_{N-1} \\
0 & 1\vspace{-2pt}%
\end{bmatrix} 
\hspace{-2pt}
\begin{bmatrix}
\kappa_0 {R}_0 & \kappa_0 {R}_1\\
\kappa_1 {F}_0 & \kappa_1 {F}_1\vspace{-1pt}%
\end{bmatrix}
=
\begin{bmatrix}
1 & \kappa_0\kappa_1^{-1}U_{N-1} \\
0 & 1\vspace{-2pt}%
\end{bmatrix} 
\hspace{-2pt}
\begin{bmatrix}
1 & \kappa_0 {R}_1\\
0 & 1\vspace{-1pt}%
\end{bmatrix}
\end{align*}
since $\kappa_1{F}_0=0$, implying   $\kappa_0{R}_0=1$, and  $\kappa_1{F}_1=1$. 
By irreducibility,  $\updatechar{-2pt}{$\sst N-2$} = 1$.
According to~\eqref{deg_lifting},  the matrix 
$\bigl[ \begin{smallmatrix} {R}_0 & {R}_1\\ {F}_0 & {F}_1\end{smallmatrix} \bigr] \eqdef \Psub{N-1}(z)$
is degree-lifting in column~$j$ iff
$\deg(P^{(N-1)}_{1j}) > \deg(P^{(N-1)}_{0j}),$
which fails for $j=0$ since $P^{(N-1)}_{10} = 0$. If it  holds for $j=1$,
\[  \deg(P^{(N-1)}_{11})  = 0 > \deg(P^{(N-1)}_{01}) = \deg({R}_1),  \]
then ${R}_1=0$. This forces $U_{N-1}=0$ and $\mathbf{U}_{N-1} = \mathbf{I}$, contradicting irreducibility of~\eqref{factor_Id}.
(The proof is similar when $\mathbf{U}_{N-1}$ is  lower-triangular.)  \qed
\begin{thm}\label{thm:Conditional}
Let $\mathbf{Q}_0(z)$ be a coprimified causal PR matrix;  
then distinct left degree-lifting factorizations of $\mathbf{Q}_0(z)$ necessarily have different  lifting signatures.
I.e., suppose we are given a left degree-lifting factorization  of  $\mathbf{Q}_0(z)$ in  standard causal lifting form and its  left lifting signature,
\begin{gather}
\mathbf{Q}_0(z) = 
\mathbf{D}_{\kappa_0,\kappa_1}\mathbf{U}_{\! N-1}(z)\,\bsy{\Lambda}_{N-1}(z)\cdots\mathbf{U}_0(z)\,\Psub{0}\,,\label{thm:Conditional:std_causal_form}\\
\liftsig =
\bigl[\sig(\Psub{N-1});\,\deg|\bsy{\Lambda}_{N-2}|,\,\sig(\Psub{N-2});
\ldots\deg|\bsy{\Lambda}_{1}|,\,\sig(\Psub{1});\,\sig(\Psub{0}) : \upchi_0 \bigr]\,.\label{thm:Conditional:lifting_sig}
\end{gather}
If another degree-lifting factorization~\eqref{thm:Conditional:std_causal_form_prime} of  $\mathbf{Q}_0(z)$  has the {same}  signature~\eqref{thm:Conditional:lifting_sig},  
\begin{align}\label{thm:Conditional:std_causal_form_prime}
\mathbf{Q}_0(z) = \mathbf{D}_{\kappa'_0,\kappa'_1}\mathbf{U}'_{\! N-1}(z)\,\bsy{\Lambda}'_{N-1}(z)\cdots\mathbf{U}'_0(z)\,\mathbf{P}'_{0}\,,
\end{align}
then corresponding matrix factors in~\eqref{thm:Conditional:std_causal_form_prime} and~\eqref{thm:Conditional:std_causal_form} are identical.
\end{thm}

\emph{Proof:} Induction on the common number of lifting steps, $N$, implicit in the lifting signature.

\emph{Case: $N=0$.}  Both  factorizations have the form
$\mathbf{Q}_0 = \mathbf{D}_{\kappa_0,\kappa_1}\Psub{0} = \mathbf{D}_{\kappa'_0,\kappa'_1}\mathbf{P}'_{\!0} $
with  empty lifting signature.  The only choices for $\Psub{0}$ and $\mathbf{P}'_{\!0}$ are $\mathbf{I}$ or  $\mathbf{J}$ so it  follows that $\Psub{0}=\mathbf{P}'_{\!0}$ and $\mathbf{D}_{\kappa_0,\kappa_1} = \mathbf{D}_{\kappa'_0,\kappa'_1}$.

\emph{Case: $N>0$.}  Assume the theorem holds for   $N-1$ lifting steps.  Equate~\eqref{thm:Conditional:std_causal_form} and~\eqref{thm:Conditional:std_causal_form_prime},
\begin{equation}\label{thm:Weak:equate_sides}\begin{split}
\mathbf{Q}_0(z) 
= \mathbf{D}_{\kappa_0,\kappa_1}\mathbf{U}_{\! N-1}(z)\bsy{\Lambda}_{N-1}(z)\Psub{N-1}(z)
= \mathbf{D}_{\kappa'_0,\kappa'_1}\mathbf{U}'_{\! N-1}(z)\bsy{\Lambda}'_{N-1}(z)\mathbf{P}'_{\!N-1}(z).
\end{split}\end{equation}
The common initial update characteristic $\upchi_0$ in~\eqref{thm:Conditional:lifting_sig} and irreducibility of~\eqref{thm:Conditional:std_causal_form} and~\eqref{thm:Conditional:std_causal_form_prime}  imply that $\bsy{\Lambda}_{N-1}$ and $\bsy{\Lambda}'_{N-1}$ have the same update characteristic.  Equality of lifting signatures and~\eqref{deg_det_Lambda_Nminus1}   imply that they have the same determinantal degree, $\deg|\bsy{\Lambda}_{N-1}| = \deg|\bsy{\Lambda}'_{N-1}|$, so $\bsy{\Lambda}_{N-1} = \bsy{\Lambda}'_{N-1}$.  Define $m_{N-1} \eqdef \deg|\bsy{\Lambda}_{N-1}|$ and move the gain matrices in~\eqref{thm:Weak:equate_sides} from left to right using~\cite[(41)]{Bris:23:Factoring-PRFBs-Causal},
\begin{equation}\label{thm:Weak:equate_sides2}\begin{split}
\mathbf{Q}_0(z) 
= \gamma_{\kappa_0,\kappa_1}\mathbf{U}_{\! N-1}(z)\bsy{\Lambda}_{N-1}(z)\mathbf{D}_{\kappa_0,\kappa_1}\Psub{N-1}(z)
= \gamma_{\kappa'_0,\kappa'_1}\mathbf{U}'_{\! N-1}(z)\bsy{\Lambda}_{N-1}(z)\mathbf{D}_{\kappa'_0,\kappa'_1}\mathbf{P}'_{\!N-1}(z).
\end{split}\end{equation}
Relabel using~\eqref{def_Ui}, \eqref{def_Li}, and~\eqref{partial_prods_quots} so that~\eqref{thm:Weak:equate_sides2}  reads
\begin{equation}\label{thm:Weak:equate_sides3}
\mathbf{Q}_0(z) 
= \mathbf{V}_0(z)\bsy{\Delta}_0(z)\mathbf{Q}_1(z)
= \mathbf{V}'_0(z)\bsy{\Delta}_0(z)\mathbf{Q}'_1(z).
\end{equation}

Suppose  the update characteristic of $\mathbf{V}_0$ and $\mathbf{V}'_0$ is $i_0=0$  and write~\eqref{thm:Weak:equate_sides3} as
%
\begin{align*}
\begin{bmatrix}
1 & V_0\\
0 & 1\vspace{-1pt}
\end{bmatrix}
\hspace{-4pt}
\begin{bmatrix}
z^{-m_{N-1}} & 0\\
0 & 1\vspace{-1pt}
\end{bmatrix}
\hspace{-4pt}
\begin{bmatrix}
\wt{R}_0 & \wt{R}_1\\
{F}_0 & {F}_1\vspace{-1pt}
\end{bmatrix}
=
\begin{bmatrix}
1 & V'_0\\
0 & 1\vspace{-1pt}
\end{bmatrix}
\hspace{-4pt}
\begin{bmatrix}
z^{-m_{N-1}} & 0\\
0 & 1\vspace{-1pt}
\end{bmatrix}
\hspace{-4pt}
\begin{bmatrix}
\wt{R}'_0 & \wt{R}'_1\\
{F}_0 & {F}_1\vspace{-1pt}
\end{bmatrix},
\end{align*}
%
where the bottom rows in  $\mathbf{Q}_1$ and $\mathbf{Q}'_1$ necessarily agree.
By hypothesis, $\Psub{N-1}$ and $\mathbf{P}'_{\!N-1}$ (and thus $\mathbf{Q}_1$ and $\mathbf{Q}'_1$)  have coprime rows and are  degree-lifting,
$\deg(\wt{R}_j),\,\deg(\wt{R}'_j) < \deg({F}_j)$ for $j\in\sig(\Psub{N-1}) = \sig(\mathbf{P}'_{\!N-1})$.
Write the determinant of $\mathbf{Q}_0(z)$ as $|\mathbf{Q}_0(z)|=\hat{a}z^{-\hat{d}_0}$; then
both  $(\wt{R}_0,\wt{R}_1)$ and $(\wt{R}'_0,\wt{R}'_1)$ are degree-reducing  causal complements  in ${F}_j$, $j\in\sig(\Psub{N-1})$,  to $({F}_0,{F}_1)$ for inhomogeneity $\hat{a}z^{-(\hat{d}_0-m_{N-1})}$.  Such complements are unique by  Corollary~\ref{cor:DRC} so $(\wt{R}_0,\wt{R}_1)=(\wt{R}'_0,\wt{R}'_1)$, implying that $\mathbf{Q}_1 = \mathbf{Q}'_1$ and therefore that $\mathbf{V}_0=\mathbf{V}'_0$. (A similar argument  applies for $i_0=1$.)

The truncated factorizations with $N-1$ lifting steps,
$\mathbf{D}_{\kappa_0,\kappa_1}\Psub{N-1} = \mathbf{Q}_1 = \mathbf{Q}'_1 
= \mathbf{D}_{\kappa'_0,\kappa'_1}\mathbf{P}'_{\!N-1}$,
are partial products of~\eqref{thm:Conditional:std_causal_form} and~\eqref{thm:Conditional:std_causal_form_prime} with the same lifting signature, obtained by truncating~\eqref{thm:Conditional:lifting_sig}.  The induction hypothesis implies that $\Psub{0} = \mathbf{P}'_{\!0}$, that  $\mathbf{D}_{\kappa_0,\kappa_1} = \mathbf{D}_{\kappa'_0,\kappa'_1}$, and that
\begin{align*}
\mathbf{U}_n(z) = \mathbf{U}'_n(z) &\mbox{ and }\bsy{\Lambda}_n(z) = \bsy{\Lambda}'_n(z),\; 0\leq n\leq N-2.
\end{align*}
Since 
$\gamma_{\kappa_0,\kappa_1}\mathbf{U}_{\! N-1} = \mathbf{V}_0 = \mathbf{V}'_0 = \gamma_{\kappa'_0,\kappa'_1}\mathbf{U}'_{\! N-1}$ 
it follows that  $\mathbf{U}_{\! N-1} = \mathbf{U}'_{\! N-1}$.  \qed

\subsection{Case Study: The Daubechies 4-tap/4-tap Paraunitary Filter Bank}\label{sec:Left:Daub44}
Table~\ref{tab:Daub44}  lists all left degree-lifting factorizations of the paraunitary Daub(4,4) wavelet filter bank~\cite{Daub88,Daub92}, which has been normalized to have  $H_0(1)=1$.   
\begin{align}\label{Daub_fourfour} 
\mathbf{H}(z) 
&\eqdef
\begin{bmatrix}
\sst \left(1+\sqrt{3} + z^{-1}(3-\sqrt{3})\right)/8	& \sst \left(3+\sqrt{3} + z^{-1}(1-\sqrt{3})\right)/8 \vspace{1pt}\\
\sst \left(\sqrt{3}-1 - z^{-1}(3+\sqrt{3})\right)/4	& \sst \left(3-\sqrt{3} + z^{-1}(1+\sqrt{3})\right)/4\vspace{-1pt}%
\end{bmatrix} \;,\quad
|\mathbf{H}(z)| = z^{-1}\,.
\end{align}
All factorizations have  3 lifting steps and are degree-lifting through $\Psub{2}$, but $\Psub{3}$ is \emph{never} degree-lifting since Daub(4,4) is an equal-length filter bank. Completeness of the list follows from Theorem~\ref{thm:Equivalence} since the  schema exhaust all possible left CCA options for  the first two lifting  steps.

\begin{table*}[t]
\vspace{-0.15in}
\begin{center}
\caption{All Left Degree-Lifting Factorizations of the Daub(4,4) Paraunitary Filter Bank}\label{tab:Daub44}
\begin{tabular}{ll}
(Schema)$\big/$[Signature] & Factorization\\
\midrule		
%
\small $\begin{matrix}
(L,0,0,\{0,1\};\,L,\{0,1\},1,0)  \\[2pt]
[\{0,1\};\,1,0;\,1 : 0] 
\end{matrix}$ 
& \small $ 
\begin{bmatrix}
 (\sqrt{3}-1)/2  &  0\\
 0  &  1+\sqrt{3} \vspace{-2pt} 
\end{bmatrix}
\hspace{-5pt}
\begin{bmatrix}
1  &  -1\\
0  &  1 \vspace{-2pt}
\end{bmatrix}
\hspace{-5pt}
\begin{bmatrix}
1  &  0\\
(2-\sqrt{3} - \zinv\sqrt{3})/4  &  1 \vspace{-2pt}
\end{bmatrix}
\hspace{-5pt}
\begin{bmatrix}
1  &  0\\
0  &  \zinv \vspace{-2pt}
\end{bmatrix}
\hspace{-5pt}
\begin{bmatrix}
1  &  \sqrt{3}\\
0  &  1 \vspace{-2pt}
\end{bmatrix} $ \vspace{1pt} \\ 
\midrule		
%
\small $\begin{matrix}
(L,0,0,\{0,1\};\,L,\{0,1\},1,1) \\[2pt]
[\{0,1\};\,1,1;\,0:0]
\end{matrix}$  
& \small $ \hspace{-2ex} 
\begin{bmatrix}
(3-\sqrt{3})/2  &  0\\
0  &  \!\!-(3+\sqrt{3})/3
\end{bmatrix}
\hspace{-5pt}
\begin{bmatrix}
1  &  1/3\\
0  &  1 \vspace{-2pt}
\end{bmatrix}
\hspace{-5pt}
\begin{bmatrix}
1  &  0\\
(-6+3\sqrt{3} - \zinv\sqrt{3})/4  &  1 \vspace{-1pt}
\end{bmatrix}
\hspace{-5pt}
\begin{bmatrix}
1  &  0\\
0  &  \zinv \vspace{-2pt}
\end{bmatrix}
\hspace{-5pt}
\begin{bmatrix}
1  &  \sqrt{3}/3 \\
0  &  1 \vspace{-2pt}
\end{bmatrix}
\hspace{-5pt}
\begin{bmatrix}
0  &  1\\
1  &  0 \vspace{-2pt}
\end{bmatrix} $ \hspace{-3ex} 
\vspace{1pt} \\ 
\midrule		
%
\small $\begin{matrix}
(L,0,1,\{0,1\};\,L,\{0,1\},0,0) \\[2pt]
[\{0,1\};\,1,0;\,1:1]
\end{matrix}$  
& \small $
\begin{bmatrix}
(1-\sqrt{3})/2  &  0\\
0  &  1+\sqrt{3} \vspace{-1pt}
\end{bmatrix}
\hspace{-5pt}
\begin{bmatrix}
1  &  0\\
1  &  1 \vspace{-2pt}
\end{bmatrix}
\hspace{-5pt}
\begin{bmatrix}
1  &  -(2+\sqrt{3} + \zinv\sqrt{3})/4\\
0  &  1 \vspace{-2pt}
\end{bmatrix}
\hspace{-5pt}
\begin{bmatrix}
\zinv  &  0\\
0  &  1 \vspace{-2pt}
\end{bmatrix}
\hspace{-5pt}
\begin{bmatrix}
1  & 0 \\
\sqrt{3}  &  1 \vspace{-2pt}
\end{bmatrix}
\hspace{-5pt}
\begin{bmatrix}
0  &  1\\
1  &  0 \vspace{-2pt}
\end{bmatrix} $ \vspace{1pt} \\ 
\midrule		
%
\small $\begin{matrix}
(L,0,1,\{0,1\};\,L,\{0,1\},0,1) \\[2pt]
[\{0,1\};\,1,1;\,0:1]
\end{matrix}$  
& \small $
\begin{bmatrix}
(3-\sqrt{3})/6  &  0\\
0  &  3+\sqrt{3} \vspace{-1pt}
\end{bmatrix}
\hspace{-5pt}
\begin{bmatrix}
1  &  0\\
-1/3  &  1 \vspace{-2pt}
\end{bmatrix}
\hspace{-5pt}
\begin{bmatrix}
1  &  (6+3\sqrt{3} - \zinv\sqrt{3})/4\\
0  &  1 \vspace{-2pt}
\end{bmatrix}
\hspace{-5pt}
\begin{bmatrix}
\zinv  &  0\\
0  &  1 \vspace{-2pt}
\end{bmatrix}
\hspace{-5pt}
\begin{bmatrix}
1  & 0 \\
\sqrt{3}/3  &  1 \vspace{-2pt}
\end{bmatrix} $ \vspace{1pt} \\ 
\midrule		
%
\small $\begin{matrix}
(L,1,0,\{0,1\};\;L,0,1,0) \\[2pt]
[\{0,1\};\,0,0;\,1:0]
\end{matrix}$  
& \small $
\begin{bmatrix}
(3+\sqrt{3})/2  &  0\\
0  &  (3-\sqrt{3})/3 \vspace{-1pt}
\end{bmatrix}
\hspace{-5pt}
\begin{bmatrix}
1  &  1/3\\
0  &  1 \vspace{-2pt}
\end{bmatrix}
\hspace{-5pt}
\begin{bmatrix}
\zinv  &  0\\
0  &  1 \vspace{-2pt}
\end{bmatrix}
\hspace{-5pt}
\begin{bmatrix}
1  &  0 \vspace{-1pt}\\
(\sqrt{3} - \zinv(6+3\sqrt{3}))/4  &  1 \vspace{-1pt}
\end{bmatrix}
\hspace{-5pt}
\begin{bmatrix}
1  &  -\sqrt{3}/3 \\
0  &  1 \vspace{-2pt}
\end{bmatrix} $ \vspace{1pt} \\ 
\midrule		
%
\small $\begin{matrix}
(L,1,0,\{0,1\};\;L,0,1,1) \\[2pt]
[\{0,1\};\,0,1;\,0:0]
\end{matrix}$
& \small $
\begin{bmatrix}
-(1+\sqrt{3})/2  &  0\\
0  &  \sqrt{3}-1
\end{bmatrix}
\hspace{-5pt}
\begin{bmatrix}
1  &  -1\\
0  &  1 \vspace{-2pt}
\end{bmatrix}
\hspace{-5pt}
\begin{bmatrix}
\zinv  &  0\\
0  &  1 \vspace{-2pt}
\end{bmatrix}
\hspace{-5pt}
\begin{bmatrix}
1  &  0 \vspace{-1pt}\\
(\sqrt{3} + \zinv(2+\sqrt{3}))/4  &  1
\end{bmatrix}
\hspace{-5pt}
\begin{bmatrix}
1  &  -\sqrt{3} \\
0  &  1 \vspace{-2pt}
\end{bmatrix}
\hspace{-5pt}
\begin{bmatrix}
0  &  1\\
1  &  0 \vspace{-2pt}
\end{bmatrix} $ \vspace{1pt} \\ 
\midrule		
%
\small $\begin{matrix}
(L,1,1,\{0,1\};\;L,0,0,0) \\[2pt]
[\{0,1\};\,0,0;\,1:1]
\end{matrix}$
& \small $ \hspace{-1ex} 
\begin{bmatrix}
(3+\sqrt{3})/6  &  0\\
0  &  \sqrt{3}-3
\end{bmatrix}
\hspace{-5pt}
\begin{bmatrix}
1  &  0\\
-1/3  &  1 \vspace{-1pt}
\end{bmatrix}
\hspace{-5pt}
\begin{bmatrix}
1  &  0\\
0  &  \zinv \vspace{-2pt}
\end{bmatrix}
\hspace{-5pt}
\begin{bmatrix}
1  &  (\sqrt{3} + \zinv(6-3\sqrt{3}))/4\\
0  &  1 \vspace{-2pt}
\end{bmatrix}
\hspace{-5pt}
\begin{bmatrix}
1  & 0 \\
-\sqrt{3}/3  &  1 \vspace{-2pt}
\end{bmatrix}
\hspace{-5pt}
\begin{bmatrix}
0  &  1\\
1  &  0 \vspace{-2pt}
\end{bmatrix} $ \hspace{-3ex} 
\vspace{1pt} \\ 
\midrule		
%
\small $\begin{matrix}
(L,1,1,\{0,1\};\;L,0,0,1) \\[2pt]
[\{0,1\};\,0,1;\,0:1]
\end{matrix}$
& \small $
\begin{bmatrix}
(1+\sqrt{3})/2  &  0\\
0  &  \sqrt{3}-1
\end{bmatrix}
\hspace{-5pt}
\begin{bmatrix}
1  &  0\\
1  &  1 \vspace{-2pt}
\end{bmatrix}
\hspace{-5pt}
\begin{bmatrix}
1  &  0\\
0  &  \zinv \vspace{-2pt}
\end{bmatrix}
\hspace{-5pt}
\begin{bmatrix}
1  &  (\sqrt{3} - \zinv(2-\sqrt{3}))/4\\
0  &  1 \vspace{-2pt}
\end{bmatrix}
\hspace{-5pt}
\begin{bmatrix}
1  & 0 \\
-\sqrt{3}  &  1 \vspace{-2pt}
\end{bmatrix} $ 
%
\end{tabular}
\end{center}
\vspace{-0.15in}
\end{table*}

\section{Conclusions}\label{sec:Conclusions}
We have derived a new causal lifting  scheme,  the \emph{Causal Complementation Algorithm} (CCA), for arbitrary causal two-channel FIR PR  filter banks.  The CCA  generalizes  lifting   based on the Extended Euclidean Algorithm (EEA) as introduced by Daubechies and Sweldens for unimodular  FIR  filter banks.  The CCA reproduces all of the  factorizations generated by the causal  EEA and yields other  factorizations not generated by the causal EEA. The key  advantage of the CCA is its ability to factor out user-specified diagonal delay matrices at arbitrary points in the factorization process, a capability not provided by the causal EEA. This not only allows one to generate novel factorizations  but also appears (empirically) capable of generating \emph{well-conditioned} lifting factorizations in cases where the causal EEA fails to do so. In contrast to the unimodular lifting scheme of Daubechies and Sweldens, the causal context ensures uniqueness of degree-reducing causal complements (Corollary~\ref{cor:DRC}),  giving the user  precise knowledge of the degree-reducing possibilities at each step in a causal lifting factorization and a constructive method for computing them (Theorem~\ref{thm:CCT}).

Detailed examples of CCA factorizations are given, including favorable computational complexity comparisons with the EEA. To manage the considerable generality of the CCA, ``left degree-lifting factorizations'' are defined as factorizations  that employ only row reductions.
The class of left degree-lifting factorizations of $\mathbf{H}(z)$ includes the causal EEA lifting factorizations computed by dividing in the columns of $\mathbf{H}(z)$. We prove (Theorem~\ref{thm:Equivalence}) that a standard causal lifting cascade is the result of a left degree-reducing CCA factorization if and only if its right partial products satisfy  certain  degree-increasing conditions, which we encode in a numerical \emph{lifting signature}. This  defines causal left degree-lifting cascades intrinsically, in terms of  partial products, independent of how the factorizations are obtained.  Having  a formal definition allows us to prove theorems about left degree-lifting factorizations (Theorems~\ref{thm:NoIdentityLiftings} and~\ref{thm:Conditional}), something not possible previously.

Work in progress includes realization theory for causal degree-lifting factorizations and specializations  incorporating  group-theoretic concepts for factoring causal  linear phase filter banks.

\section*{Acknowledgments} 
This research did not receive any specific grant from funding agencies in the public, commercial, or not-for-profit sectors.
%
\appendix\section{Condition Numbers for Filter Banks}\label{app:Condition}
Let $\mathbf{A}$ be an invertible bounded linear operator on Hilbert space, and consider a linear system $\mathbf{A}\bsy{x}=\bsy{b}$.  Propagation of additive noise, $\mathbf{A}(\bsy{x}+\delta\bsy{x})=\bsy{b}+\delta\bsy{b}$, is bounded by the \emph{condition number} of $\mathbf{A}$~\cite{HornJohnson:85:Matrix-Analysis,GolubVanLoan:2013:Matrix-Computations,Strang:2016:Intro-Linear-Algebra},
\begin{align}\label{condition_number}
\frac{\|\delta\bsy{x}\|}{\|\bsy{x}\|} \leq\cond(\mathbf{A})\frac{\|\delta\bsy{b}\|}{\|\bsy{b}\|},
\text{\ \ where\ \ }\cond(\mathbf{A})\eqdef \|\mathbf{A}\|\|\mathbf{A}^{-1}\|,\;\;\|\mathbf{A}\|\eqdef \sup_{\|\bsy{x}\|=1}\|\mathbf{A}\bsy{x}\| .
\end{align}
If $\bsy{x}_n$ is the solution to a sequence (cascade) of  problems,
$\mathbf{A}_1\bsy{x}_1=\bsy{x}_0$, $\mathbf{A}_2\bsy{x}_2=\bsy{x}_1$,...,
\eqref{condition_number}  implies 
\begin{align}\label{condition_number_cascade_bound}
\frac{\|\delta\bsy{x}_n\|}{\|\bsy{x}_n\|} \leq
\cond(\mathbf{A}_n)\dotsm\cond(\mathbf{A}_1)\frac{\|\delta\bsy{x}_0\|}{\|\bsy{x}_0\|}.
\end{align}
In finite dimensions the matrix  norms of $\mathbf{A}$ and $\mathbf{A}^{-1}$ are given by 
\begin{align}\label{operator_norm}
\|\mathbf{A}\|_{\text{mat}}\eqdef \sup_{|\bsy{x}|=1}|\mathbf{A}\bsy{x}| = \sigma_{\text{max}}(\mathbf{A})\text{\ \ and\ \ }
\|\mathbf{A}^{-1}\|_{\text{mat}} = 1/\sigma_{\text{min}}(\mathbf{A}),
\end{align}
where $\sigma_{\text{max}}(\mathbf{A})$ and $\sigma_{\text{min}}(\mathbf{A})$ are the largest and smallest \emph{singular values} of   $\mathbf{A}$~\cite{HornJohnson:85:Matrix-Analysis,GolubVanLoan:2013:Matrix-Computations,Strang:2016:Intro-Linear-Algebra}. 
The singular values are the positive square roots of the eigenvalues of $\mathbf{A}^{\!\dagger}\mathbf{A}$ for adjoint $\mathbf{A}^{\!\dagger}\eqdef\overline{\mathbf{A}}^T$.  
The matrix-valued impulse response, $\mathbf{h}(n)$, of a  filter bank is the convolution kernel of a bounded linear operator  on vector-valued signals  $\bsy{x}\in\ell^2(\mathbb{Z},\mathbb{C}^2)$.  The discrete-time Fourier transform is a unitary mapping of $\ell^2(\mathbb{Z},\mathbb{C}^2)$ onto $L^2(\mathbb{T},\mathbb{C}^2)$, where $\mathbb{T}\eqdef\{z: |z|=1\}$.  The operator norm of  $\mathbf{h}$ on $\ell^2(\mathbb{Z},\mathbb{C}^2)$,
\[  \|\mathbf{h}\| \;\eqdef\; \sup_{\|\bsy{x}\|_{\ell^2}=1}\|\mathbf{h}*\bsy{x}\|_{\ell^2}\,,  \]
equals the operator norm of its matrix-valued transfer function $\mathbf{H}$ acting as a multiplier on $L^2(\mathbb{T},\mathbb{C}^2)$,
\begin{align}
\|\mathbf{h}\|  
\,=\, \|\mathbf{H}\| 
\,\eqdef\, \sup_{\|\bsy{X}\|_{L^2}=1}\|\mathbf{H}(z)\bsy{X}(z)\|_{L^2}\;.\nonumber
\end{align}
By modifying the frequency-domain characterization of  translation-invariant bounded linear operators on $L^2(\mathbb{R}^n)$~\cite[Theorem~I.3.18]{SteinWeiss:71:Fourier-Analysis-Euclidean}
for vector-valued discrete-time signals, one can show that $\|\mathbf{H}\|$ equals the supremum over $\mathbb{T}$ of the matrix  norms  of the $2\times 2$ transfer matrices $\mathbf{H}(z)$ acting on $\mathbb{C}^2$,
\begin{align}
\|\mathbf{H}\| 
\,=\, \sup_{|z|=1}\|\mathbf{H}(z)\|_{\text{mat}} 
\,=\, \sup_{|z|=1}\sigma_{\text{max}}(\mathbf{H}(z))\text{ by~\eqref{operator_norm}.} \label{transfer_matrix_norm}
\end{align}

Formula~\eqref{transfer_matrix_norm} yields the following useful theorem, which  lets us  bound
the condition numbers of lifting factorizations via~\eqref{condition_number_cascade_bound}.  This sharpens an approach taken in~\cite[\S6]{ZhuWicker:12:NearestNeighborLifting}.

\medskip
\noindent\textbf{Theorem~A.1.}
\begin{it}
Let $\mathbf{S}(z)$ be a FIR lifting matrix with lifting filter $S(z)$, and denote the norm of $S$ in $L^{\infty}(\mathbb{T})$ by $\|S\|_{\infty}\eqdef\sup\limits_{\ssst |z|=1}|S(z)|$.  
The condition number of $\mathbf{S}$ acting on $L^2(\mathbb{T},\mathbb{C}^2)$ is given by
\begin{align}\label{lifting_step_cond_nr}
\cond(\mathbf{S}) \,=\, 1 + \|S\|_{\infty}^2/2 + \sqrt{\|S\|_{\infty}^2 + \|S\|_{\infty}^4/4}\;,
\end{align}
which satisfies the bounds
\begin{align}\label{cond_nr_bounds}
1 + \|S\|^2_{\infty} \,\leq\, \cond(\mathbf{S}) \,\leq\, 2 + \|S\|^2_{\infty}\;.
\end{align}
\end{it}
\emph{Proof:} 
The characteristic polynomial of $\mathbf{S}^{\dagger\!}(z)\mathbf{S}(z)$ in both the lower- and upper-triangular cases is
\begin{align}\label{char_poly}
\det\left(\lambda\mathbf{I} - \mathbf{S}^{\dagger\!}(z)\mathbf{S}(z)\right)
\,=\,
\lambda^2 - \bigl(2 + |S(z)|^2\bigr)\lambda + 1
\,=\,
\det\left(\lambda\mathbf{I} - \bigl(\mathbf{S}^{-1}(z)\bigr)^{\!\dagger}\mathbf{S}^{-1}(z)\right)\,,
\end{align}
so $\mathbf{S}(z)$ and $\mathbf{S}^{-1}(z)$ have the same singular values $\sigma_i=\sqrt{\lambda_i}$.
The roots of~\eqref{char_poly} are
\begin{align}\label{eigenvalues}
\lambda = 1 + |S(z)|^2/2 \pm\sqrt{|S(z)|^2 + |S(z)|^4/4}\;.
\end{align}
Since $\sigma_{\text{max}}(\mathbf{S}^{-1}(z)) = \sigma_{\text{max}}(\mathbf{S}(z))$, it follows from~\eqref{condition_number}, \eqref{transfer_matrix_norm} and~\eqref{eigenvalues}  that
\begin{align*}
\cond(\mathbf{S}) 
\,\eqdef\, \|\mathbf{S}\|\|\mathbf{S}^{-1}\| 
\,=\, \sup_{|z|=1}\sigma^2_{\text{max}}(\mathbf{S}(z))
\,=\, \sup_{|z|=1}\Bigl(1 + |S(z)|^2/2 + \sqrt{|S(z)|^2 + |S(z)|^4/4}\Bigr),
\end{align*}
which implies~\eqref{lifting_step_cond_nr}.  The lower  bound in~\eqref{cond_nr_bounds} is obtained by dropping the $\|S\|_{\infty}^2$ term under the radical sign in~\eqref{lifting_step_cond_nr}; the upper bound follows by adding 1 under the radical sign.  \qed

\rem
For $S(z)=\sum s(n)z^{-n}$  we have the upper bound $\|S\|_{\infty} \leq \sum |s(n)|$.
The examples in  this paper all have first-order lifting filters, $S(z) = a + bz^{-1}$, 
for which this bound is sharp:
\begin{align}
\|S\|_{\infty}^2 \,=\, \sup_{t\in[0,2\pi)} \bigl|a + be^{-it}\bigr|^2
\,&=\, \sup_{t\in[0,2\pi)} \Bigl( |a|^2 + 2\Re\bigl(\bar{a}b e^{-it}\bigr) + |b|^2\Bigr) \nonumber\\
&=\, |a|^2 + \sup_{t\in[0,2\pi)}\! 2\Re\bigl(|\bar{a}b|e^{i\alpha} e^{-it}\bigr) + |b|^2\text{\quad where }\alpha=\arg(\bar{a}b) \nonumber\\
&=\, \bigl(|a| + |b|\bigr)^2\text{\quad when }t=\alpha\,.\label{S_infty}
\end{align}

\bibliography{CMBstring,CMBpubs,acad-press,elsevier,IEEE,Math-Soc,Misc,Oxbridge,prentice-hall,springer,standards,Theses,CMBcrossref}
\end{document}